\documentclass[12pt]{article}
\pdfoutput=1
\usepackage{booktabs}
\usepackage{amsmath}
\usepackage{amssymb}
\usepackage{hhline}
\usepackage[scale={.75,.75}]{geometry}
\usepackage{url}
\usepackage{caption}
\usepackage[numbers, sort&compress]{natbib}
\usepackage{epsfig}
\usepackage{lscape}
\usepackage{mathrsfs}
\usepackage{multirow}
\usepackage{color}
\usepackage{verbatim}
\usepackage{nicefrac} 
\usepackage{upgreek}
 \usepackage{bbm}
\usepackage{psfrag}
\usepackage{hyperref}

\newcommand{\T}{{\rm T}}

\newcommand{\p}{\textbf{p}}
\renewcommand{\k}{\textbf{k}}
\renewcommand{\l}{\textbf{l}}

\newcommand{\V}{\mathcal{V}}

\begin{document}
\date{\mbox{ }}

\title{ \vspace{-1cm}
%
\begin{flushright}
{\scriptsize \tt NIST-PHY-2015001, TUM-HEP-992/15}  
\end{flushright}
{\bf \boldmath 
 Effective Action for Cosmological Scalar Fields at Finite Temperature
 }
\\ [8mm]
}

\author{Yeuk-Kwan E. Cheung$^{a}$, Marco Drewes$^{b}$, 
Jin U Kang$^{a, c}$, Jong Chol Kim$^{a, c}$ \\  \\
{\normalsize \it$^a$  Department of Physics,  Nanjing University,}\\
{\normalsize \it 22 Hankou Road, Nanjing, China 210093}\\
{\normalsize \it$^b$ Physik Department T70, Technische Universit\"at M\"unchen,} \\{\normalsize \it James Franck Stra\ss e 1, D-85748 Garching, Germany}\\
{\normalsize \it$^c$ Department of Physics,  Kim Il Sung University,}\\
{\normalsize \it RyongNam Dong, TaeSong District, Pyongyang, DPR Korea}\\
}

\maketitle

\thispagestyle{empty}

\begin{abstract}
  \noindent Scalar fields appear in many theories beyond the Standard Model of particle physics. In the early universe, they are exposed to extreme conditions, including high temperature and rapid cosmic expansion.
Understanding their behavior in this environment is crucial to understand the implications for cosmology. 
We calculate the finite temperature effective action for the field expectation value in two particularly important cases, for damped oscillations near the ground state 
and for scalar fields with a flat potential.
We find that the behavior in both cases can in good approximation be described by a complex valued effective potential that yields Markovian equations of motion. 
Near the potential minimum, we recover the solution to the well-known Langevin equation.
For large field values we find a very different behavior, 
and our result for the damping coefficient differs from the expressions frequently used in the literature.  We illustrate our results in a simple scalar model, for which we give analytic approximations for the effective potential and damping coefficient.
We also provide various expressions for loop integrals at finite temperature that are useful for future calculations in other models.
\end{abstract}

\newpage
\begin{footnotesize}
\tableofcontents
\end{footnotesize}
\newpage
\section{Introduction}\label{intro}
\subsection{Scalar fields and the early universe}

Scalar fields play important roles in particle physics and cosmology. 
They appear in many theories beyond the Standard Model of particle physics (BSM). For instance, the low-energy effective description of string theory and other theories involving extra dimensions generally include numerous scalar fields, called moduli, that parameterize the properties of the compactified internal dimensions (see e.g. \cite{Douglas:2015aga, Denef:2007pq} for review). 
Scalar fields can drive cosmic inflation (inflaton) \cite{Starobinsky:1980te,Guth:1980zm,Linde:1981mu, ArmendarizPicon:1999rj}, explain the strong CP-problem (axion) \cite{Peccei:1977hh}, induce phase transitions \cite{Profumo:2007wc, Ashoorioon:2009nf, Patel:2011th, Espinosa:2011ax}, be Dark Matter candidates \cite{McDonald:1993ex, Burgess:2000yq, Bento:2000ah, Arias:2012az} or Dark Energy \cite{Wetterich:1987fm, ArmendarizPicon:2000dh, Kang:2007vs, Copeland:2006wr}. 
The recent discovery of a Higgs boson \cite{Aad:2012tfa, Chatrchyan:2012ufa} confirms that they indeed exist in nature.

In cosmology involving scalar fields, it is crucial to understand their time evolution in the early universe, where they are exposed to extreme conditions including high temperature, large energy density and rapid cosmic expansion. 
Their evolution in a time-dependent background provided by the primordial plasma and the cosmic expansion is a nonequilibrium process. 
The initial conditions for this process depend on the model with the specific scalar fields, and the mechanism that set the initial conditions is unknown in almost all models. 
Many observational hints point towards cosmic inflation, but alternative models such as bouncing universe scenarios (see \cite{Novello:2008ra, Brandenberger:2012zb} for review) have been also studied, though there are still controversies \cite{Kallosh:2007ad, Battefeld:2014uga}. The bouncing models also invoke scalar fields, typically inspired by string theory, to resolve the cosmic singularity by a bouncing phase in the very early universe \cite{Loewenfeld:2009aw, Li:2013bha}. 
In general, the field value at initial time is far away from the potential minimum.
If the effective potential is sufficiently steep, the field will relax to its minimum very quickly. However, there are numerous examples for scalar fields with a flat potential that evolve slowly compared to the time scale related to the propagation and interactions of individual particles, including the moduli, inflaton and axions. 
With the present work, focusing on the finite-temperature effects in a thermal bath, we aim to make progress towards a quantitative understanding of the nonequilibrium dynamics of scalar fields in the nontrivial backgrounds of the early universe.

\subsection{Motivations and assumptions}
Let us consider a real scalar quantum field $\phi$ with a symmetry under $\phi\rightarrow -\phi$ that couples to other degrees of freedom, which we collectively refer to as $\mathcal{X}_i$. At this stage we do not make any assumptions about the spin or interactions of the fields $\mathcal{X}_i$.
If we restrict ourselves to renormalizable operators, then the most general Lagrangian reads
\begin{equation}\label{L0}
\mathcal{L}=
\frac{1}{2}\partial_\mu\phi\partial^\mu \phi
-V(\phi)
-\phi^2 {\rm O} 
[\mathcal{X}_i]
+\mathcal{L}_{\mathcal{X}}
\end{equation}
with
\begin{equation}\label{V}
V(\phi)=\frac{m_\phi^2}{2}\phi^2+\frac{\lambda_\phi}{4!}\phi^4.
\end{equation}
Here ${\rm O}[\mathcal{X}_i]$ is a sum of operators that depend on combinations of the fields $\mathcal{X}_i$ only and $\mathcal{L}_{\mathcal{X}}$ is a Lagrangian that specifies their masses and interactions amongst each other.
The generalization to non-renormalizable operators is straightforward.
If we neglect the interactions with $\mathcal{X}_i$ for a moment, then the zero mode of $\phi$ in a Friedmann-Robertson-Walker spacetime would classically follow the equation of motion
\begin{eqnarray}
\ddot{\phi}+3H\dot{\phi} + \partial_\phi V(\phi)
= 0,\label{FieldEqOfMotClassical}
\end{eqnarray}
where $H$ is the Hubble constant. 
The full quantum field $\phi$  can be decomposed into 
\begin{equation}\label{EtaDef}
\phi=\varphi + \eta.
\end{equation} 
Here $\varphi$ is the expectation value of the field $\phi$, while the field $\eta$ contains quantum fluctuations, such as single particle excitations. 
If we switch on the interactions with the background medium and take into account radiative corrections,
one could expect $\varphi$ to follow an equation of motion of the form 
\begin{eqnarray}
\ddot{\varphi}+(3H+\Gamma_\varphi)\dot{\varphi} 
+ \partial_\varphi \V(\varphi)
= 0.\label{FieldEqOfMot}
\end{eqnarray}
Here $\V(\varphi)$ is an effective potential for $\varphi$ that includes radiative and thermal corrections, while $\Gamma_\varphi$ is the rate at which energy is dissipated from $\varphi$ to other degrees of freedom.
The equation (\ref{FieldEqOfMot}) looks physically intuitive, but should be rigorously justified when being applied. 
It cannot always be valid because the nonequilibrium dynamics of interacting quantum fields involves non-Markovian memory kernels. 
It is the goal of this work to derive an effective equation of motion of the form (\ref{FieldEqOfMot}) from first principles, and to determine $\V(\varphi)$ and $\Gamma_\varphi$ explicitly in an illustrative simple model. Together with the effective kinetic equation for the $\eta$-propagators derived in \cite{Drewes:2012qw}, this allows to understand the dynamics of the scalar field in terms of Markovian equations.
Our derivation of (\ref{FieldEqOfMot}) uses a somewhat different method than previous approaches in \cite{Morikawa:1986rp,Greiner:1998vd,Yokoyama:2004pf,Boyanovsky:2004dj,Anisimov:2008dz,Boyanovsky:2015xoa,Mukaida:2013xxa}, but yields a result that is consistent with previous works in the limit of small field values, where a comparison can be made. The explicit expression for $\Gamma_\varphi$ at large field values, however, is considerably different from those found in the literature.

In order to obtain a Markovian equation of motion for $\varphi$, we have to make several approximations.
As usual in the derivation of effective kinetic equations \cite{Berges:2005md}, these rely on a separation of time scales.
In closed systems, such a separation is automatically given if one assumes a sufficiently weak coupling. 
In this case the rates at which particles scatter are much smaller than their effective masses. This implies that macroscopic properties of the system change much slower than the time scale associated with individual scatterings. 
It also ensures that the system can be described as a dilute gas of well-defined quasiparticles 
with effective masses  $M_\eta$, $M_{\mathcal{X}_i}$ and widths $\Gamma_\eta$, $\Gamma_{\mathcal{X}_i}$ with $\Gamma_\eta\ll M_\eta$, $\Gamma_{\mathcal{X}_i}\ll M_{\mathcal{X}_i}$.
Here $M_\eta$ is the pole mass of the resummed $\eta$-propagator (the same for $M_{\mathcal{X}_i}$) 
and  in general effective mass and width  both depend on momentum\footnote{In thermal equilibrium the dispersion relations and widths of quasiparticles can be read off from the pole structure of the real time propagators \cite{LeB}. This definition can be generalized to situations far from equilibrium, see e.g. \cite{Drewes:2012qw}. The propagators in a medium can in principle have a complicated pole structure due to the appearance of collective excitations, such as luons \cite{Drewes:2013bfa}. We ignore this subtlety here.}.

In the early universe, Hubble expansion brings in an additional time scale. 
In the present work we make two basic assumptions
\begin{itemize}
\item[1)] The effective masses of all particles that couple to $\varphi$ are larger than the rate of Hubble expansion, i.e. $H< M_\eta, M_{\mathcal{X}_i} $.\footnote{This assumption is usually not fulfilled during inflation in the most popular models. It can, however, apply in several other applications that we have in mind, such as the late time behaviour of moduli, warm inflation scenarios and, more general, the behaviour of an order parameter in a hot plasma that is not in de Sitter space.}
\item[2)] The interactions in the primordial plasma are sufficiently weak to guarantee $\Gamma_\eta, \Gamma_\varphi \ll M_\eta, M_\varphi$ (and analogous for all other fields $\mathcal{X}_i$). Here $M_\varphi^2$ is the local curvature of $\V(\varphi)$. 
\item[3)] The change in the effective masses and widths on microscopic times scales $\sim 1/M_\eta, 1/M_{\mathcal{X}_i}$ should be small compared to these quantities themselves, i.e. $\dot{M}_\eta\ll M_\eta^2$, $\dot{\Gamma}_\eta\ll \Gamma_\eta M_\eta$ etc.
\end{itemize}
For practical purposes we make two additional assumptions that are not required for the derivation of a Markovian equation of motion, but lead to considerable simplifications
\begin{itemize}
\item[4)] $\phi$ is the only field with a non-zero expectation value, i.e. $\langle\mathcal{X}_i\rangle=0$. 
\item[5)] The thermodynamical state of the constituents $\mathcal{X}_i$ of the primordial plasma can be characterized by a single parameter, an effective temperature $T$. This is usually justified if $\Gamma_{\mathcal{X}_i}>H$.
\end{itemize}
\emph{Assumption 1)} allows to use Minkowski-space propagators in loop diagrams. 
Although $H$ is absolutely crucial in the kinetic equation for $\varphi$ on macroscopic time scales, it can be neglected when determining the transport coefficients in this equation from microphysics.
Without this assumption the computation of loops is technically very difficult \cite{Berera:2004kc, Brunier:2004sb, Garbrecht:2013coa, Gautier:2013aoa, Herranen:2015aja}.
Physically this assumption can be interpreted in terms of a separation of time scales: 
The microphysical time scales in processes described by loop diagrams (which are related to the inverse of the masses) must be much faster than the macroscopic Hubble expansion, such that the background metric is approximately constant during the duration of individual processes. 
\emph{Assumption 2)} is simply necessary for any perturbative treatment to be valid.
It also ensures that the effective masses (which include contributions that depend on $\varphi$ and $T$, which evolve with time) are approximately constant on the microscopic time scale that is relevant in loop integrals.
\emph{Assumption 3)} is the physical ``adiabaticity assmption'' that the properties of (quasi)particles do not change significantly within the duration of individual microscopic processes.
\emph{Assumption 4)} is for simplicity only, and it is straightforward to relax it. 
\emph{Assumption 5)} implies that $\mathcal{X}_i$ have reached some degree of kinetic equilibration. Though thermalization in general is a complicated problem \cite{Ellis:1987rw,Dodelson:1987ny,Enqvist:1990dp,Enqvist:1993fm,McDonald:1999hd,Davidson:2000er,Berges:2004ce,Berges:2010zv,Mazumdar:2013gya,Harigaya:2013vwa,Harigaya:2014waa}, this assumption seems at least qualitatively justified in the case under investigation here. If $\phi$ has a flat effective potential, then this means that it must have rather weak interactions (weaker than the interactions of the $\mathcal{X}_i$ amongst each other). This implies that the time scales $1/\Gamma_\varphi$, $1/H$ on which $\varphi$ evolves are much longer than the time scales $1/\Gamma_{\mathcal{X}_i}$ on which the $\mathcal{X}_i$ relax to local thermal equilibrium. 
Making the same assumption for $\eta$ is far less justified. For simplicity, we will nevertheless assume that the occupation numbers of $\eta$ modes can be characterized by the same effective temperature as the rest of the plasma. This assumption, which has been made in almost all past studies (often without mentioning), allows to use equilibrium propagators for all fields.

The setup defined by the assumptions 1)-5) is not sufficient for a complete understanding of scalar fields in the early universe. In particular assumptions 1) and 5) considerably constrain its applicability. Moreover, the properties of the primordial plasma in realistic models of particle physics is far more complicated than the toy model presented in section \ref{model}.
However, the analytic expressions we find are derived more systematically and consistently than any comparable results in the literature we are aware of. Moreover, in the course of their derivation we find analytic estimates for various integrals in finite-temperature field theory that will be very useful for calculations in more realistic models.
Finally, in spite of the comments above, there are at least two cosmological problems in which the assumptions 1)-5) are widely believed to be justified: warm inflation \cite{Berera:2008ar} and the fate of moduli at late times \cite{Kane:2015jia}.

\subsection{Overview of this article}

This paper is organized as follows. In section \ref{NQFT} we briefly review the elements of nonequilibrium quantum field theory to set up the theoretical framework and notations for the following sections. In section \ref{Effective} we derive the approximate effective action and the equation of motion for the expectation value $\varphi$ in a thermal bath.  These results are illustrated in a simple scalar model in section \ref{model}, and we present  approximate analytic estimates for effective potentials and dissipation coefficients. 
In section \ref{discussion} we summarize the results of this work and discuss some related issues in comparison with previous studies.  Section \ref{conclusion} is devoted to conclusions and outlook.  In the appendices we calculate the spectral self-energies from setting-sun diagrams, which are needed for the computation of the dissipation coefficients in section \ref{model}.

\section{Elements of nonequilibrium quantum field theory} \label{NQFT}

The standard methods to calculate S-matrix elements in particle physics 
are not suitable to describe systems far from equilibrium at large density
because there is no well-defined notion of asymptotic states, 
the properties of elementary excitations may significantly differ from those of particles in vacuum and classical particle number in general is not a suitable quantity to characterize the system. 

However, all observables can be expressed in terms of time-dependent correlation functions of the quantum fields without reference to asymptotic states or free particles. 
In most practical applications, all relevant information is contained in the one- and two-point functions.
The expectation value $\varphi$ can be identified with the one-point function, 
\begin{equation}\label{VarphiDef}
\varphi\equiv\langle\phi\rangle.
\end{equation}
That is, the average 
\begin{equation}\label{AverageDef}
\langle\ldots\rangle={\rm Tr}(\varrho \ldots)
\end{equation}
 is taken over quantum as well as statistical fluctuations encoded in the density operator $\varrho$.
We will in the following always assume that $\varphi=0$ in the thermodynamic ground state and study how the system relaxes to this state if $\varphi\neq0$ at initial time. 
There are two independent two-point functions. Common choices for these are the connected Wightman-functions
\begin{eqnarray}
\Delta_\eta^>(x_1,x_2)&\equiv& \langle\phi(x_1)\phi(x_2)\rangle -\varphi(x_1)\varphi(x_2), \\
\Delta_\eta^<(x_1,x_2)&\equiv& \langle\phi(x_2)\phi(x_1)\rangle - \varphi(x_1)\varphi(x_2)
\end{eqnarray}
or their linear combinations
\begin{eqnarray} \label{identify-delta}
\Delta_\eta^-(x_1,x_2)\equiv i\left(
\Delta_\eta^>(x_1,x_2)-\Delta_\eta^<(x_1,x_2)
\right) \ , \ \Delta_\eta^+(x_1,x_2)\equiv \frac{1}{2}\left(
\Delta_\eta^>(x_1,x_2)+\Delta_\eta^<(x_1,x_2)
\right).
\end{eqnarray}
These have a clear physical interpretation. The Fourier transform
\begin{equation}\label{rhodef}
\rho_\eta(p;x_1+x_2)\equiv -i\int d^4(x_1-x_2) \ e^{i p(x_1-x_2)}\Delta_\eta^-(x_1,x_2)
\end{equation}
of $\Delta^-$ is the \emph{spectral density} $\rho_\eta$, the poles of which in $p_0$ determine the spectrum of quasiparticles in a medium. The Fourier transform of $\Delta^+$, on the other hand, characterizes the occupation numbers of different field modes and can be interpreted as a field theoretical generalization of the classical particle distribution function, see e.g. discussion in \cite{Drewes:2012qw}. 
In most practical applications, $\varphi$, $\Delta^+$ and $\Delta^-$ contain all relevant informations.

\subsection{The Closed-Time-Path formalism}
\begin{figure}
	\center
	\includegraphics[width=12cm]{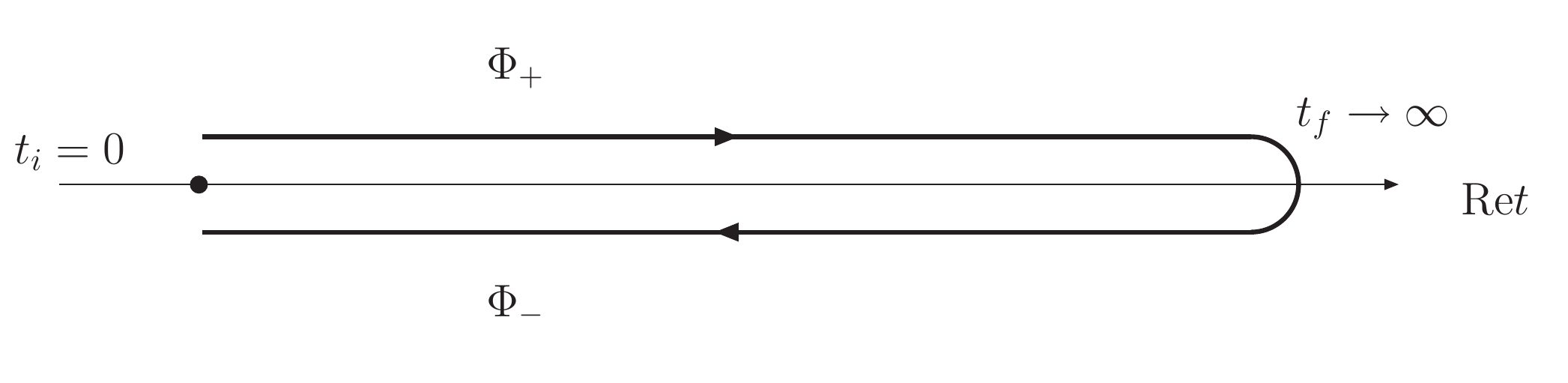}
\caption{The contour $\mathcal{C}$ in the complex time plane.\label{contour}}
\end{figure}

A convenient framework to obtain equations of motion for $\varphi$, $\Delta^+$ and $\Delta^-$ is offered by the Closed-Time-Path (CTP) formalism of nonequilibrium quantum field theory \cite{Schwinger:1960qe,Bakshi:1962dv,Bakshi:1963bn,Keldysh:1964ud}. 
In this formalism, correlation functions are defined with time arguments on the contour shown in figure~\ref{contour},
which starts from an initial time $t_i+i\epsilon$, runs parallel to the real axis to $t_f+i\epsilon$, turns around to $t_f-i\epsilon$ and returns to $t_i-i\epsilon$. We consider the limit $t_f\rightarrow\infty$, $\epsilon\rightarrow 0$ and $t_i\rightarrow -\infty$, where boundary conditions at finite time $t_0$ can be imposed by external sources localized at $t_0$ \cite{Drewes:2012qw}. 
In the following we define the quantities required for the present analysis. A more detailed introduction to the CTP-formalism can be found in \cite{Berges:2004yj,Chou:1984es}.

The standard strategy to calculate correlation functions is to split the field $\phi$ on the contour into $\phi^+$ and $\phi^-$, which denote the field with time argument on the ``forward'' and ``backward'' part of $\mathcal{C}$, see e.g. \cite{Chou:1984es}. 
One can define a  Feynman propagator $\langle \T_{\mathcal{C}}\,\phi(x_1)\phi(x_2) \rangle$, where $\T_{\mathcal{C}}$ indicates time ordering along the contour path $\mathcal{C}$.
It can be decomposed into the correlation functions for these fields as 
\begin{eqnarray}
\Delta_\eta^{++}(x_1,x_2)&\equiv& \langle\T\phi^+(x_1)\phi^+(x_2)\rangle -\varphi(x_1)\varphi(x_2), \\
\Delta_\eta^{+-}(x_1,x_2)&\equiv& \langle\phi^-(x_2)\phi^+(x_1)\rangle -\varphi(x_1)\varphi(x_2), \\
\Delta_\eta^{-+}(x_1,x_2)&\equiv& \langle\phi^-(x_1)\phi^+(x_2)\rangle -\varphi(x_1)\varphi(x_2), \\
\Delta_\eta^{--}(x_1,x_2)&\equiv& \langle\bar{\T}\phi^-(x_1)\phi^-(x_2)\rangle -\varphi(x_1)\varphi(x_2), 
\end{eqnarray}
where $\T$ is the usual time-ordering and $\bar{\T}$ is anti-time-ordering (because the backwards part of the contour runs from positive to negative infinity). 
Considering that time arguments on the forward branch are always ``earlier'' along the contour than those on the backward branch, one can easily identify
\begin{eqnarray}\label{identify}
\Delta^{+-}(x_1,x_2)=\Delta^<(x_1,x_2) \ ,  \ \Delta^{-+}(x_1,x_2)=\Delta^>(x_1,x_2).
\end{eqnarray}
The same decomposition can be performed for the self-energies, see appendix of \cite{Drewes:2010pf}.
Then the combination $\phi_c\equiv(\phi^+ + \phi^-)/2$ is identified with a physical field and $\phi_\Delta\equiv\phi^+ - \phi^-$ is treated as a response field. 
After integrating out the bath fields $\mathcal{X}_i$ and $\phi_\Delta$, one obtains an effective action for $\phi_c$, from which one can obtain its equation of motion. It can be expanded in powers of $\phi_c$. At lowest non-trivial order one finds the Langevin type equation (\ref{Langevin}), 
 including higher powers of $\phi_c$ in the effective action leads to nonlinear interactions between the different field modes (``multiplicative noise''), see e.g.\ \cite{Yokoyama:2004pf,Boyanovsky:2015xoa}.
By taking quantum and thermal averages over products of $\phi_c$, one can then calculate all correlation functions for $\phi$.

\subsection{Equation of motion for the background field $\varphi$}
In the present context we are, however, mainly interested in the one-point function $\varphi$. We can therefore simplify the procedure and directly write down  an effective action for $\varphi$ (instead of $\phi$) on the closed time path $\mathcal{C}$ \cite{Dolan:1973qd,Jackiw:1974cv,Calzetta:1986ey,Calzetta:1986cq},
\begin{eqnarray}
\Upgamma&=&
\sum_{n=1}^\infty\int_{\mathcal{C}}d^4x_1\ldots d^4x_n  
\frac{1}{n!}\Upgamma^{(n)}(x_1,\ldots,x_n)\varphi(x_1)\ldots\varphi(x_n).\label{Seff}
\end{eqnarray}
Here the label $_{\mathcal{C}}$ indicates that the time integral is to be taken along $\mathcal{C}$.  
The $\Upgamma^{(n)}(x_1,\ldots,x_n)$  are the 
amputated \emph{vertex functionals}. 
The expansion in (\ref{Seff}) is around the minimum at $\varphi=0$,
and the coefficients $\Upgamma^{(n)}$ are taken at $\varphi=0$ in the sense that they contain the field $\eta$ defined in (\ref{EtaDef}).
However, for a consistent expansion in the small couplings requires the use of full (resumed) propagators in the loops because the $T$- and $\varphi$-dependent corrections to  effective masses bring powers of the couplings into the denominator of the propagators (similar to the situation of gauge theories at high $T$ \cite{Braaten:1989mz}).
This is done in the 2PI formalism of non-equilibrium quantum field theory, see e.g. \cite{Berges:2004yj}.
Hence, the $\Upgamma^{(n)}$ implicitly depend on $\varphi$ via the effective masses.

With (\ref{V}) in mind, we  expand (\ref{Seff}) up to fourth power.
$\Upgamma^{(1)}$ is eliminated by the stationarity condition for vanishing external sources, 
\begin{equation}\label{StationarityCondition}
\frac{\delta\Upgamma}{\delta\varphi}=0. 
\end{equation}
$\Upgamma^{(2)}$ is simply the inverse dressed propagator
$-i\Upgamma^{(2)}(x_1,x_2)=\delta_{\mathcal{C}}(x_1-x_2)\left[\partial_{x_1}^2+m_\phi^2\right]+\Pi_{\mathcal{C}}(x_1,x_2)$.
$\Upgamma^{(3)}$ vanishes due to the symmetry $\phi\leftrightarrow-\phi$ in (\ref{V}) and $\Upgamma^{(4)}$ is the amputated four-point function. 
Due to energy-momentum conservation we can replace 
\begin{eqnarray}
\int_{\mathcal{C}}d^4x_1\ldots d^4x_4  
\frac{1}{4!}\Upgamma^{(4)}(x_1,\ldots,x_4)\varphi(x_1)\ldots\varphi(x_4)\nonumber\\
\rightarrow 
\int_{\mathcal{C}}d^4x_1d^4x_2  
\frac{1}{4!}\varphi(x_1)^2[\lambda_\phi + \tilde{\Pi}_{\mathcal{C}}(x_1,x_2)]\varphi(x_2)^2,
\end{eqnarray}
i.e. express the effective action in terms of an effective potential $\mathcal{V}[\varphi]\equiv-\delta^{(4)}(0)\Upgamma[\varphi]$.
The function $\tilde{\Pi}$ is, for example, given by the diagram in figure \ref{fish}.

We now perform a spatial Fourier transform to the mixed representation with arguments $(t,\p)$ and decompose $\varphi$ into modes $\p$, which in general are coupled to each other by the integrals in (\ref{Seff}).
We are interested in the case, in which only $\p=0$ mode of $\varphi$ significantly differs from zero, and we can thus perform the spatial integrals analytically. 
Using the stationarity condition (\ref{StationarityCondition}) we can find an equation of motion for $\varphi$ on $\mathcal{C}$, 
\begin{eqnarray}\label{NonMarkovian}
(\partial_t^2+m_\phi^2)\varphi(t) + \frac{\lambda_\phi}{3!}\varphi(t)^3 
+\int_{\mathcal{C}}dt' \Pi_{\mathcal{C}}(t,t')\varphi(t') 
+ \frac{1}{3!}\varphi(t)\int_{\mathcal{C}}dt' \tilde{\Pi}_{\mathcal{C}}(t,t')\varphi(t')^2=0.
\end{eqnarray}
Here we have suppressed the momentum index and all functions are to be evaluated at $\p=0$. 
Note that this equation does not contain a Brownian noise term $\xi$ on the RHS, as one might have expected.
The reason is that this term comes from thermal fluctuations, and the expectation (\ref{AverageDef}) that is taken in the definition (\ref{VarphiDef}) of $\varphi$ includes an average over thermal fluctuations. A noise therm $\xi$ does appear in the (Langevin type) equation of motion (\ref{Langevin}) for $\phi$ if one only traces over the bath degrees of freedom \cite{Morikawa:1986rp,Greiner:1998vd,Yokoyama:2004pf,Boyanovsky:2004dj,Anisimov:2008dz,Boyanovsky:2015xoa}. 
While these fluctuations cancel out in the equation for the one-point function $\varphi$, they crucially affect $n$-point correlation functions of $\phi$ with $n>1$.

\subsection{Equations of motion for the fluctuations $\eta$ and perturbation theory}
In order to calculate the coefficients $\Pi$ and $\tilde{\Pi}$ in (\ref{NonMarkovian}) one has to evaluate loop integrals with $\eta$ and $\mathcal{X}_i$ running in the loop. This requires knowledge of the propagators.
Their equations of motion are the Kadanoff-Baym equations (see \cite{KBE}), which can be derived in a similar manner as (\ref{NonMarkovian}), see e.g. \cite{Anisimov:2008dz}, 
\begin{eqnarray}
(\square_{1}+m_\eta^{2})\Delta_\eta^{-}(x_{1},x_{2}) &=& 
-\int d^{3}\textbf{x}'\int_{t_{2}}^{t_{1}} d t' 
\Pi_\eta^{-}(x_{1},x')\Delta_\eta^{-}(x',x_{2})\;,\label{KBE1}\\
(\square_{1}+m_\eta^{2})\Delta_\eta^{+}(x_{1},x_{2}) &=&
-\int d^{3}\textbf{x}'\int_{t_{i}}^{t_{1}} dt'
\Pi_\eta^{-}(x_{1},x')\Delta_\eta^{+}(x',x_{2})\nonumber\\
&& +\int d^{3}\textbf{x}'\int_{t_{i}}^{t_{2}} dt' 
\Pi_\eta^{+}(x_{1},x')\Delta_\eta^{-}(x',x_{2})\;\label{KBE2},
\end{eqnarray}
where $m_\eta$ is a vacuum mass of $\eta$-quanta\footnote{Note that $m_\eta$ is in general different from $m_\phi$ since it obtains $\varphi$-dependence due to self-interaction of $\phi$: For example, $m_\eta^2 = m_\phi^2 + \frac{\lambda_\phi}{2} \varphi^2$ 
for $\frac{\lambda_\phi}{4!} \phi^4$ self-interaction.}
and the self-energies are defined, equivalently to (\ref{identify-delta}), as
\begin{eqnarray}
\Pi_\eta^-(x_1,x_2)\equiv 
\Pi_\eta^>(x_1,x_2)-\Pi_\eta^<(x_1,x_2)
 \ , \ \Pi_\eta^+(x_1,x_2)\equiv -\frac{i}{2}\left(
\Pi_\eta^>(x_1,x_2)+\Pi_\eta^<(x_1,x_2)
\right).
\end{eqnarray}
The usual retarded self-energy can be identified with
\begin{equation}
\Pi^R_\eta(x_1,x_2)=\theta(t_1-t_2)\Pi_\eta^-(x_1,x_2).
\end{equation}

\paragraph{Approximate Markovian equations}  - Correlation functions and self-energies for the bath fields $\mathcal{X}_i$ can be defined analogously.
In general, the coupled second order integro-differential equations (\ref{KBE1}) and (\ref{KBE2}) are difficult to solve. However, in a weakly coupled theory, and in a background that changes slowly compared to the time scale of individual particle scatterings, the system can be described as a gas of quasiparticles. 
In a homogeneous medium, the spectral density  defined in (\ref{rhodef}) only depends on time, $\rho_\eta(p;x)=\rho_\eta(p;t)$.
Let $\hat{\Omega}_\p(t)$ be a pole of $\rho_\eta(p;t)$, with 
\begin{eqnarray}
\Omega_\eta(t)\equiv{\rm Re}\hat{\Omega}_\eta(t) \ {\rm  and} \ \Gamma_\eta(t)\equiv 2{\rm Im}\hat{\Omega}_\eta(t).
\end{eqnarray}
Both are in general time-dependent due to the interaction with the medium.
In weakly coupled theories one observes the hierarchy
\begin{equation}\label{quasiparticle}
\Gamma_\eta(t)\ll\Omega_\eta(t) .
\end{equation}
In Minkowski space this is enough to justify a ``quasiparticle approximation'', i.e. to understand most properties of the system by picturing it as a gas of weakly coupled quasiparticles.
In a time-dependent background we must make the additional assumption that the above quantities change adiabatically, leading to a more precise formulation of our \emph{assumption 3)},  
\begin{eqnarray}\label{adiabaticitycondition}
 \dot{\Omega}_\eta\ll \Omega_\eta^2, \Omega_{\mathcal{X}}\Omega_\eta \ 
{\rm and} 
\  \dot{\Gamma}_\eta\ll \Gamma_\eta \Omega_\eta, \Gamma_\eta \Omega_{\mathcal{X}}
\end{eqnarray}
Under these assumptions $\rho_\eta(p;t)$ at fixed $t$ features peaks of width $\sim\Gamma_\eta(t)$ at energies $p_0\simeq\pm\Omega_\eta(t)$, which can be interpreted as {\it quasiparticle}-resonances,
 and loop integrals are typically dominated by the regions near these poles.
Here (and in the following) we omit the spatial momentum index.
In this situation, one approximately finds \cite{Drewes:2012qw}
\begin{eqnarray}
\Delta_\eta^-(t_1,t_2)&\simeq& \frac{
{\rm sin}\left(
\int_{t_2}^{t_1}dt'\Omega_\eta(t')
\right)
e^{
-\frac{1}{2}\left|
\int_{t_2}^{t_1}dt' \Gamma_\eta(t')
\right|
}
}{
\sqrt{
\Omega_\eta(t_1)
\Omega_\eta(t_2)
}
}, \label{DeltaMinus}\\
\Delta_\eta^+(t_1,t_2)&\simeq&
 \frac{
{\rm cos}\left(
\int_{t_2}^{t_1}dt'\Omega_\eta(t')
\right)
e^{
-\frac{1}{2}\left|
\int_{t_2}^{t_1}dt' \Gamma_\eta(t')
\right|
}
}{
\sqrt{2
\Omega_\eta(t_1)
\Omega_\eta(t_2)
}
}\left(
1+2f(
t
)
\right)\big|_{t={\rm min}(t_1,t_2)}. \label{DeltaPlus}
\end{eqnarray}
Here $f(t)$ is a generalized distribution functions, which follows the Markovian equation of motion
\begin{eqnarray}
\partial_tf(t)=[ 1 + f(t)]
 \Gamma_\eta^<(t)
- f(t)\Gamma_\eta^>(t)
\end{eqnarray}
with the ``gain'' and ``loss'' terms\footnote{Here we ignore subtleties about finite initial time in definitions. }
\begin{equation}
\Gamma_\eta^\gtrless(t)\equiv i\frac{\Pi_\eta^\gtrless(p,t)}{2\Omega_\eta(t)}\Big|_{p_0=\Omega_\eta(t)},
\end{equation}
see also \cite{Garbrecht:2011xw}.

\paragraph{Special case - thermal equilibrium}   - If the background medium is in thermal equilibrium, the functions $\Delta_\eta^-$ and $\Pi_\eta^-$ are independent of the coordinate $x_1+x_2$ \cite{Anisimov:2008dz}. 
Moreover, the Fourier-transforms  of the self-energies in the coordinate $x_1-x_2$ are related by the Kubo-Martin-Schwinger (KMS) relation

\begin{eqnarray}
\Pi_\eta^<(p)=e^{-p_0/T}\Pi_\eta^>(p),
\end{eqnarray}
or equivalently
\begin{eqnarray}\label{KMS}
\Pi_\eta^+(p)=-i\left(\frac{1}{2}+f_B(p_0)\right)\Pi_\eta^-(p),
\end{eqnarray}
where $f_B(p_0)=(e^{p_0/T}-1)^{-1}$ is the Bose-Einstein distribution.
In this case, $\rho_\eta$ reads \cite{Anisimov:2008dz}
\begin{eqnarray}\label{spectralfunction2}
\rho_{\eta}(p)={-2{\rm Im}\Pi^R_{\eta}(p)+2p_0\epsilon\over 
(p_0^2-m_\eta^2-\p^2-{\rm Re}\Pi^R_{\eta}(p))^2+({\rm Im}\Pi^R_{\eta}(p)+p_0\epsilon)^2}, 
\end{eqnarray}
where $\epsilon$ is an infinitesimal parameter.
Note that the self-energies $\Omega_\eta$ and $\Gamma_\eta$ all depend on $T$.
It is clear from (\ref{spectralfunction2}) that, if condition (\ref{quasiparticle}) is fulfilled, the quasiparticle dispersion relation (or ``mass shell'') $\Omega_\eta$ is essentially fixed by ${\rm Re}\Pi^R_{\eta}(p)$ via the condition 
\begin{equation}\label{dispersionrelation}
p_0^2-\p^2-m_\eta^2-{\rm Re}\Pi^R_{\eta}(p)=0, 
\end{equation}
while ${\rm Im}\Pi^R_{\eta}(p)$ gives the thermal width,
\begin{equation}
\Gamma_\eta=-\mathcal{Z}\frac{{\rm Im}\Pi_\eta^R(\Omega_\eta)}{\Omega_\eta}
=\mathcal{Z}\frac{i\Pi_\eta^-(\Omega_\eta)}{2\Omega_\eta}
=\Gamma_\eta^>-\Gamma_\eta^<,
\end{equation}
which is just the difference of gain and loss term.
${\rm Re}\Pi^R_\eta(p)$ contains a zero temperature divergence that can be absorbed into the physical mass in the same way as in vacuum \cite{Boyanovsky:2004dj,Anisimov:2008dz}. 
In the following we interpret $m_\phi$ as the physical mass in vacuum, with ${\rm Re}\Pi^R_\eta(p)$ being finite. The effective mass $M_\eta$ can be defined as $\Omega_\eta$ for vanishing spacial momentum $\p=0$. 
It is very useful to fit a Breit-Wigner function to (\ref{spectralfunction2}) near the pole, which is parametrized by $\Gamma_\eta$ and $\Omega_\eta$, as 
\begin{equation}\label{BWphi}
\rho_{\eta}^{\rm BW}(p)=2\mathcal{Z}\frac{p_0\Gamma_{\eta}}{(p_0^2-\Omega_{\eta}^2)^2+(p_0\Gamma_{\eta})^2} + \rho_{\eta}^{\rm cont}(p).
\end{equation}
Here $\rho_{\eta}^{\rm cont}(p)$ is the continuous part of $\rho_{\eta}(p)$.
To obtain the correct residue, we introduced the parameter
\begin{equation}
\mathcal{Z}=\left[1-\frac{1}{2\Omega_\eta}\frac{\partial {\rm Re}\Pi^R_\eta(p)}{\partial p_0}\right]^{-1}_{p_0=\Omega_\eta}
.\end{equation}
For weak coupling, it is often sufficient to use the ``zero width limit'', 
\begin{equation}\label{rhofree}
\rho_\eta^{\rm 0}(p)=2\pi\mathcal{Z}{\rm sign}(p_0)\delta(p_0^2-\Omega_\eta^2) + \rho_{\eta}^{\rm cont}(p)
.\end{equation}
Other useful relations that follow directly from the definitions of the correlators are
\begin{eqnarray}
\Pi_\eta^-(p)&=&2i {\rm Im}\Pi_\eta^R(p),\label{UsefulRelation1}\\
\Delta_\eta^<(p)=f_B(p_0)\rho_\eta(p) \ &,& \ \Delta_\eta^>(p)=[1+f_B(p_0)]\rho_\eta(p),\label{UsefulRelation2}
\end{eqnarray}
and equivalently for the self-energies of other fields.
Note that all the above functions depend on $T$, which we have not made explicit here to simplify the notation. 
Correlators and self-energies for all other fields are defined analogously.

\paragraph{Perturbation theory} - The perturbative expansion for correlation functions in a medium can be performed in terms of Feynman diagrams, analogous to the vacuum case. One difference is that one has to use the propagators $\Delta^\gtrless$ or, often more conveniently, their transforms in Wigner space \cite{Garbrecht:2011xw}. Knowledge of $\Delta^\gtrless$ allows to determine all other correlation functions, in particular
\begin{eqnarray}
\Delta_\eta^{++}(x_1,x_2)   &=& \Delta_\eta^{+}(x_{1},x_{2}) 
- \frac{i}{2}{\rm sign}(t_1-t_2)\Delta_\eta^{-}(x_{1},x_{2}), \label{feynmanprop}\\
\Delta_\eta^{--}(x_1,x_2)   &=& \Delta_\eta^{+}(x_{1},x_{2}) 
+ \frac{i}{2}{\rm sign}(t_1-t_2)\Delta_\eta^{-}(x_{1},x_{2}),\\
\Delta_\eta^{-+}(x_1,x_2)   &=& \Delta_\eta^{+}(x_{1},x_{2}) 
- \frac{i}{2}\Delta_\eta^{-}(x_{1},x_{2}),\label{groesserinliste} \\
\Delta_\eta^{+-}(x_1,x_2)   &=& \Delta_\eta^{+}(x_{1},x_{2}) 
+ \frac{i}{2}\Delta_\eta^{-}(x_{1},x_{2}).\label{kleinerinliste}
\end{eqnarray}
Another difference from the vacuum case is the fact that $\phi^+$ and $\phi^-$ formally have to be treated as two different fields. There are no vertices that directly connect a $\phi^+$ line to a $\phi^-$ line, i.e.\ the vertices are always either of ``+'' type or ``-'' type, but the propagators (\ref{groesserinliste}) and (\ref{kleinerinliste}) mix the fields and can connect a ``-'' vertex to a ``+'' vertex and vice versa, see e.g. \cite{LeB} for a detailed discussion.

\section{Effective action and equation of motion for $\varphi$} \label{Effective}
The known equations (\ref{DeltaMinus}), (\ref{DeltaPlus}) and (\ref{feynmanprop})-(\ref{kleinerinliste}) allow to describe the dynamics of the field fluctuations in terms of the Markovian equations, which can be physically interpreted as effective Boltzmann equations for quasiparticles in the plasma. In the following sections we derive an approximate Markovian equation for the field $\varphi$, so that the system can  be completely described in terms of Markovian dynamics.

\subsection{Small field values: Brownian motion and Langevin dynamics}\label{SmallField}
We consider the case in which all degrees of freedom except $\varphi$ are in thermal equilibrium.
If $\phi$ at initial time is very close to its thermodynamic ground state, then the effect that this deviation has on the thermal bath (``back reaction'') is negligible. This in particular requires that $\varphi$ be small enough that we can use a quadratic approximation to $\V(\varphi)$, and the effective masses in the plasma are dominated by vacuum and thermal masses (rather than coupling to the background $\varphi$). Then the self-energies have the properties (\ref{KMS}). 
In this case the behavior of $\phi$ can be understood in the terminology of Brownian motion. 
Such systems have been studied by a number of authors \cite{Morikawa:1986rp,Greiner:1998vd,Yokoyama:2004pf,Boyanovsky:2004dj,Anisimov:2008dz,Boyanovsky:2015xoa,Miyamoto:2013gna,Bartrum:2014fla,Rigopoulos:2013exa}, and it was found that the zero mode in this case follow a Langevin equation \cite{Boyanovsky:2004dj,Anisimov:2008dz},
\begin{eqnarray}
\left(\partial_t^2 + M_\eta^2\right)\phi(t)+\int_0^t dt' \Pi_\eta^-(t-t')\phi(t')=\xi(t). \label{Langevin}
\end{eqnarray}
Here we have absorbed the local part of the $\eta$-self-energy into a slowly varying  $M_\eta^2$ on the left hand side, which therefore depends on $T$ and $\varphi$. We have also assumed that the dependence of $\Pi_\eta^-(t,t')$ on $t+t'$ is negligible. 
The validity of these approximations is restricted to the regime where the amplitude of $\varphi$-oscillations is sufficiently small that the effective masses are not dominated by the $\varphi$-dependent contributions. Otherwise the adiabaticity condition (\ref{adiabaticitycondition}) is violated. This is consistent because we consider small oscillations near the minimum. Note that we do not need to require $\dot{\Gamma}_\eta/\Gamma_\eta \ll H$ or $\dot{\Omega}_\eta/\Omega_\eta \ll H$ because the integral in (\ref{Langevin}) is cut off due to the finite support of $\Pi_\eta^-(t,t')$.
The size of this support is roughly given by the inverse of the (quasi)particle energies in the loop, 
but it is difficult to clearly identify the relevant scale in general \cite{Gautier:2012vh}.  
The integral leads to dissipation, and $\xi$ is a Gaussian noise term with 
\begin{eqnarray}
\langle\xi(t)\rangle=0 \quad {\rm and}\quad \langle\xi(t)\xi(t')\rangle=-\Pi_\eta^+(t-t')\delta^{(3)}(0).\label{GaussianNoise}
\end{eqnarray}
The relations (\ref{KMS}) and (\ref{GaussianNoise}) ensure that the fluctuation-dissipation theorem holds in (\ref{Langevin}).
Strictly speaking the field in (\ref{Langevin}) is not the original field $\phi$ on the contour $\mathcal{C}$, but the combination $\phi_c=(\phi^+ + \phi^-)/2$, where $\phi^+$ and $\phi^-$ denote the field with time argument on the ``forward'' and ``backward'' part of the contour, as indicated in figure~\ref{contour}. 
We ignore this subtlety here, as both notions give the same equation of motion for $\varphi$.

The Langevin equation (\ref{Langevin}) can be solved analytically by use of a Laplace transformation \cite{Boyanovsky:2004dj,Anisimov:2008dz},
\begin{eqnarray}\label{LangevinSolution}
\phi(t)=\phi(t=0)\dot{\Delta}^-(t)+ \dot{\phi}(t=0)\Delta^-(t)+\int_0^t\Delta^-(t-t')\xi(t'),
\end{eqnarray}
where 
\begin{equation}
\Delta^-(t)=\int d^3\textbf{x}\Delta^-(x)
\end{equation}
can be analytically obtained by Fourier-transforming (\ref{spectralfunction2}) in the Breit-Wigner approximation as
\begin{equation}
\Delta^-(t)\simeq \mathcal{Z}\frac{{\rm sin}(\Omega_\eta t)}{\Omega_\eta}e^{-\Gamma_\eta t}.
\end{equation}
By taking the expectation value (cf. (\ref{VarphiDef})) of $\phi$ in (\ref{LangevinSolution}), we find (neglecting the $p_0$-dependence of ${\rm Re}\Pi_\eta^R$, setting $\Omega_\eta=M_\eta$ for the zero mode and neglecting terms suppressed by $\Gamma_\eta/\Omega_\eta$)
\begin{eqnarray}\label{LangevinSol}
\varphi(t)=\left[\varphi(t=0){\rm cos}(M_\eta t) + \dot{\varphi}(t=0)\frac{{\rm sin}(M_\eta t)}{M_\eta}\right] e^{-\Gamma_\eta t}.
\end{eqnarray}
The solution (\ref{LangevinSol}) describes damped oscillations of $\varphi$ with the plasma frequency $M_\eta$ near the potential minimum. 
In spite of the integral in (\ref{Langevin}), non-Markovian effects do not play any role for $\varphi$ because $\Gamma_\eta$ and $M_\eta$ are in good approximation independent of $\varphi$, and we have assumed that the temperature remains constant on time scales $\sim1/\Gamma_\eta$. This is realistic if $\Gamma_\eta>H$, and if the initial $\varphi$ is sufficiently small that the energy release into the plasma due to its dissipation is not significant. 
In this case an equation of the form (\ref{FieldEqOfMot}) holds, with
\begin{eqnarray}
\V(\varphi)&\simeq& \frac{M_\eta^2}{2} \varphi^2, \label{LangevinPotential}\\  
\Gamma_\varphi&\simeq&\Gamma_\eta.\label{LangevinGamma}
\end{eqnarray} 
That is, $\V(\varphi)$ can be approximated by a parabola and is obtained from the free classical Lagrangian if one simply replaces the mass $m_\phi$ by the effective mass $M_\eta$.
The damping rate of $\varphi$ is given by the thermal quasiparticle width of $\eta$; it is obtained by evaluating ${\rm Im}\Pi_\eta^R(p)$ at the quasiparticle pole $p_0=M_\eta$. Finally, each mode of $\phi$ couples to the bath, but there are no interactions of the different modes with each other.
This approximation only holds for small deviations from the ground state, i.e.\ small field values $\varphi$. 
In cosmology this situation can be realized in the late phase of \emph{cosmic reheating} after inflation. In this context, the temperature dependence of the rate $\Gamma_\varphi$ has been studied in detail in \cite{Drewes:2013iaa}, and it was found in \cite{Drewes:2014pfa} that this temperature dependence can have a drastic effect on the thermal history of the universe during reheating.

For larger deviations from the ground state, the problem becomes much more involved because the quadratic approximation of the effective action near $\varphi=0$ is not sufficient. If it is still possible to describe the dynamics in terms of an effective potential $\V(\varphi)$ and (\ref{FieldEqOfMot}), then $\V(\varphi)$ certainly must involve higher powers of $\varphi$. This leads to ``multiplicative noise terms'' in the generalization of (\ref{Langevin}) \cite{Boyanovsky:2015xoa, Yokoyama:2004pf}, and the effective masses generally depend on $\varphi$. Finally, we do not expect the simple relation (\ref{LangevinGamma}) to hold, though it is frequently used in the literature without justification. 
In section \ref{EffectiveActionSec} we show that one can nevertheless find an equation of motion of the form (\ref{FieldEqOfMot}) if $\varphi$ changes slowly compared to the microphysical scales, but the expressions for $\V(\varphi)$ and $\Gamma_\varphi$ are in general more complicated.

\subsection{Large field values: Slow rolling}\label{EffectiveActionSec}
We continue to assume that all degrees of freedom except $\varphi$ are in equilibrium (or their thermodynamic state can at least effectively be characterized by an effective temperature), but we allow $\varphi$ to take large values, where the quadratic approximation (\ref{LangevinPotential}) of the effective potential is expected to break down.
If $\phi$ has only feeble interactions (which is required if the effective potential should be flat), then it is justified to assume a separation of scales
\begin{equation}\label{SeparationOfScales}
H, \dot{\varphi}/\varphi \ll T, \varphi, M_\eta
.\end{equation}
This implies that $\varphi$ changes adiabatically, and we can approximate 
\begin{eqnarray}\label{LocalInTime}
\varphi(t')^n\simeq \varphi(t)^n + n (t'-t) \dot{\varphi}(t)\varphi(t)^{n-1}.
\end{eqnarray}
The change of the functions $\Pi_{\mathcal{C}}(t,t')$ and $\tilde{\Pi}_{\mathcal{C}}(t,t')$ 
with respect to the time coordinate $t+t'$ occurs on the same macroscopic time scales as $\dot{\varphi}/\varphi$ and $H$. Hence (\ref{SeparationOfScales}) ensures that (\ref{adiabaticitycondition}) is fulfilled and we can expand $\Pi_{\mathcal{C}}(t,t')$ and $\tilde{\Pi}_{\mathcal{C}}(t,t')$ in this coordinate analogously to the expansion (\ref{LocalInTime}). For the present purpose it is sufficient to keep the zeroth term in this expansion, i.e.\ to assume that there is no explicit dependence on $t+t'$, meaning that $\Pi_{\mathcal{C}}(t,t') \simeq \Pi_{\mathcal{C}}(t-t')$ and $\tilde{\Pi}_{\mathcal{C}}(t,t') \simeq \tilde{\Pi}_{\mathcal{C}}(t-t')$. This will be justified later after (\ref{EOM}).

We can split the contour integral in the equation of motion \eqref{NonMarkovian} into
\begin{equation}\label{ContourSplitting}
\int_{\mathcal{C}}dt'=\int_{-\infty+i\epsilon}^{\infty+i\epsilon}dt'+\int_{\infty-i\epsilon}^{-\infty-i\epsilon}dt'.
\end{equation}
With (\ref{LocalInTime}), the only quantities under the integrals that depend on $t'$ are $\Pi_{\mathcal{C}}(t-t')$ and $\tilde{\Pi}_{\mathcal{C}}(t-t')$.
Since we are interested in physical times, $t$ always lies on the ``forward'' part of the contour $\mathcal{C}$.
If $t'$ also lies on the forward part (first term in (\ref{ContourSplitting})), then $\Pi_{\mathcal{C}}(t-t')$ is to be identified with the usual time-ordered self-energy (often referred to as $\Pi^{++}(t-t')$ in thermal field theory).
If  $t'$ lies on the backward part (second term in (\ref{ContourSplitting})), then $\Pi_{\mathcal{C}}(t-t')$ is to be identified with a ``Wightman type'' self-energy without time ordering  $\Pi^<(t-t')$, as $t'$ is then always ``later'' along the contour than $t$, see e.g. \cite{Chou:1984es,LeB}.
Their difference can be identified with the usual retarded self-energy, i.e.  $\Pi^{++}(t-t')-\Pi^<(t-t')=\Pi^R(t-t')$. 
This allows to combine the two terms in (\ref{ContourSplitting}) into a single integral
\begin{equation}\label{MasterOfPuppets}
\int_{\mathcal{C}}dt' \Pi_{\mathcal{C}}(t-t')=\int_{-\infty}^\infty dt' \Pi^R(t-t')
\end{equation}
and likewise for  $\tilde{\Pi}_{\mathcal{C}}(t-t')$.
The RHS of (\ref{MasterOfPuppets}) is nothing but the Fourier transform of $\Pi^R(t-t')$ evaluated at energy $\omega=0$. 
Some care should be taken at this point. The limit $\omega\rightarrow 0$ is obtained due to the linear approximation (\ref{LocalInTime}), which is controlled by the separation (\ref{SeparationOfScales}) between macroscopic and microscopic time scales.
It is probably best to think of $\omega$ as a parameter that characterizes this separation, i.e. $\omega\sim\dot{\varphi}/\varphi$, where $\dot{\varphi}/\varphi$ should be identified with the macroscopic scales $H$ or $\Gamma$.
With these considerations we finally obtain a Markovian approximation to the equation of motion (\ref{NonMarkovian}), 
\begin{eqnarray}
&&\left[\partial_t^2+m_\phi^2 
+\left(1-\omega\partial_\omega\right)\Pi_\varphi^R(-\omega)\right]_{\omega=0}\varphi(t) 
+\frac{1}{3!}\left[\lambda_\phi+
\left(1-\omega
\partial_\omega\right)\tilde{\Pi}_\varphi^R(-2\omega)
\right]_{\omega=0}\varphi(t)^3\nonumber\\
&& \hspace{2cm}-i\dot{\varphi}(t)\left[\partial_\omega \Pi_\varphi^R(-\omega)
+\frac{1}{3!}\varphi(t)^2\partial_\omega\tilde{\Pi}_\varphi^R(-2\omega)
 \right]_{\omega=0}=0.\label{EOM}
\end{eqnarray}
Here we have given the retarded self-energies $\Pi_\varphi^R$ and $\tilde{\Pi}_\varphi^R$ an index $_\varphi$ to distinguish them for the self-energies for $\mathcal{X}_i$.
The parameter $\omega$ should be thought of as much smaller than all microphysical scales and can be set to zero as long as $\omega < H < \Gamma_\eta < T$, which always holds under the previous assumptions. 
Since the real part of the retarded self-energy is always symmetric and the imaginary part antisymmetric, we can easily identify 
\begin{eqnarray}
\lim_{\omega\rightarrow0}
\Pi_\varphi^R(\omega)&=&
\lim_{\omega\rightarrow0}
{\rm Re}\Pi_\varphi^R(\omega),\label{realpart}\\
\lim_{\omega\rightarrow0}
\partial_\omega \Pi_\varphi^R(-\omega)&=&
-i\lim_{\omega\rightarrow0}
\frac{{\rm Im}\Pi_\varphi^R(\omega)}{\omega},\label{imaginarypart}
\end{eqnarray}
and likewise for $\tilde{\Pi}_\varphi^R$
\footnote{From these expressions it is clear that all coefficients of $\varphi$, $\varphi^3$  and $\dot{\varphi}$ in \eqref{EOM} are real. Therefore the effective equation of motion \eqref{EOM} is manifestly real.}.

With this limit and by comparing \eqref{EOM} with \eqref{FieldEqOfMot}, we can express the effective potential and dissipation coefficient in terms of $\Pi_\varphi^R$ and $\tilde{\Pi}_\varphi^R$.  
The first line in (\ref{EOM}) describes the evolution of $\varphi$ in an effective potential with the quantum corrections to the classical potential given by $\Pi_\varphi^R$ and  $\tilde{\Pi}_\varphi^R$:
\begin{eqnarray}
\partial_\varphi\V(\varphi)&=&
\lim_{\omega\rightarrow 0}\Big\{\left[m_\phi^2 
+\left[1-\omega\partial_\omega\right]\Pi_\varphi^R(-\omega)\right]\varphi 
+\frac{1}{3!}\left[\lambda_\phi+
\left[1-\omega
\partial_\omega\right]\tilde{\Pi}_\varphi^R(-2\omega)
\right]\varphi^3\Big\}  \nonumber \\
&=& \left[m_\phi^2 
+\Pi_\varphi^R(\omega)|_{\omega=0}\right]\varphi 
+\frac{1}{3!}\left[\lambda_\phi+
\tilde{\Pi}_\varphi^R(\omega)|_{\omega=0}
\right]\varphi^3 . \label{VSlowRoll}
\end{eqnarray}
The second line describes dissipative effects, i.e.\ the damping of $\varphi$ due to the interactions with the  bath of $\mathcal{X}_i$-particles:
\begin{eqnarray}
\Gamma_\varphi &=&
-i \lim_{\omega\rightarrow 0}
\Big[\partial_\omega \Pi_\varphi^R(-\omega)
+\frac{1}{3!}\varphi(t)^2\partial_\omega\tilde{\Pi}_\varphi^R(-2\omega)\Big]  \nonumber \\
&=& i 
\Big[\partial_\omega \Pi_\varphi^R(\omega)|_{\omega=0}
+\frac{1}{3}\varphi(t)^2\partial_\omega\tilde{\Pi}_\varphi^R(\omega)|_{\omega=0}\Big]. \label{GammaSlowRoll}
\end{eqnarray}
These results allow to justify why we could neglect their dependence on $t+t'$.
The terms containing derivatives $\partial_{t+t'}$ acting on $\Pi_\varphi^R$ and $\tilde{\Pi}_\varphi^R$ are suppressed with respect to terms without derivatives due to the separation of macroscopic and microscopic scales. 
For instance, $\partial_{t+t'}\tilde{\Pi}_\varphi^R(t,t')$ is suppressed by three small parameters, two powers of a coupling constant and a time derivative. All terms in the first line of (\ref{EOM}), which give the effective potential (\ref{VSlowRoll}), contain two or less small parameters, hence $\partial_{t+t'}\tilde{\Pi}_\varphi^R(t,t')$ can be regarded as a higher order correction. 
In spite of this, we have to keep the terms in the second line in (\ref{EOM}), which in principle are also suppressed by a time derivative and two powers of a coupling constant. 
The reason is that these form the leading order contribution to the dissipation coefficient (\ref{GammaSlowRoll}), while in the effective potential (\ref{VSlowRoll}) they would just act as higher order correction. 
In this sense, (\ref{EOM}) is a consistent expansion in gradients and the small coupling constants.

These results (\ref{VSlowRoll}) and (\ref{GammaSlowRoll}) should be compared to (\ref{LangevinPotential}) and (\ref{LangevinGamma}).
One obvious (and expected) difference is the appearance of higher powers of $\varphi$ in $\V(\varphi)$. 
Another crucial difference is that the dissipation coefficient (\ref{GammaSlowRoll}) is evaluated at $\omega=0$ and not at $\omega=M_\eta$, as in (\ref{LangevinGamma}). 
This supports the interpretation of $\omega$ as the rate at which $\varphi$ changes: During oscillations around the minimum, this rate is given by the plasma frequency $M_\eta$; if the field slowly rolls down a flat potential, then this rate is much smaller than all other scales (practically $\omega\simeq 0$ within loop diagrams).

Indeed, if we drop the terms $\propto\varphi^3$ in (\ref{NonMarkovian}) and use (\ref{ContourSplitting}) and (\ref{MasterOfPuppets}), we find
\begin{eqnarray}\label{NewLangevin}
(\partial_t^2+m_\phi^2)\varphi(t) 
+\int dt' \Pi_\phi^R(t-t')\varphi(t')=0.\nonumber
\end{eqnarray}
This equation is formally the same as (\ref{Langevin}) with $\xi=0$, so we can read off the solution from (\ref{LangevinSolution}),
\begin{eqnarray}\label{NewLangevinSolution}
\varphi(t)=\varphi(t=0)\dot{\Delta}^-(t)+ \dot{\varphi}(t=0)\Delta^-(t).
\end{eqnarray}
Again using the Breit-Wigner approximation for (\ref{spectralfunction2}), we can bring this into the explicit form (\ref{LangevinSol}). Hence, we have recovered the behavior of Brownian motion for small $\varphi$ from the more general expression (\ref{NonMarkovian}).

\section{A simple scalar model} \label{model}
\begin{figure}
	\center
	\includegraphics[width=12cm]{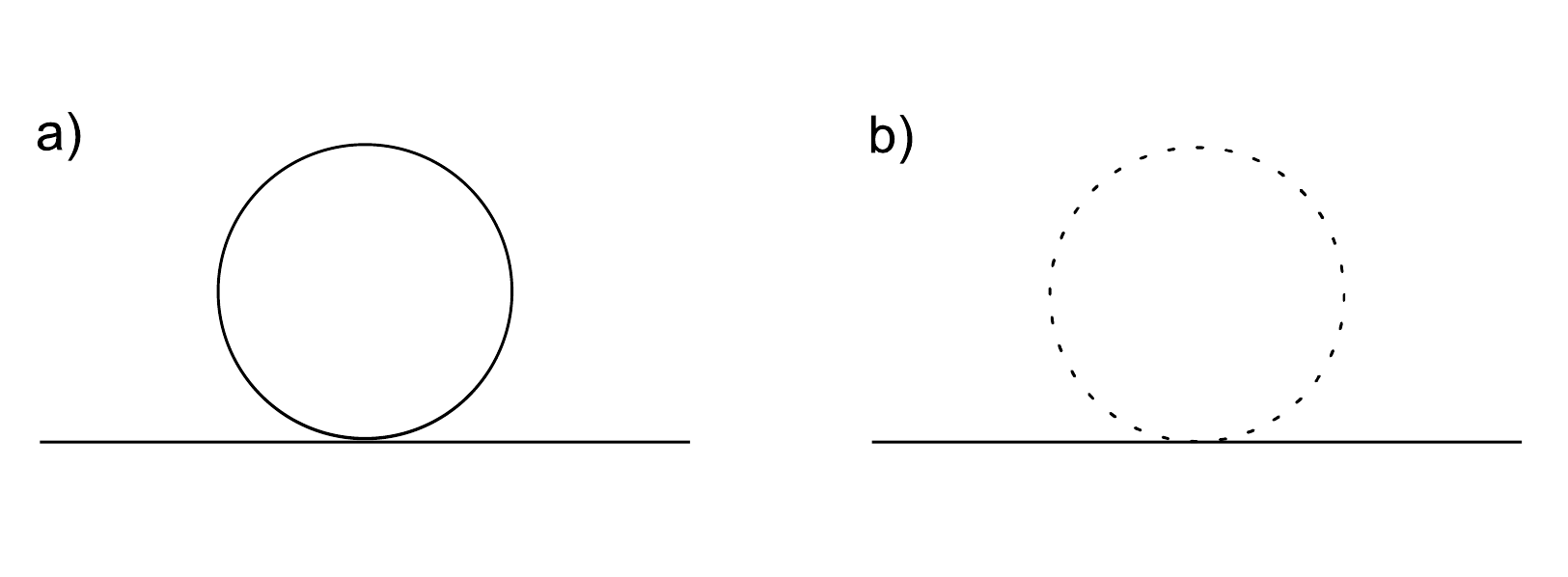} 
\caption{Self-energy diagrams contributing to mass correction at leading order. Solid line corresponds to $\eta$ and dashed line to $\chi$. Note that the functionals $\Pi_\varphi^R$ and $\tilde{\Pi}_\varphi^R$ should strictly be represented by amputated diagrams. The external lines are added for representative purposes and in the effective action they are to be replaced by $\varphi$.  \label{tad}}
\end{figure}

\begin{figure}
	\center
	\includegraphics[width=12cm]{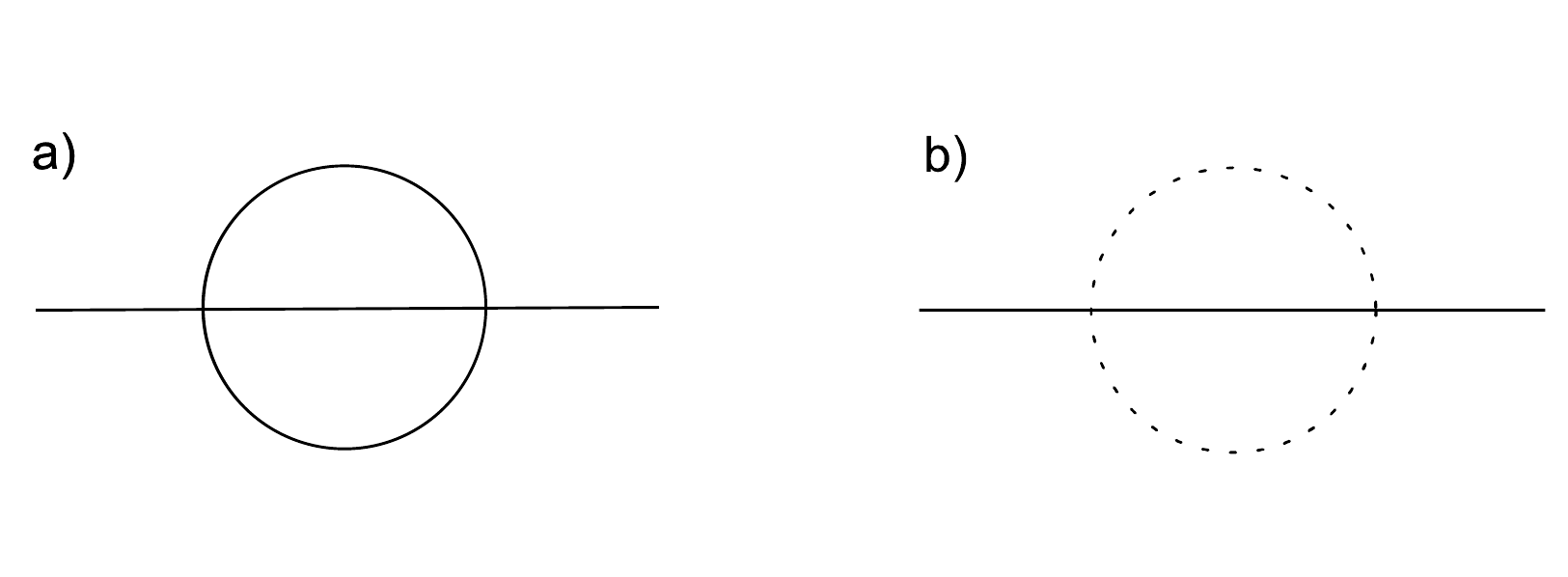}
\caption{Feynman diagrams contributing to $\Pi_\varphi^R$. Solid line corresponds to $\eta$ and dashed line to $\chi$.  \label{setting}}
\end{figure}

In the following we illustrate our results in a simple scalar model, in which the ``bath'' consists only of one other real scalar field $\chi$,
\begin{equation}\label{L}
\mathcal{L}=
\frac{1}{2}\partial_\mu\phi\partial^\mu \phi
-\frac{m_\phi^2}{2}\phi^2-\frac{\lambda_\phi}{4!}\phi^4
+\frac{1}{2}\partial_\mu\chi\partial^\mu \chi
-\frac{m_\chi^2}{2}\chi^2-\frac{\lambda_\chi}{4!}\chi^4 
-\frac{h}{4}\phi^2\chi^2 .
\end{equation}
In order to obtain explicit expressions for $\V(\varphi)$ and $\Gamma_\varphi$ from  (\ref{VSlowRoll}) and (\ref{GammaSlowRoll}), we have to evaluate the loop diagrams given in figure \ref{tad}, \ref{setting} and \ref{fish}.

Here we focus on the case of adiabatically changing large $\varphi$ discussed in section~\ref{EffectiveActionSec}. For the case of oscillations near the potential minimum, $\Gamma_\varphi$ has been calculated in detail in \cite{Drewes:2013iaa}.

\subsection{The effective potential $\V(\varphi)$}
From (\ref{realpart}) it is clear that $\V(\varphi)$ is obtained from the real part of retarded self-energies. 
To establish a consistent perturbation theory, these have to be evaluated with resummed thermal propagators. 
For the present purpose it is sufficient to evaluate these in the pole approximation of (\ref{rhofree}) for $\eta$ and $\chi$.
The potential in \eqref{L} in terms of $\eta$ and $\chi$ reads  
\begin{equation}
V(\eta,\chi) = \frac{m_\eta^2}{2} \eta^2 + \frac{\lambda_\phi\, \varphi}{3!} \eta^3 + \frac{\lambda_\phi}{4!} \eta^4
+ \frac{(m_\chi^2+h \,\varphi^2/2)}{2}\chi^2 + \frac{\lambda_\chi}{4!} \chi^4 + \frac{h\, \varphi}{2} \eta \chi^2 + \frac{h}{4} \eta^2 \chi^2
\end{equation}
with $m_\eta^2 = m_\phi^2 +\frac{\lambda_\phi}{2}\varphi^2$.


\subsubsection{The self-energy $\Pi_\varphi^R$}
The real part of $\Pi_\varphi^R|_{\omega=0}$ is dominated by the well-known local diagrams due to the vertices $\lambda_\phi \eta^4/4!$ and 
$h\eta^2\chi^2/4$, see figure \ref{tad}. 
It reads
\begin{eqnarray}
\Pi_\varphi^R&=&\frac{\lambda_\phi}{2}\int\frac{d\p^3}{(2\pi)^3}\frac{1}{2\Omega_\eta}[1+2f_B(\Omega_\eta)]
+\frac{h}{2}\int\frac{d\p^3}{(2\pi)^3}\frac{1}{2\Omega_\chi}[1+2f_B(\Omega_\chi)]\nonumber\\
&=&
\frac{\lambda_\phi}{2}\frac{M_\eta^2}{(4\pi)^2}\left[N_\epsilon + \T(M_\eta^2,\mu^2)\right] + \frac{h}{2}\frac{M_\chi^2}{(4\pi)^2}\left[N_\epsilon + \T(M_\chi^2,\mu^2)\right]\nonumber\\
&&+\int\frac{d\p^3}{(2\pi)^3}\left[\frac{\lambda_\phi}{2\Omega_\eta} f_B(\Omega_\eta)
+\frac{h}{2\Omega_\chi}f_B(\Omega_\chi)\right]   \label{Tad-Pi}
\end{eqnarray}
with 
\begin{equation}
\T(M^2,\mu^2)= 1 - {\rm log}\frac{M^2}{\mu^2},   N_\epsilon=\frac{2}{\epsilon} + {\rm log}4\pi - \gamma,
\end{equation}
where we have used dimensional regularization.
The divergent  part ($N_\epsilon$ terms) can be eliminated by renormalisation of $m_\phi$ and coupling constants at zero temperature, 
using $\overline{\rm MS}$ scheme.\footnote{Note that the temperature-dependent divergent terms, which come from the thermal corrections contained in $M_\eta$ and $M_\chi$ multiplying $N_\epsilon$, are sufficiently taken care of by the temperature-independent counterterms from $\lambda_\phi$ and $h$ in the third line of \eqref{Tad-Pi}, which in fact represents the thermal corrections.} 
In the following we assume that this has been done and that vacuum masses and coupling constants are the renormalized parameters at $T=0$. 
In this work we are mainly interested in thermal corrections, which are most relevant if the temperature is larger than the effective masses (otherwise the thermal corrections are negligible since they are Boltzmann suppressed due to the appearance of the distribution functions $f_B$ in the thermal propagators). 
For $ M_\eta, M_\chi \ll T$, the finite part (i.e. the $\T(M^2, \mu^2)$ terms), remaining after subtraction of the divergences, can be ignored since it is subdominant compared to the $f_B$-part (the third line of \eqref{Tad-Pi}).\footnote{This can be seen as follows. Let's choose $\mu$ to be $\mu=M_i\equiv {\rm Max}[M_\eta, M_\chi]$ evaluated at initial value of $\varphi$. Then $M_\eta, M_\chi< M_i  $ for all the times. Now we will show that  $|\lambda_\phi\, M_\eta^2\, \T(M_\eta^2,M_i^2)| \ll \lambda_\phi\, T^2$ and $|h\, M_\chi^2\, \T(M_\chi^2,M_i^2)| \ll h\, T^2$ 
when $ M_\eta, M_\chi \ll T$ (so $M_i \ll T$). To this end it suffices to show 
$\left|\lambda_\phi\, M_\eta^2\, \log\frac{M_\eta^2}{M_i^2}\right| \ll \lambda_\phi\, T^2$ and $\left|h\, M_\chi^2\, \log\frac{M_\chi^2}{M_i^2}\right| \ll h\, T^2$. This is clearly the case since, for example, 
$ \left| M_\eta^2\, \log\frac{M_\eta^2}{M_i^2}\right|= \left|M_i^2 \frac{M_\eta^2}{M_i^2}\, \log\frac{M_\eta^2}{M_i^2} \right| <  M_i^2\ll T^2$ for $\frac{M_\eta^2}{M_i^2}<1$ (note that $|\varepsilon \log \varepsilon|\leq e^{-1}$ for $ 0< \varepsilon \leq 1 $). \label{finite}
} 
Due to the momentum-independence there is no need for a finite $T$ wave function renormalisation.
For $M_\eta, M_\chi \ll T$, one can perform the integrals analytically, for example,
\begin{eqnarray} 
\int\frac{d\p^3}{(2\pi)^3}
\frac{h}{2\Omega_\chi}f_B(\Omega_\chi)\simeq \frac{h}{24}T^2. 
\end{eqnarray}
The same applies to the diagram with $\eta$-loop.  We thus obtain for $M_\eta, M_\chi \ll T$
\begin{equation}
\Pi_\varphi^R \simeq \frac{(h+\lambda_\phi)}{24} T^2.
\end{equation}
There are also contributions from the setting sun diagrams in figure \ref{setting}, but they are of higher order in $h$ and $\lambda_\phi$. 

\subsubsection{The vertex functional $\tilde{\Pi}_\varphi^R$}

We now turn to the vertex functional, which is given by the ``fish diagrams'' a) in figure \ref{fish}. 
The $f_B$-independent part of these diagrams is again divergent; for vanishing external momentum it is given by (including the analogous fish diagram with $\eta$-loop) 
\begin{equation}
\frac{h^2}{2(4\pi)^2}\left[N_\epsilon-{\rm log}\frac{M_\chi^2}{\mu^2}\right]
+\frac{\lambda_\phi^2}{2 (4\pi)^2}\left[N_\epsilon-{\rm log}\frac{M_\eta^2}{\mu^2}\right].
\end{equation}
Since we are only interested in the case of vanishing external momentum, we can simply absorb the $N_\epsilon$ terms into the renormalisation of the vacuum couplings $h$ and $\lambda_\phi$. The remaining finite part ($\log$ terms) can be ignored for $ M_\eta, M_\chi \ll T$ in comparison with $f_B$-part given below, in the similar way to the case of $\Pi_\varphi^R$.\footnote{Setting again $\mu=M_i\equiv {\rm Max}[M_\eta, M_\chi]$ evaluated at initial value of $\varphi$, it is straightforward to see that 
$\left|\log\frac{M_\eta}{M_i}\right|= \left|\frac{M_i}{M_\eta} \frac{M_\eta}{M_i} \log\frac{M_\eta}{M_i}\right| < \frac{M_i}{M_\eta} \ll \frac{T}{M_\eta}$, where we used the same tricks as in footnote \ref{finite}. The same can be shown for $M_\chi$.
}

The $f_B$-part  from  the fish diagrams is given by the the standard expression (see e.g.\ section 4.2.2 in~\cite{LeB}).
Let us focus on the $\chi$-loop with $T\gg M_\chi$,
\begin{eqnarray}
\tilde{\Pi}_\varphi^R|_{\omega=0}=-\frac{h^2}{4} \frac{4\pi}{(2\pi)^3}\int_{M_\chi}^\infty d\Omega_\chi
\frac{\sqrt{\Omega_\chi^2-M_\chi^2}}{\Omega_\chi^2}f_B(\Omega_\chi)
-\frac{\lambda_\phi^2}{4} \frac{4\pi}{(2\pi)^3}\int_{M_\eta}^\infty d\Omega_\eta
\frac{\sqrt{\Omega_\eta^2-M_\eta^2}}{\Omega_\eta^2}f_B(\Omega_\eta). \nonumber \\
\end{eqnarray}
The above integral can not be done analytically, but it is easy to see that it is strongly dominated by small $\Omega$ when $M_\eta, M_\chi \ll T$. Therefore we can approximate $f_B(\Omega)\simeq T/\Omega$ by expanding the exponential to get   
\begin{eqnarray}
-\frac{h^2}{4} \frac{4\pi}{(2\pi)^3}\int_{M_\chi}^\infty d\Omega_\chi
\frac{\sqrt{\Omega_\chi^2-M_\chi^2}}{\Omega_\chi^2}f_B(\Omega_\chi)\simeq-\frac{h^2}{4} \frac{4\pi}{(2\pi)^3}\int_{M_\chi}^\infty d\Omega_\chi
\frac{T\sqrt{\Omega_\chi^2-M_\chi^2}}{\Omega_\chi^3}
= -\frac{h^2}{32\pi}\frac{T}{M_\chi} . \nonumber \\
\end{eqnarray}
The contribution from the $\eta$-loop can be obtained from the above expression by replacing $\chi\rightarrow \eta$ and $h\rightarrow \lambda_\phi$, and altogether we obtain for $M_\eta, M_\chi \ll T$
\begin{equation}
\tilde{\Pi}_\varphi^R|_{\omega=0} \simeq -\frac{h^2}{32\pi}\frac{T}{M_\chi}-\frac{\lambda_\phi^2}{32\pi}\frac{T}{M_\eta}. \label{RePiTilde}
\end{equation}

\subsubsection{The full potential}
\begin{figure}
            \hspace{-1cm}
	\includegraphics[width=18cm]{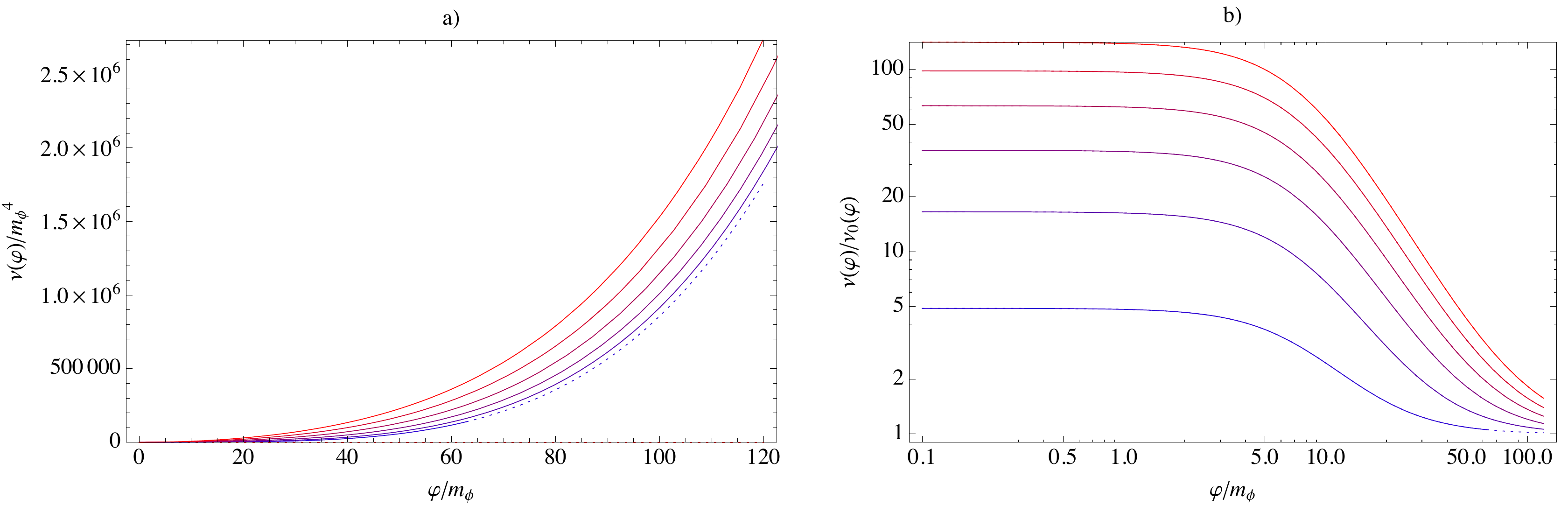}
	\caption{The full effective potential $\V(\varphi)$ as a function of $\varphi$ for various temperatures, normalized to $m_\phi^4$ (figure a)) and to zero temperature value $\V_0(\varphi) \equiv \V(\varphi)|_{T=0}$ (figure b)).  Here we set $m_\chi= m_\phi/5$, $\lambda_\phi=\lambda_\chi=1/5$ and $h=1/30$. Different values of temperature are indicated by the color coding  with the pure blue line being for the lowest temperature $T=20 m_\phi$.  When the color varies from blue towards red, the temperature increases with a step $\Delta T=20 m_\phi$. The dotted line is used to denote the regime where the condition $M_\eta, M_\chi <T$ is not satisfied. }
	\label{potential}
\end{figure}

If the temperature is much smaller than the effective masses, then thermal corrections are negligible. In this case radiative corrections are mainly interesting if some mass squares are negative, leading to symmetry breaking; otherwise they just lead to small modifications of the potential. 
If $T$ is comparable to the effective masses, then the loop integral in general cannot be solved analytically. For the interesting case $M_\eta, M_\chi\ll T$, we can use the above results in \eqref{VSlowRoll} and find
\begin{eqnarray}
\partial_\varphi\V(\varphi)=\left(m_\phi^2+\frac{(h+\lambda_\phi)}{24}T^2\right)\varphi
+\frac{1}{3!}\left(\lambda_\phi
-\frac{h^2}{32\pi}\frac{T}{M_\chi}
-\frac{\lambda_\phi^2}{32\pi}\frac{T}{M_\eta}
\right)\varphi^3.\label{Vbeforeint}
\end{eqnarray} 
In order to check for which range of field values the assumption $M_\eta, M_\chi\ll T$ can be justified, we can estimate the effective masses as
\begin{eqnarray}
M_\eta^2&=&m_\phi^2+\left(h+\lambda_\phi\right)\frac{T^2}{24}+\frac{\lambda_\phi}{2}\varphi^2, \label{Meta} \\
M_\chi^2&=&m_\chi^2+\left(h+\lambda_\chi\right)\frac{T^2}{24}+\frac{h}{2}\varphi^2. \label{Mchi}
\end{eqnarray}
This leads to the conditions 
\begin{eqnarray}
\varphi^2<\frac{2\left[\left(1-\frac{(h+\lambda_\phi)}{24}\right)T^2-m_\phi^2\right]}{\lambda_\phi} \ , \ 
\varphi^2<\frac{2\left[\left(1-\frac{(h+\lambda_\chi)}{24}\right)T^2-m_\chi^2\right]}{h}.
\end{eqnarray}
These conditions imply that the $\varphi$-dependence of the effective masses in (\ref{Vbeforeint}) is not necessarily negligible. Taking this into account, we integrate (\ref{Vbeforeint}) and obtain
\begin{equation} \label{FullPotential}
\V(\varphi) = \frac{M_{\eta 0}^2 \varphi^2}{2} + \frac{\lambda_\phi \varphi^4}{4!} 
+ \frac{T}{288 \pi} \Big[4 (M_\eta^3+M_\chi^3- M_{\eta 0}^3-M_{\chi 0}^3 )-3 \varphi^2(\lambda_\phi M_\eta + h M_\chi) \Big], 
\end{equation}
where 
\begin{eqnarray}
M_{\eta 0}^2&\equiv& M_\eta(\varphi=0)^2=m_\phi^2+\left(h+\lambda_\phi\right)\frac{T^2}{24}, \\
M_{\chi 0}^2&\equiv& M_\chi(\varphi=0)^2=m_\chi^2+\left(h+\lambda_\chi\right)\frac{T^2}{24}.
\end{eqnarray}
Figure \ref{potential} shows the effective potential \eqref{FullPotential} as a function of $\varphi$ for various temperatures. 

\subsection{The damping coefficient $\Gamma_\varphi$}
Damping coefficient (\ref{GammaSlowRoll}) depends on $\varphi$ through the effective masses and has to be evaluated at $\omega=0$.

\subsubsection{Contribution from $\Pi_\varphi^R$
} \label{Pi_varphi}
Using (\ref{imaginarypart}), we can determine the coefficient $\lim_{\omega\rightarrow0}\partial_\omega \Pi_\varphi^R(-\omega)$ by calculating ${\rm Im}\Pi^R(p)$ from diagrams given in figure \ref{setting} for vanishing $p$. The leading order contribution comes from the setting sun diagram  b) in figure \ref{setting} since diagram a) gives vanishing result for kinematic reasons.\footnote{
There are two kinds of processes contained in cuts through the diagram in figure \ref{setting} a) for $\Gamma_\eta=0$, decays $\phi\rightarrow \phi\phi\phi$ and scatterings $\phi\phi\rightarrow\phi\phi$ (and their inverse). The former are clearly kinematically forbidden. For vanishing external four-momentum, the latter effectively correspond to $\phi\rightarrow\phi\phi$ decays and inverse decays, which are also kinematically forbidden on-shell. Finite width corrections are of higher order in this case. 
} The calculation of the loop integral is rather technical and is summarized in appendix~\ref{appendixA}.  For $M_\chi \ll M_\eta \ll T$, the result reads (cf. \eqref{Pi_A})  
\begin{equation} \label{ImPiEstimate}
\partial_\omega \Pi_\varphi^R(\omega)|_{\omega=0}\approx -i \frac{h^2\,T^2}{(4 \pi)^3 M_\eta} \log\left(\frac{M_\eta}{M_\chi}\right).
\end{equation}

\subsubsection{Contribution from the vertex  $\tilde{\Pi}_\varphi^R$}  \label{tildePi}

\begin{figure}
	\center
	\includegraphics[width=12cm]{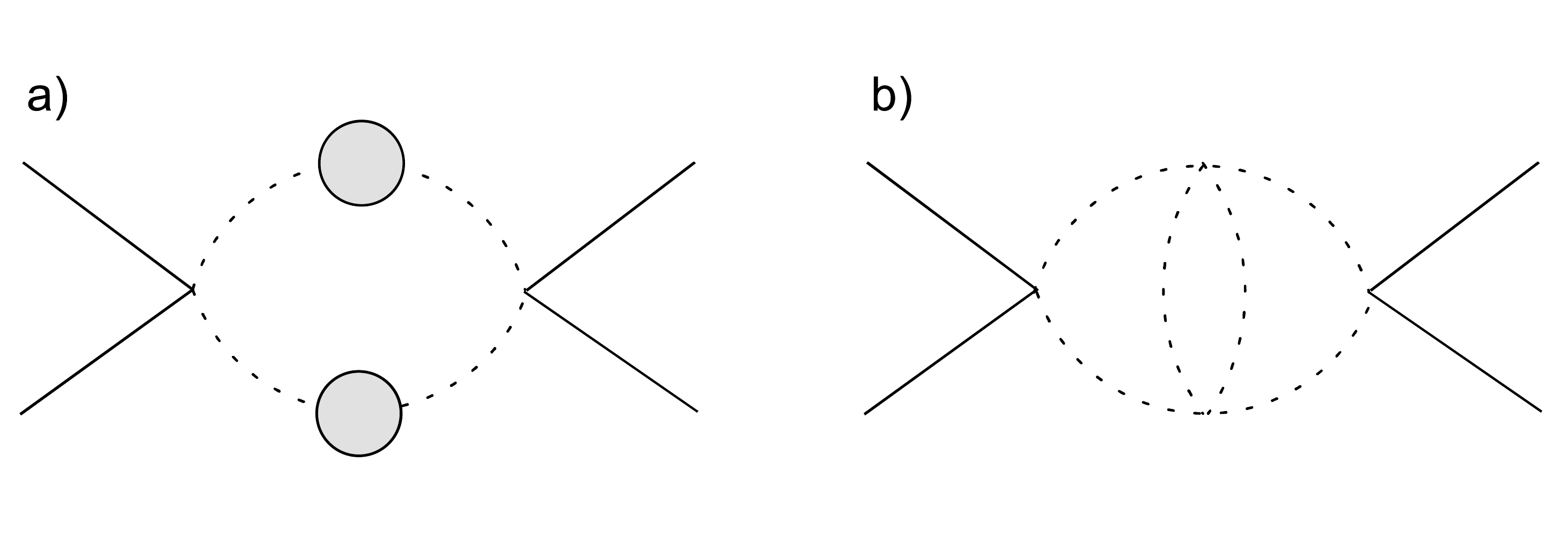}
\caption{Feynman diagrams contributing to $\tilde{\Pi}_\varphi^R$ with only $\chi$-propagators in loops. Dashed line corresponds to $\chi$. Note that there are analogous diagrams which involve $\eta$-propagators in loops.  \label{fish}}
\end{figure}

Let us focus on diagram a) in figure \ref{fish}, which gives dominant contribution\footnote{See section \ref{Interp} for the argument that the vertex diagram b) in figure~\ref{fish} is subdominant.}.
The contribution from the analogous diagram with $\eta$-loop can be obtained by proper replacements of coupling constants and masses. 
In the zero-width approximation (\ref{rhofree}),
the imaginary part ${\rm Im}\tilde{\Pi}_\varphi^R$ vanishes for any finite positive $\omega<2M_\chi$ 
and diverges for $\omega=0$.\footnote{These two facts can be seen from figure \ref{support} b) and c), which depict the supports of the spectral densities in the expresion \eqref{loopintegral} for ${\rm Im}\tilde{\Pi}_\varphi^R$, as discussed in section \ref{FishDifference}.}
One can understand this from kinematic reasons when recalling that the imaginary part can be interpreted in terms of microphysical processes when applying the optical theorem at finite $T$ \cite{Drewes:2013iaa}, see figure \ref{NewOne}. 
A non-zero $\partial_\omega\tilde{\Pi}_\varphi^R|_{\omega\rightarrow 0}$ can be obtained by including the finite width of the $\chi$-propagators in the loop, which parameterizes the effect of scattering processes with more vertices \cite{Drewes:2010pf,Drewes:2013iaa}, see figure \ref{ScatteringExample}.
This aspect has been pointed out in \cite{BasteroGil:2010pb,BasteroGil:2012cm}. 

\begin{figure}
	\center
	\includegraphics[width=12cm]{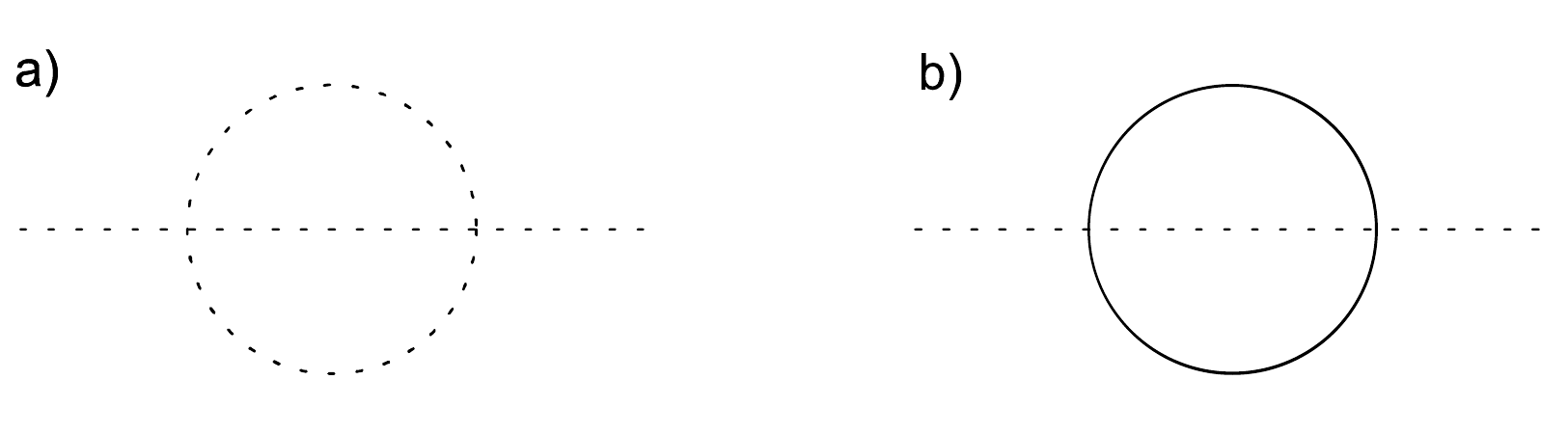}
\caption{Feynman diagrams for self-energy of $\chi$ giving finite width of the $\chi$-propagators in the loop in figure \ref{fish} a). Dashed line corresponds to $\chi$ and solid line to $\eta$.  \label{chi_width}}
\end{figure}

The loop in the ``fish diagram'' a) in figure \ref{fish} can be expressed as  \cite{Drewes:2010pf}
\begin{eqnarray}
{\rm Im}\tilde{\Pi}_\varphi^{R}(\omega)&=&\frac{1}{2i}\tilde{\Pi}^{-}_\varphi(\omega)=\frac{1}{2i}f_{B}^{-1}(\omega)\tilde{\Pi}_\varphi^{<}(\omega)\nonumber\\
&=&-\frac{h^{2}}{2}f_{B}^{-1}(\omega)\int\frac{d^{4}p}{(2\pi)^{4}}\Delta^{<}_\chi(p_{0})\Delta^{ >}_\chi(p_{0}-\omega)\nonumber\\
&=&-\frac{h^{2}}{2}f_{B}^{-1}(\omega)\int\frac{d^{4}p}{(2\pi)^{4}}\Delta_\chi^{<}(p_{0})\Delta_\chi^{ <}(\omega-p_{0})\nonumber\\
&=&-\frac{h^{2}}{2}\int\frac{d^{4}p}{(2\pi)^{4}}\left(1+f_B(p_{0})+f_B(\omega-p_{0})\right)\rho_\chi(p_{0})\rho_\chi(\omega-p_{0})  \label{loopintegral}
.\end{eqnarray}
For notational simplicity we continue to suppress the dependence on spatial momentum; the external spatial momentum is vanishing, and for all quantities under the integral it is given by the loop momentum.
Here we have first used (\ref{UsefulRelation1}) and the KMS relation (\ref{KMS}) to express ${\rm Im}\tilde{\Pi}_\varphi^{R}$ in terms of $\tilde{\Pi}_\varphi^<$.
Using the Feynman rules sketched after equation (\ref{kleinerinliste}), which are derived in detail in \cite{LeB}, this function can be calculated from an integral over $\Delta_\chi^\gtrless$ alone. With (\ref{UsefulRelation2}), this can finally be expressed as an integral over a product of spectral densities, which is particularly easy to be evaluated in the approximation (\ref{rhofree}). In the analogous expressions (\ref{SettingSunA}) and (\ref{SettingSunB}) for the setting sun diagrams one can indeed use the approximation (\ref{rhofree}).
In (\ref{loopintegral}), on the other hand, we cannot neglect the finite width and have to  use the \emph{pole approximation} to (\ref{spectralfunction2}), see e.g.~\cite{Drewes:2013iaa} for a discussion,
\begin{equation}\label{onepolerho}
\rho_{\chi}(p_0)
\simeq
\frac{
-2
{\rm Im}\Pi_{\chi}^R(p_0)
}{(p_0-\hat{\Omega}_{\chi})(p_0+\hat{\Omega}_{\chi})(p_0-\hat{\Omega}_{\chi }^*)(p_0+\hat{\Omega}_{\chi}^*)
}, 
\end{equation}
where $\hat{\Omega}_\chi \equiv \Omega_\chi + \frac{i}{2}\Gamma_\chi$. 
For $\omega\ll\Gamma_\chi$ the quasiparticle resonances at $p_0\simeq \Omega_\chi$ and $p_0\simeq \Omega_\chi - \omega$ of (\ref{onepolerho})  in (\ref{loopintegral}) overlap in the entire integration region, and the integral  is strongly dominated by this ``pole contribution''. 
Inserting (\ref{onepolerho}), we obtain 
\begin{eqnarray}
{\rm Im}\tilde{\Pi}^{R}_\varphi(\omega)&=&\frac{h^2}{2}\int\frac{d|\textbf{p}|}{2(2\pi)^{2}}\textbf{p}^{2}
2{\rm Re}\Bigg[
\frac{
{\rm Im}\Pi^{R}_\chi(\omega-\Omega_\chi)
\left(1+f_{B}(\Omega_\chi)+f_{B}(\omega-\Omega_\chi)\right)}{\hat{\Omega}_\chi\left(\left((\omega-\hat{\Omega}_\chi)^{2}-\Omega_\chi^{2}\right)^{2}+\left({\rm Im}\Pi^{R}_\chi(\omega-\hat{\Omega}_\chi)
\right)^{2}\right)}\nonumber\\
&&
+\frac{
{\rm Im}\Pi^{R}_\chi(\omega+\Omega_\chi)
\left(
f_{B}(\Omega_\chi)
-f_{B}(\omega+\Omega_\chi)
\right)}{\hat{\Omega}_\chi^{*}\left(\left((\omega+\hat{\Omega}_\chi^{*})^{2}-\Omega_\chi^{2}\right)^{2}+\left({\rm Im}\Pi^{R}_\chi(\omega+\hat{\Omega}_\chi^{*})
\right)^{2}\right)}\Bigg].
\label{OffShell}
\end{eqnarray}
We now take the limit $\omega\rightarrow 0$: We insert (\ref{OffShell}) into the relation 
$\lim_{\omega\to0}\partial_\omega \tilde{\Pi}_\varphi^R(\omega)=\lim_{\omega\to0} \frac{i{\rm Im}\tilde{\Pi}_\varphi^R(\omega)}{\omega}$  with approximations
\begin{eqnarray}\label{HutAb}
{\rm Im}\Pi^{R}_\chi(\omega-\hat{\Omega}_\chi)\simeq {\rm Im}\Pi^{R}_\chi(\omega-\Omega_\chi) \ , \ {\rm Im}\Pi^{R}_\chi(\omega+\hat{\Omega}_\chi^*)\simeq {\rm Im}\Pi^{R}_\chi(\omega+\Omega_\chi)
\end{eqnarray}
and then replace ${\rm Im}\Pi^{R}_\chi(\omega\pm\Omega_\chi)|_{\omega=0}\rightarrow \mp\Gamma_\chi\Omega_\chi$ in the denominator 
to obtain 
\begin{eqnarray}
\lim_{\omega\rightarrow0}\partial_\omega \tilde{\Pi}_\varphi^R(\omega) &=& \lim_{\omega\rightarrow0} \frac{i {\rm Im}\tilde{\Pi}^R_\varphi(\omega)}{\omega} \nonumber\\
 &=& - i\frac{40 h^2}{\pi^2T}\int d|\p| \frac{\p^2\Omega_\chi^2}{\left(\Gamma_\chi^4+68\Gamma_\chi^2\Omega_\chi^2+256\Omega_\chi^4\right)\Gamma_\chi\left({\rm cosh}(\Omega_\chi/T)-1\right)}\nonumber\\
&\approx&-i\frac{40 h^2}{\pi^2T}\int d|\p| \frac{\p^2\Omega_\chi^2}{256\Omega_\chi^4\Gamma_\chi\left({\rm cosh}(\Omega_\chi/T)-1\right)}.
\label{tildePieq}
\end{eqnarray}
The calculation of the momentum-dependent thermal width $\Gamma_\chi$ from the two loop diagram in figure \ref{chi_width} is generally difficult. The full $\Gamma_\chi$ is sum of contributions from diagram a) and b) in figure \ref{chi_width}, i.e. $\Gamma_\chi=\Gamma_\chi^{(a)}+\Gamma_\chi^{(b)}$. 
The calculation of  the imaginary part of diagram a) in figure \ref{chi_width}  is presented in detail in appendix \ref{On-shell}
and $\Gamma_\chi^{(a)}$ can be approximated $M_\chi \ll T$ as (cf. \eqref{On-shell-Pichi})  
\begin{equation} \label{gamma_chi_a}
\Gamma_\chi^{(a)}\simeq \frac{\gamma_\chi^{(a)}}{\Omega_\chi} \ {\rm with} \ 
\gamma_\chi^{(a)}\equiv \frac{\lambda_\chi^2 T^2}{256 \pi^3}.
\end{equation}
The main momentum dependence is due to the time dilatation; the factor $1/\Omega_\chi$ is due to the extended lifetime of a relativistic $\chi$-particle in the bath rest frame. The contribution from the diagram  b) in figure \ref{chi_width} should have essentially the same behavior  as that of a) as long as $M_\chi, M_\eta \ll T$ because the loop is dominated by hard momenta $\gg M_\chi, M_\eta$ in this case. Therefore one can obtain the contribution $\Gamma_\chi^{(b)}$ by replacement $\lambda_\chi\rightarrow h$ in \eqref{gamma_chi_a} and by taking into account a proper symmetry factor of the loop:
\begin{equation} \label{gamma_chi_b}
\Gamma_\chi^{(b)}\simeq \frac{\gamma_\chi^{(b)}}{\Omega_\chi} \ {\rm with} \ 
\gamma_\chi^{(b)}\equiv \frac{3 h^2 T^2}{256 \pi^3}.
\end{equation}   
The integral then reads
\begin{eqnarray}
\lim_{\omega\rightarrow0}\partial_\omega \tilde{\Pi}_\varphi^R(\omega)\approx
-i\frac{40 \pi h^2}{T^3(\lambda_\chi^2+ 3 h^2)}\int d|\p| \frac{\p^2}{\Omega_\chi\left({\rm cosh}(\Omega_\chi/T)-1\right)}.
\end{eqnarray}
This integral is strongly dominated by the region $\Omega_\chi\ll T$. We expand the integrand in $\Omega_\chi/T$ up to second order and integrate from $\Omega_\chi=M_\chi$ up to $\Omega_\chi=T$ to obtain
\footnote{It should be pointed out that for $0<\omega<\Gamma_\chi$ and with the previous assumptions, the integrand in (\ref{OffShell}) formally exhibits a sharp peak at $\Omega_\chi\approx \gamma_\chi^{2/3}/(2\omega^{1/3})=\Gamma_\chi^2/(8 \omega)$. In this region the integrand can be approximated by the Breit-Wigner curve. In the zero width limit the contribution from this region can be estimated as  
\begin{equation}\label{unreal}
\frac{
h^2
\left(
2
+\frac{4M_\chi^2}{(\gamma_\chi\omega)^{2/3}}
-\frac{\gamma^{2/3}}{\omega^{4/3}}
\right)
}{
3072\pi T\left(
{\rm cosh}\left(
\frac{\gamma^{2/3}/\omega^{1/3}+\omega}{2T}
\right)-1
\right)
}.
\end{equation}
Though this contribution smoothly vanishes in the limit $\omega\rightarrow 0$, it is much bigger than (\ref{ImtildePiresultat}) for most finite $0<\omega<\Gamma_\chi$ even if $\omega\ll \Gamma_\chi$. This appears to be at odds with our intuitive argument that the result of (\ref{OffShell}) should be independent of $\omega$ if $|\omega|\ll \Gamma_\chi$. 
The contribution (\ref{unreal}) is not physical and we artificially introduced by the approximation (\ref{HutAb}), which is only valid in the region $\Omega_\chi\ll \Gamma_\chi^2/\omega$. 
}
\begin{eqnarray}
\partial_\omega \tilde{\Pi}_\varphi^R(\omega)|_{\omega=0}\approx
 -i\frac{80 \pi h^2 }{T (\lambda_\chi^2+3 h^2)} \log\left(\frac{T}{M_\chi}\right).\label{ImtildePiresultat}
\end{eqnarray}
This is for the fish diagram a) in figure \ref{fish} with $\chi$-loop. The contribution from the analogous fish diagram with $\eta$-loop can be obtained from above expression by the replacement $M_\chi\rightarrow M_\eta$ and $h^2/(\lambda_\chi^2+3 h^2)\rightarrow \lambda_\phi^2/(\lambda_\phi^2+3 h^2)$.   

\subsubsection{The full damping coefficient}
\begin{figure}
          \hspace{-1cm}
	\includegraphics[width=18cm]{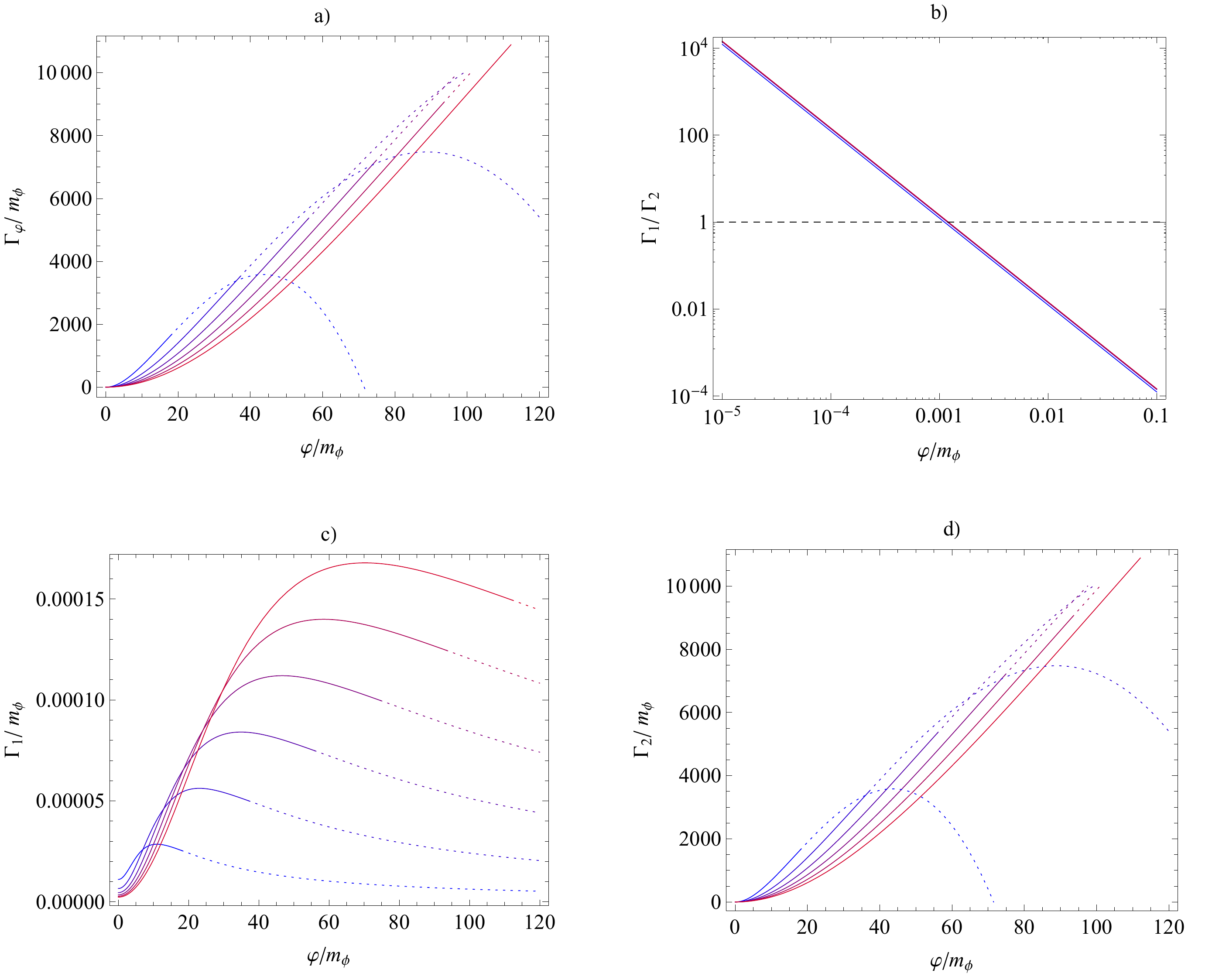}
	\caption{
	The damping coefficients as a function of $\varphi$ for various temperatures, normalized to $m_\phi$. In these plots the choice of the parameter values and the color coding of lines for different temperatures are the same as in figure \ref{potential} except that the dotted lines here denote the regime where the condition $M_\chi<M_\eta <T$ is not satisfied. Figure a) shows the full damping coefficient $\Gamma_\varphi$. In figure b) we plot the ratio $\Gamma_1/\Gamma_2$ with $\Gamma_1$ and $\Gamma_2$ being contributions from  $\Pi_\varphi^R$ (setting-sun diagram) and $\tilde{\Pi}_\varphi^R$ (fish diagram), respectively (i.e. the first and the second term in \eqref{FullGamma}). This plot shows that $\Gamma_1$ is dominant only for very small $\varphi$ and otherwise $\Gamma_2$ is the main contribution to the damping.  $\Gamma_1$ and $\Gamma_2$ are separately plotted in figure c) and d).}
	\label{Gamma}
\end{figure}

Using above results in \eqref{GammaSlowRoll}, we obtain the full damping coefficient for $M_\chi\ll M_\eta \ll T$
\begin{equation} \label{FullGamma}
\Gamma_\varphi \approx \frac{h^2\,T^2}{(4 \pi)^3 M_\eta} \log\left(\frac{M_\eta}{M_\chi}\right)+\frac{\varphi^2}{3} \frac{80 \pi}{T} 
\left[\left(\frac{h^2}{\lambda_\chi^2+3h^2}\right) \log\left(\frac{T}{M_\chi}\right)+\left(\frac{\lambda_\phi^2}{\lambda_\phi^2+3h^2}\right) \log\left(\frac{T}{M_\eta}\right) \right],
\end{equation} 
which is plotted in figure \ref{Gamma} as a function of $\varphi$ for various temperatures.
 
\subsubsection{Physical interpretation of the coefficients} \label{Interp}
\begin{figure}
	\center
	\includegraphics[scale=0.6]{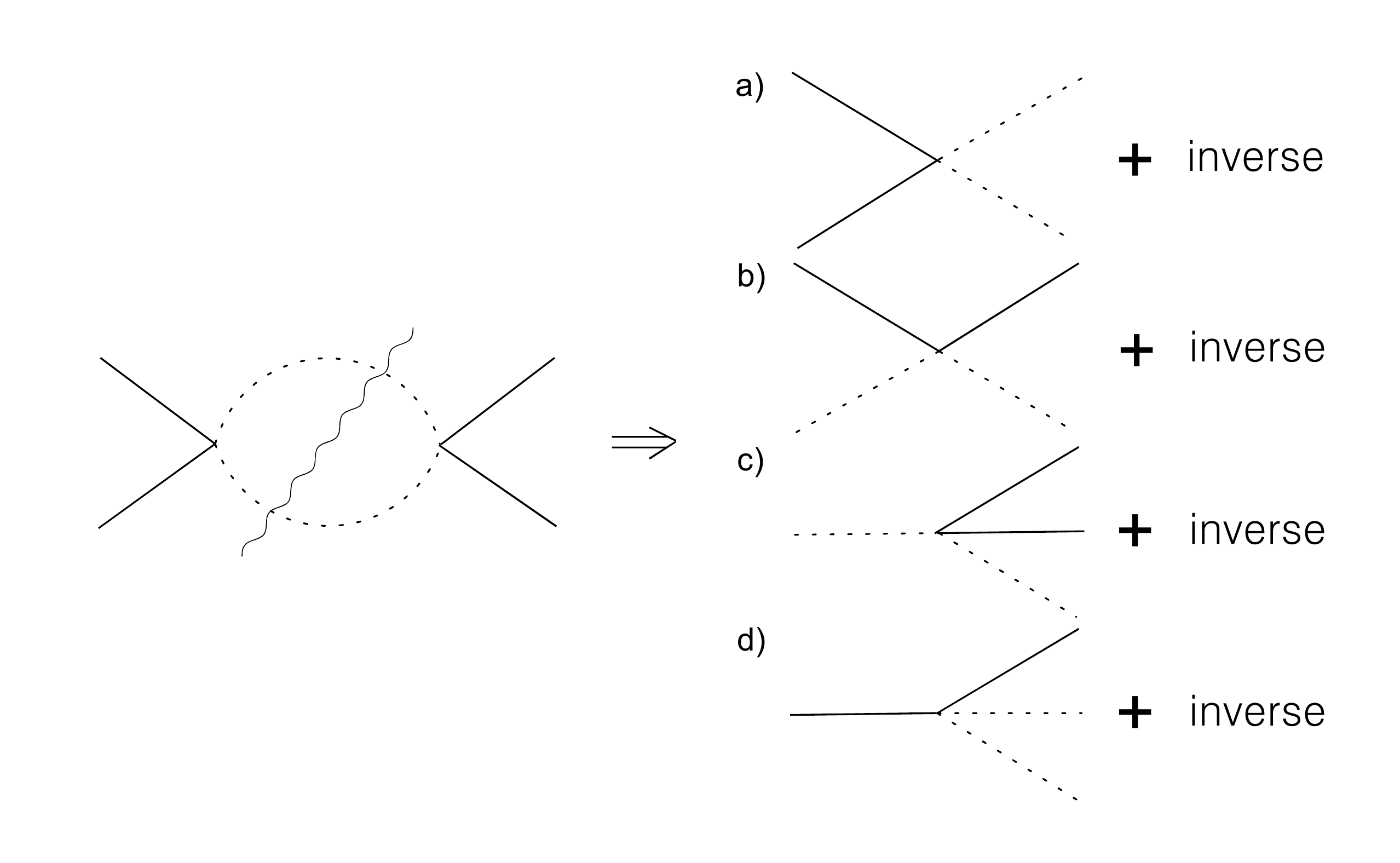}
	\caption{Cut through the fish diagram with zero width for the propagators. Solid lines correspond to $\phi$ and dashed lines to $\chi$. Due to the optical theorem at finite $T$, the imaginary part of the fish diagram can be interpreted in terms of physical processes in which the external legs and cut propagators appear as initial and final states \cite{Kobes:1986za,Kobes:1985kc,Bedaque:1996af,Landshoff:1996ta,Gelis:1997zv}. The diagrams on the right show those processes in which initial and final state both contain particles, as only these can contribute in the present calculation. Note that all $\chi$-particles here are on shell. 
Diagram a) corresponds to the process $\phi\phi \rightarrow \chi\chi$, in which two $\phi$-quanta out of $\varphi$-condensate become two $\chi$-quanta. If the total four-momenta of the two $\phi$-quanta is $(\omega,\textbf{0})$ with $\omega < 2 M_\chi$, this process is kinematically not allowed for on-shell $\chi$-particles. 
Diagram b) represents the scattering $\chi\phi\rightarrow\chi\phi$ between a $\chi$-quantum 
and a single $\phi$-quantum in the condensate  (in both, initial and final state). If the sum of four-momenta of the $\phi$-quanta  is $(\omega,\textbf{0})$ with any $\omega\neq0$, then it is clear that such process is also forbidden for on-shell $\chi$-particle due to the energy-momentum conservation. 
Diagram c) represents a decay process $\chi\rightarrow\chi\phi\phi$. Here the initial state contains a single $\chi$-quasiparticle with non-vanishing four-momentum. In the final state, there is also a $\chi$-quasiparticle 
and two $\phi$-quanta with combined four-momentum $(\omega,\textbf{0})$ have been added to the $\varphi$-condensate. This process is kinematically forbidden on-shell for any $\omega\neq0$, but the cross section is divergent for $\omega=0$.
Similarly, in diagram d) an $\phi$-quantum in the $\varphi$-condensate ``decays'' into an $\phi$ and two $\chi$-quasiparticles. 
The resulting total cross section is vanishing for any finite $\omega<2 M_\chi$ and diverges at $\omega=0$.
These intuitive interpretations hold at the level of individual quanta: The $\varphi$-condensate is a superposition of infinitely many $\phi$-quantum states with different particle numbers. However, if one decomposes the condensate into eigenstates of the free Hamiltonian (with well-defined particle number), then the same argument can be drawn for any multi-particle state.
}
	\label{NewOne}
\end{figure}

\begin{figure}
	\center
	\includegraphics[scale=0.5]{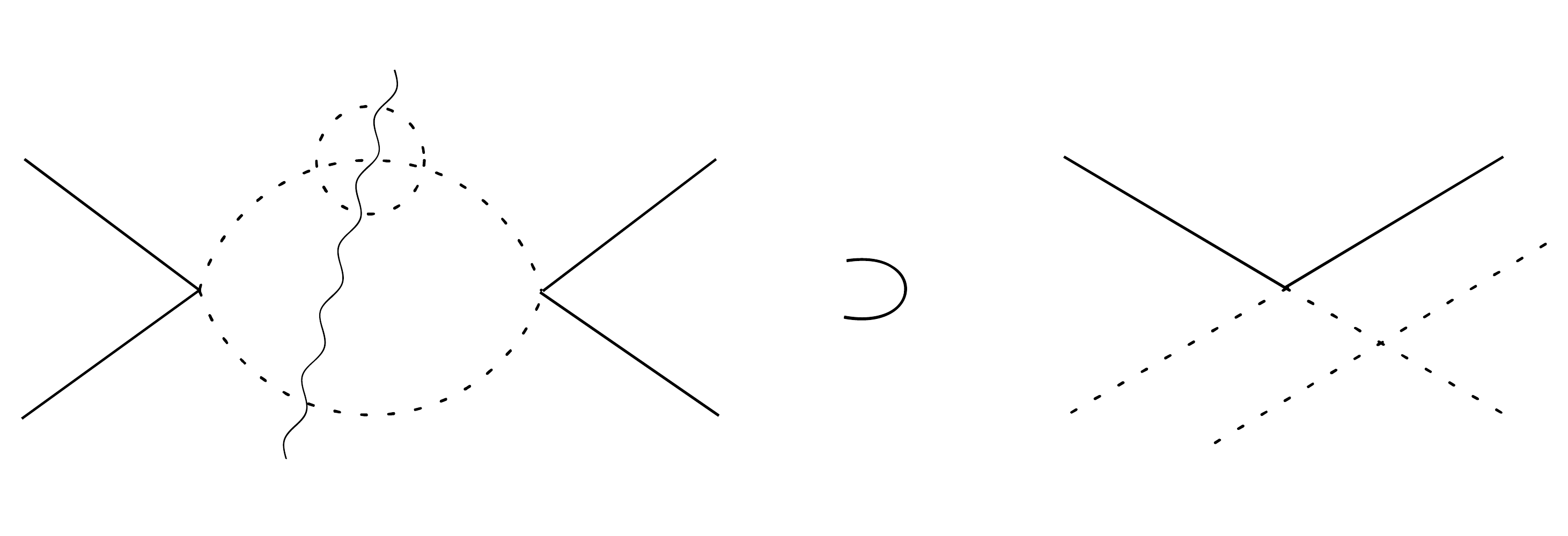}
	\caption{Cut through the fish diagram with a finite width ($\Gamma_\chi\neq0$) for the $\chi$-propagator in the loop. If one evaluates this fish diagram, then it includes contribution from process $\eta\chi\chi\rightarrow\eta\chi\chi$, which is mediated by an off-shell intermediate $\chi$-quasiparticle. This process is allowed and finite for $\omega=0$.}
	\label{ScatteringExample}
\end{figure}

The parametric dependence of (\ref{ImPiEstimate}) can easily be understood physically; the rate for $2\rightarrow2$ scatterings is proportional to $h^2$ and grows $\propto T^2$ due to the higher density of scattering partners, just as a quasiparticle damping rate would. The coefficient (\ref{ImtildePiresultat}) shows a more unusual behavior: most notably it decreases with temperature and with $\lambda_\chi$, both of which may seem counter-intuitive if one has the scattering interpretation in mind.
It should, however, be pointed out that neither the limit $T\rightarrow0$ nor $\lambda_\chi\rightarrow 0$ can be applied to the approximate analytic formula (\ref{ImtildePiresultat}) because it is only valid under the assumptions $M_\chi < T$ and  $\dot{\varphi}/\varphi<\Gamma_\chi\propto\lambda_\chi^2$, in which case the limit $\omega\rightarrow0$ is only justified.
In addition, the limit $\lambda_\chi\rightarrow 0$, which makes $\Gamma_\chi\propto\lambda_\chi^2$ vanish, is not allowed for following reasons. In the approximation $\Gamma_\chi=0$, the cut through the diagram in figure \ref{NewOne} at finite $T$ includes various processes, which are obviously kinematically impossible for any finite $\omega<2 M_\chi$ as explained in the caption. 
Moreover, for $\omega=0$ the rate has an unphysical divergence, which can be traced back to the infrared divergence of the Bose-Einstein distribution for the quasiparticles.\footnote{If we were dealing with real quasiparticles in the initial and final states (rather than those with vanishing external four-momentum in the condensate), then this divergence would be regularized by the thermal mass; there would be no such case as $\omega=0$ because the self-energies are evaluated at the quasiparticle mass shell. In a plasma the quasiparticles at rest are massive due to the thermal mass even if their vacuum mass is zero, such as for photons. In the limit $T\rightarrow0$ the thermal mass disappears and only the vacuum mass remains (which can be chosen to be zero), but the divergence from the Bose-Einstein distribution also disappears.} 
The divergence in the limit $\omega\rightarrow0$ is regularized by the finite width $\Gamma_\chi\neq 0$ of the intermediate $\chi$-particle in the contributing processes, see figure~\ref{ScatteringExample}. Therefore $\Gamma_\chi$ needs to be kept non-vanishing.

\begin{figure}
	\hspace{-1cm}
	\includegraphics[width=18cm]{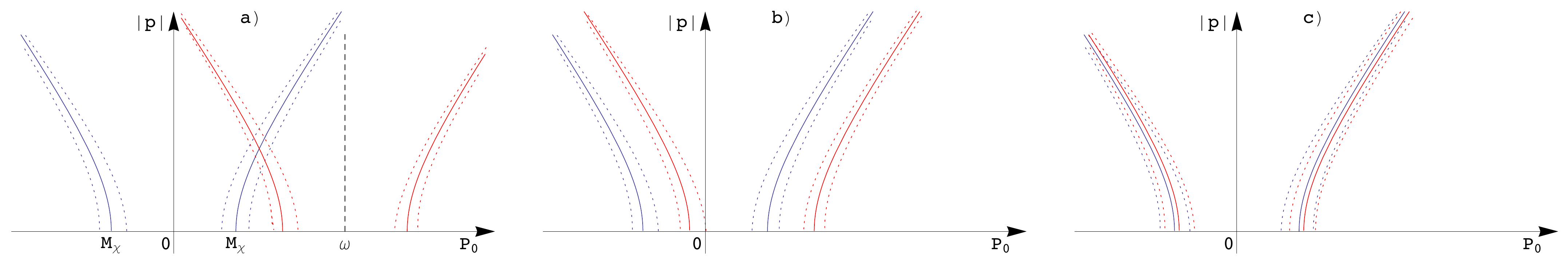}
	\caption{Support of the spectral densities $\rho_\chi(p_0)$ and $\rho_\chi(p_0-\omega)$  in $p_0$-$|\p|$ plane. 
	The solid blue and red lines correspond to the mass-shell  $p_0^2-|\p|^2 = M_\chi^2$ and $(p_0-\omega)^2-|\p|^2 = M_\chi^2$, respectively. 
	The dotted lines represent the approximate support of the spectral densities if the finite widths $\Gamma_\chi$ are included.  Figure a), b) and c) correspond to the cases of $\omega > 2 M_\chi \gg \Gamma_\chi$, $2 M_\chi>\omega > \Gamma_\chi$ and $2 M_\chi \gg \Gamma_\chi> \omega $, respectively. In figure a), the region where two mass-shell curves intersect gives rise to the non-vanishing contribution from kinematically allowed processes of the on-shell $\chi$-particles, such as decay and scattering.}
		\label{support}
\end{figure}

Though not infinite, the cross section is resonantly enhanced with finite $\Gamma_\chi$ as $1/\Gamma_\chi$, and it becomes larger for smaller $\Gamma_\chi\propto\lambda_\chi^2$. This explains the $\lambda_\chi^2$ in the denominator of (\ref{ImtildePiresultat}).
The resonant enhancement of the ``self-energy correction'' in figure~\ref{fish} a) is not present in the ``vertex correction'' in figure~\ref{fish} b). This allows us to neglect the vertex correction, which is of the same loop order as the self-energy correction.
Quantitatively there is not much difference between the rate at finite $\omega\ll\Gamma_\chi$ and $\omega=0$, as one can see from figure~\ref{support}.
The resonant enhancement would be weaker if we were dealing with fermions with gauge interactions in the loop (instead of $\chi$). The reason is that the damping $\Gamma_\chi$ in the scalar model only appears at two-loop order (i.e. from the setting sun diagram in figure \ref{setting}), while the mass correction is of order $\sqrt{\lambda_\chi}$ (from the tadpole in figure \ref{tad}). This makes $\Gamma_\chi/\Omega_\chi$ very small for the scalar model.

Finally, in section~II of \cite{Bartrum:2014fla} it has been claimed that all dissipative effects vanish if the light field $\chi$ is in a vacuum state.  This appears to be confirmed by our result for $\Pi_\varphi^R$.  Physically this conclusion seems, however, counter-intuitive. The basic laws of statistical mechanics imply that the energy in an interacting physical system should relax to a state of equilibrium, which implies equipartition of the energy amongst all degrees of freedom. If dissipation were absent for $T=0$, then all energy would forever remain trapped in $\varphi$ if the initial state of the system contains no $\chi$-particles.   The reason why we find vanishing dissipation rates for $T\rightarrow0$ in the present calculation lies in the approximation $\omega=0$. Note that ${\rm Im}\tilde{\Pi}_\varphi^R(\omega)$ in (\ref{OffShell}) gives a non-zero contribution for finite $\omega$ and $T=0$.  The precise determination of the physical damping rate at $T=0$ would, however, require some extra work and a more consistent treatment of renormalization. 

\begin{figure}
	\hspace{-1cm}
	\includegraphics[width=18cm]{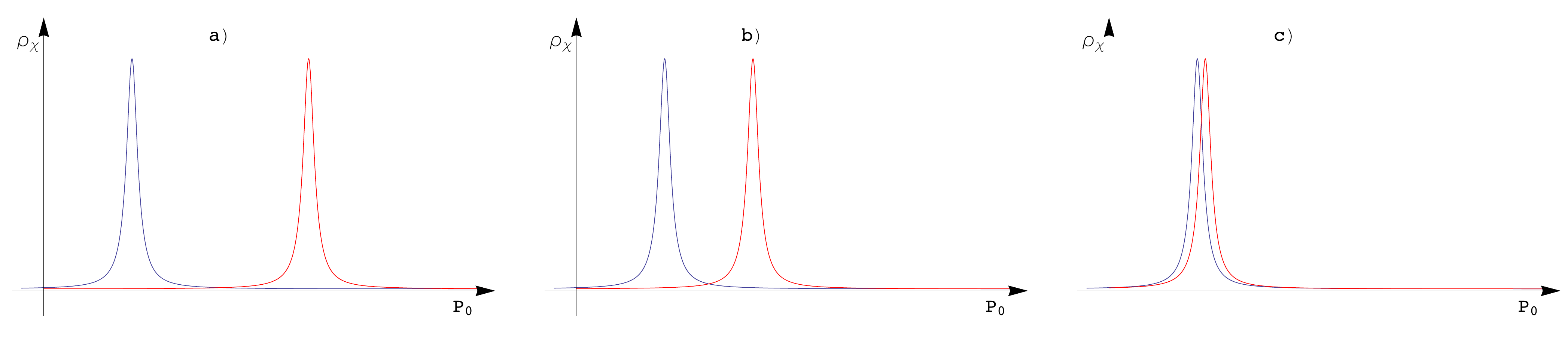}
	\caption{Spectral densities $\rho_\chi(p_0)$ (blue line) and $\rho_\chi(p_0-\omega)$ (red line) as a function of $p_0$ for fixed $|\p|$  (e.g. for $|\p|=0$). Figure a), b) and c) represent the same cases as in figure \ref{support}. }
	\label{rho}
\end{figure}

\section{Discussion} \label{discussion}
\subsection{Main results}
\paragraph{Effective equation of motion for the field $\varphi$} - In section \ref{Effective} we have shown from first principles that the expectation value $\varphi$ of a scalar field in a medium approximately follows the Markovian equation of motion (\ref{EOM}), which is of the type \eqref{FieldEqOfMot}, if the coupling to the medium is weak and the properties of the medium change sufficiently slowly. That is, the system can be characterized by a complex valued  effective potential. Its real part $\V(\varphi)$ is the usual finite temperature effective potential, and the imaginary part $\Gamma_\varphi$ is the dissipation coefficient. They depends on temperature and the field expectation  value $\varphi$. 
The equation of motion (\ref{EOM}) and the general expressions for $\V(\varphi)$ in (\ref{VSlowRoll}) and $\Gamma_\varphi$ in (\ref{GammaSlowRoll})
are amongst the main results of this work. The expression (\ref{GammaSlowRoll}) extends our calculation of $\Gamma_\varphi$ in \cite{Drewes:2013iaa}, which is valid near the potential minimum, to the case of large field values in the slow-roll phase.
They appear to be consistent with the equation of motion for the field $\phi$ found in the literature (see e.g.~\cite{Yokoyama:2004pf,Boyanovsky:2015xoa} and references therein). However, our derivation is much shorter, as we directly seek an equation of motion for $\varphi$.
Together with the expressions (\ref{DeltaMinus}) and (\ref{DeltaPlus}) for the propagators, the equation of motion \eqref{FieldEqOfMot} with $\V(\varphi)$ in (\ref{VSlowRoll}) and $\Gamma_\varphi$ in (\ref{GammaSlowRoll}) allows to describe the nonequilibrium dynamics of a slow-rolling scalar field entirely in terms of Markovian equations.

For small values of $\varphi$, and if the medium is composed of a sufficiently large thermal bath, it is well-known that the field $\phi$ is exposed  to Brownian motion, while its expectation value $\varphi$ performs damped oscillations and relaxes to its minimum on a time scale that is given by the thermal quasiparticle width in the plasma.  
We have recovered this behavior as a limiting case in \eqref{NewLangevin}.
Far away from the potential minimum the behavior is very different.

\paragraph{Effective potential and dissipation coefficient in a scalar theory} - In section \ref{model} we have applied the formulae derived in section~\ref{EffectiveActionSec} to a simple scalar model.  
Analogue results for the small field case discussed in section \ref{SmallField} are given in \cite{Drewes:2013iaa}.
The finite temperature effective potential and damping coefficient are approximately given by the analytic expressions \eqref{FullPotential} 
(equivalently \eqref{Vbeforeint}) and \eqref{FullGamma}. 
Our results appear to be in some tension with the previous literature, which we discuss in section~\ref{Tension}.

\paragraph{Loop integrals at finite temperature} - In the appendices we have provided general expressions and analytic estimates for nontrivial two-loop integrals for setting-sun diagrams at finite temperature. This includes the setting-sun diagram at vanishing external four-momentum (appendix \ref{vanishing_p}) and thermally on-shell cases (appendix \ref{chi4}). In appendix \ref{Angle} we have given analytic results for the angle integrals in setting-sun diagrams. These formulae will be very useful for further calculations in finite-temperature scalar field theory.

\subsection{Comparison with previous results}\label{Tension}

Our results for the damping coefficient $\Gamma_\varphi$ differ from those obtained in the previous literature in two ways.

\subsubsection{$\omega=0$ versus $\omega=M_\eta$} \label{omega}
A main difference between the damping coefficient (\ref{LangevinGamma}) during oscillations near the potential minimum 
and the expression (\ref{GammaSlowRoll}) for large field values in a slow-roll phase is that the diagrams in the former are evaluated on the quasiparticle mass shell, while in the latter they have to be evaluated at vanishing external energy (i.e. at vanishing external four-momentum, which is off shell). In spite of this, the coefficient (\ref{LangevinGamma}) is frequently used in the literature in all regimes, which is clearly incorrect.
This point has previously been realized by some authors, see e.g. \cite{Yokoyama:1998ju,Yokoyama:2004pf,BasteroGil:2010pb}. 
In spite of this realization, the authors of \cite{Yokoyama:1998ju,Yokoyama:2004pf} used the on-shell damping rate (\ref{LangevinGamma}) in their calculations, probably for simplicity. The authors of \cite{BasteroGil:2010pb}, on the other hand, calculated $\Gamma_\varphi$ with vanishing external four-momentum.

The fact that the damping coefficient in (\ref{GammaSlowRoll}) is evaluated at vanishing external four-momentum raises the questions to which degree our results are sensitive to various known infrared problems of thermal field theory \cite{Linde1980289, Braaten:PhysRevLett.74.2164, Pisarski:RevModPhys.53.43}, and whether further resummations and the inclusion of vertex (and ladder) diagrams are necessary to obtain physically consistent results. For example, in the calculation of $\tilde{\Pi}_\varphi^R$ we neglected the vertex diagram shown in figure \ref{fish} b). This can be justified in the present case because it is not subject to the resonant enhancement $1/\Gamma_\chi$ of the fish diagram in figure \ref{fish} a), cf. (\ref{tildePieq}). In the calculation of $\Pi_\varphi^R$, the main argument in favor of our approach is that the thermal masses in \eqref{Meta} - \eqref{Mchi} already appear at order $\sqrt{\lambda_\chi}$, $\sqrt{\lambda_\phi}$ and $\sqrt{h}$, while the thermal width is of higher order in the couplings. This tends to suppress all sorts of finite width and vertex corrections. However, this handwaving argument is of course not a strict proof, and we postpone the clarification of these issues to future work.


\subsubsection{Evaluation of the fish diagram}\label{FishDifference}
The dissipation coefficient $\Gamma_\varphi$ at leading order is given by the imaginary part of the retarded self-energy from the fish diagram, ${\rm Im}\tilde{\Pi}_\varphi^R(\omega)$.
The fact that it should be evaluated at vanishing external four-momenta (i.e. $\omega \rightarrow 0$ in \eqref{loopintegral})  
was previously pointed out in \cite{BasteroGil:2010pb,BasteroGil:2012cm}. 
However, the results obtained there at first sight appear to differ from ours. In the following we try to understand the origin of the difference.
Again recalling the optical theorem, we can identify the region in the integration volume of \eqref{loopintegral} where both spectral densities are thermally on-shell (i.e. at the intersection of two mass-shell curves in figure \ref{support} a)) as the non-vanishing contribution from kinematically allowed processes (e.g. decay and scattering) of the on-shell $\chi$-quasiparticles. 
From the kinematic considerations given in figure~\ref{NewOne}, it is clear that such a region exists for $\omega>2M_\chi$. 
For $\Gamma_\chi\ll\Omega_\chi$ it strongly dominates the integral \eqref{loopintegral}, see e.g.~\cite{Drewes:2010pf,Drewes:2013iaa} for a detailed discussion of this integral. 
In this case one can use the zero-width approximation $\rho_{\chi }(p_0)\simeq2\pi{\rm sign}(p_0)\delta(p_0^2-\Omega_\chi^2)$ to evaluate the integral. 
For $\Gamma_\chi < \omega <2M_\chi$ (see figure \ref{support} b)), the use of this approximation is not allowed. In this case the integral is dominated by the regions in which one of the $\rho_{\chi }$ is on-shell (see figure \ref{rho} b)).
Within these regions, the part, where the distribution functions $f_B$ have their maxima, gives the biggest contribution.
For the case $\omega \ll \Gamma_\chi$ under consideration in this work, the pole regions overlap in the entire integration volume, as depicted in figures~\ref{support} c) and figure \ref{rho} c).
This leads to a strong enhancement, and the integral is generally dominated by this pole region $p_0\simeq \Omega_\chi$. In this case we can use the expression (\ref{onepolerho}) for the spectral densities to obtain the results (\ref{OffShell}) and \eqref{tildePieq}.

The setup discussed in \cite{BasteroGil:2010pb,BasteroGil:2012cm} is an interesting exception, in which the integral \eqref{loopintegral} for $\omega \ll \Gamma_\chi$
is dominated by the off-shell regions in spite of the alignment of the two quasiparticle peaks illustrated in figures~\ref{support} c) and figure \ref{rho} c). The reason is that the contribution from the poles becomes Boltzmann suppressed for $M_\chi\gg T$. For sufficiently large $M_\chi$, the integral in \eqref{loopintegral} for $\omega\rightarrow 0$ is dominated by the $p_0\simeq 0$ region. 
In the model under consideration here, all thermal corrections to the $\chi$-propagators become suppressed in this limit, and one essentially recovers the vacuum limit. 
By adding another fields $\sigma$ with effective mass $M_\sigma\ll T \ll M_\chi$ and an interaction term $\chi\sigma^2$, the authors of \cite{BasteroGil:2010pb,BasteroGil:2012cm} have constructed an interesting scenario in which the loop integral is dominated by off-shell regions, but thermal corrections are not negligible. If the $\chi$-self-energy is dominated by the interactions with $\sigma$, then its temperature dependence is crucial and affects $\Gamma_\varphi$. 
This is the origin of the qualitatively different behaviour of the results found in those articles, as compared to ours. The different regimes are discussed in more detail in \cite{BasteroGil:2012cm}. They appear to be consistent with our results where the range of applicability overlaps.
Practically the integral \eqref{loopintegral} can be evaluated using the approximation 
\begin{equation}
\label{BerreraReplacement}
\rho_{\chi } \rightarrow 
-\frac{2{\rm Im}\Pi_\chi^R
}{M_\chi^4}.
\end{equation}
This approach was referred to as the low-momentum approximation.
Unfortunately this introduces a strong sensitivity to the infrared behaviour of the self-energy, which for a $\chi\sigma^2$ coupling is rather complicated and not fully understood \cite{Drewes:2013bfa}.
The strongly hierarchical arrangement $M_\sigma\ll T$ and $T, M_\eta \ll M_\chi$ corresponds to a rather specific corner in parameter space, but is cosmologically very interesting because it has been argued \cite{Berera:2001gs} that it provides a viable ``two-stage mechanism'' for warm inflation \cite{Berera:2008ar}.




\section{Conclusions and outlook} \label{conclusion}
We have studied the effective action and equation of motion for the expectation value $\varphi$ of a scalar field $\phi$ in a dense medium from first principles of nonequilibrium quantum field theory. We focused on two cosmologically important processes, damped oscillations near the ground state and a slow-roll phase in a flat potential. In a series of controlled approximations, we showed that these processes can be described in terms of Markovian effective equations of motion for both, the occupation numbers in the plasma and the field expectation value. This allows to describe the system in terms of a complex valued effective potential. The real part $\mathcal{V}(\varphi)$ is the usual effective potential at finite temperature, which includes radiative and thermodynamic corrections to the scalar potential. The imaginary part $\Gamma_\varphi$ is the dissipation coefficient that leads to damping and particle production.

When applying our results to the damped oscillations near the potential minimum, we found that our method reproduces the well-known results that have been obtained by studying the Brownian motion of $\phi$ near the ground state. As expected, the effective potential and damping coefficient for $\varphi$ coincide with the thermal mass and width of quasiparticles in the plasma. However, far away from the minimum we found a very different behavior. In particular, during a slow-roll phase, the loop integrals that determine the coefficients in the effective action have to be evaluated at vanishing external four-momentum and in resummed perturbation theory. This shows that the common practice, which uses the thermal mass and width of quasiparticles as the order-of-magnitude estimates for the thermal corrections to the effective potential and dissipation coefficient, is clearly incorrect in this situation.

We illustrated our results in a simple scalar model. 
In this context, we provided explicit expressions and analytic estimates for the setting-sun and fish diagrams at finite temperature. These will be very useful for future computations in scalar field theories at finite temperature.

Our results mark an important step forward towards a quantitative understanding of the evolution of scalar fields in the early universe. For instance, they can be applied to study the fate of moduli, curvatons, axions and other scalars with a flat potential during and after reheating. They may also be used to models of warm inflation. Finally, the scalar field may also represent an order parameter in applications outside the domain of cosmology.
Our analysis is, to the best of our knowledge, the most systematic and comprehensive one of the subject to date. However, it still relies on several restrictive assumptions. 
We have used thermal propagators in Minkowski space to evaluate loop integrals. This can only be justified if the particles in the loop have reached a local kinetic equilibrium, and when the thermal masses in the plasma are larger than the rate of Hubble expansion. Both of these assumptions should be relaxed in future work. Finally, the model in which we illustrate our results is a simple toy model in which the primordial plasma is composed only of another real scalar field. In most realistic applications, the scalar field in question couples directly or indirectly to fermions with non-Abelian gauge interactions. 
Some progress towards an inclusion of medium effects on the effective potential has recently been made e.g. in \cite{Mukaida:2012bz,Mukaida:2012qn,Drewes:2013iaa,Enqvist:2013qba,Mukaida:2013xxa,Miyamoto:2013gna,Mazumdar:2013gya,Harigaya:2013vwa,Bartrum:2014fla,Lerner:2015uca,Ho:2015jva,Boyanovsky:2015xoa}, and phenomenological implications have been discussed in \cite{Martin:2014nya,Kitajima:2014xna,Meyers:2013gua,Drewes:2014pfa,Dev:2014tla,Amin:2014eta,Cook:2015vqa,Domcke:2015iaa,Erickcek:2015jza,Rehagen:2015zma}, but it is still a long way to go to a complete quantitative understanding of scalar fields in the early universe in realistic models.


\section*{Acknowledgement}
This research project has been supported in parts by the Jiangsu Ministry of Science and Technology under contract BK20131264. We also acknowledge 985 Grants from the Chinese Ministry of Education, and the Priority Academic Program Development for Jiangsu Higher Education Institutions (PAPD). The work of M.D. has been supported by the Gottfried Wilhelm Leibniz program of the Deutsche Forschungsgemeinschaft (DFG) and by the DFG cluster of excellence Origin and Structure of the Universe. J.Kang, J.Kim and M.D. thank Fernando Quevedo, Kumar Narain, Paolo Creminelli, Giovanni Villadoro and Basudeb Dasgupta for the hospitality and useful discussions during the visits to the ICTP, where part of this work was done. We would also like to thank Michael J. Ramsey-Musolf and Arjun Berera  for stimulating discussions.

\begin{appendix}

\section{The self-energy $\Pi^-_{\varphi}$ from setting-sun diagram} \label{appendixA}
\begin{figure}
	\center
	\includegraphics[scale=0.2]{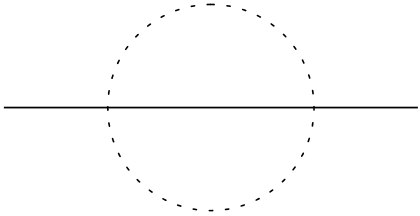}
	\caption{Setting sun diagram for $\eta^2\chi^2$ interaction. Solid line is $\eta$, while dashed line being $\chi$.}
	\label{fig:diagphi2sig2}
\end{figure}
The main goal of this appendix is to  evaluate the self-energy in \eqref{imaginarypart} from the setting-sun diagram shown in figure  \ref{fig:diagphi2sig2}, using the zero-width spectral density (\ref{rhofree}) neglecting $\rho^{\rm cont}$. 
For earlier discussions of these diagrams, see e.g. \cite{Parwani:1991gq,Gleiser:1993ea,Buchmuller:1997yw,Berera:1998gx,Moss:2006gt}.
We consider the spectral self-energy $\Pi^-_{\varphi}$, which is  related to the imaginary part of the retarded self-energy. 
\subsection{General expression}\label{appendixA1}
The general expression of $\Pi^-_{\varphi \p}(p_0)= 2 i {\rm Im}\Pi^R_{\varphi \p}(p_0)$ is given by 
\begin{eqnarray}
\Pi^-_{\varphi \p}(p_0)&=& -\frac{i h^2}{2} \int \frac{d^4q}{(2 \pi)^4} \frac{d^4k}{(2 \pi)^4} \frac{d^4 l}{(2 \pi)^4} (2 \pi)^4 \delta^{(4)}(p-q-k-l) 
\rho_{\eta}(q) \rho_{\chi}(k) \rho_{\chi}(l) \nonumber \\
&& \times \Big[\left(1+f_B(q_0)\right) \left(1+f_B(k_0)\right) \left(1+f_B(l_0)\right)
-f_B(q_0)f_B(k_0)f_B(l_0)\Big], \label{SettingSunA} 
\end{eqnarray}
which can be derived analogously to (\ref{loopintegral}).
After performing an integral over $q$ for $\eta$-propagator, we obtain (suppressing subscript $\varphi$ for $\Pi^-$)
\begin{eqnarray} \label{Pi_phi}
\Pi^-_{\p }(p_0)=- i \big(\mathcal{D}_{\p}(p_0)+\mathcal{S}_{\p 1}(p_0)+\mathcal{S}_{\p 2}(p_0)
-\mathcal{D}_\p(-p_0)-\mathcal{S}_{\p 1}(-p_0)-\mathcal{S}_{\p 2}(-p_0)\big)
\end{eqnarray}
with
\begin{eqnarray}
\mathcal{D}_{\p}(p_0)&=&\pi\frac{h^2}{8 (2\pi)^5}\theta(p_0)
\int_{M_\chi}^{p_0-M_\chi-M_\eta} d\Omega_{\l} \int_{M_\chi}^{\Omega_1^-} d\Omega_{\k} \, F_d(\Omega_{\k},\Omega_{\l},-A_1)\,\mathcal{I}(a_1,b,c), \label{D1} \\
\mathcal{S}_{\p 1}(p_0)&=&\pi\frac{h^2}{8 (2\pi)^5}
\int_{M_\chi}^\infty d\Omega_{\l} \int_{\Omega_1^+}^\infty d\Omega_{\k}\ F_{s1}(\Omega_{\k},\Omega_{\l},A_1)\,\mathcal{I}(a_1,b,c), \label{S1} \\ 
\mathcal{S}_{\p 2}(p_0)&=&2 \pi\frac{h^2}{8(2\pi)^5}\theta(p_0)
\int_{M_\chi}^\infty d\Omega_{\l} \int_{\Omega_2}^\infty d\Omega_{\k}\, F_{s2}(\Omega_{\k},\Omega_{\l},A_2)\,\mathcal{I}(a_2, b,c). \label{S2} 
\end{eqnarray}
Here
\begin{eqnarray}
F_d(\Omega_{\k},\Omega_{\l},A)&=&
\big(1+f_B(\Omega_{\k})\big) \big(1+f_B(\Omega_{\l})\big)
\big(1+f_B(A)\big)-
f_B(\Omega_{\k}) f_B(\Omega_{\l}) f_B(A), \\
F_{s1}(\Omega_{\k},\Omega_{\l},A)&=&
\big(1+f_B(\Omega_{\k})\big) \big(1+f_B(\Omega_{\l})\big)
f_B(A)-
f_B(\Omega_{\k}) f_B(\Omega_{\l})  \big(1+f_B(A)\big), \\
F_{s2}(\Omega_{\k},\Omega_{\l},A)&=&
f_B(\Omega_{\k}) \big(1+f_B(\Omega_{\l})\big)
\big(1+f_B(A)\big)
- \big(1+f_B(\Omega_{\k})\big) f_B(\Omega_{\l})  f_B(A), \\
\Omega_1^\pm&=&{\rm max}\big[p_0-\Omega_{\l}\pm M_\eta,M_\chi\big], \\
\Omega_2&=&{\rm max}\big[\Omega_{\l}-p_0 + M_\eta,M_\chi\big], \\
A_1&=&\Omega_{\k}+\Omega_{\l}-p_0, \label{A1} \\
A_2&=&\Omega_{\k}-\Omega_{\l}+p_0, \label{A2} \\
\mathcal{I}(a,b,c)&=&\int_{-1}^{1}dx \int_{-1}^1 dy\, \frac{\theta\left(I(x,y,z)\right)}{\sqrt{I(x,y,z)}}, \label{I_A}\\
z &=& a+ b\, y + c\, x, \\
I(x,y,z)&=&(1-x^2)(1-y^2)-(z-xy)^2,  \\
a_j&=& \frac{ A_j^2 - \p^2-\k^2-\l^2-M_\eta^2}{2 |\k| |\l|} \,\, {\rm with} \,\, j=1,2\,,\,\, b= \frac{|\p|}{|\k|}, \,\, c= \frac{|\p|}{|\l|}, \label{abc2}
\end{eqnarray}
The $\mathcal{D}$ terms represent decay and inverse decay $\eta\leftrightarrow \eta \chi\chi$. Note that the integration limits in \eqref{D1} imply that $\mathcal{D}_{\p}(p_0)$ vanishes for $p_0<2M_\chi+M_\eta$, as expected from energy-momentum conservation in the processes $\eta\leftrightarrow \eta \chi\chi$ that are impossible for on-shell particles.
$\mathcal{S}_{\p 1}(p_0)$ and $\mathcal{S}_{\p 2}(p_0)$ correspond to the damping by scattering processes (``Landau damping''), 
$\eta \eta\leftrightarrow\chi\chi$ and $\eta \chi \leftrightarrow\eta \chi$, respectively.  
The variables $x$ and $y$ are the cosines of the two nontrivial angles, i.e. $x=\frac{\p \cdot \k}{|\p| |\k|}, y=\frac{\p \cdot \l}{|\p| |\l|}$, so that $\mathcal{I}(a,b,c)$ represents integral over angles. 

The integrals cannot be performed analytically in general. For the special case $p\rightarrow 0$ under consideration, we can, however, find analytic approximations, see appendix \ref{vanishing_p} . 

\subsection{Analytic approximation for $|\p|=0$ and $\omega\rightarrow 0$ \label{vanishing_p}
} \label{Pi(omega)_phi}
Here we derive approximate estimates of $\Pi^-_{0}(\omega)$ for zero spatial momentum and arbitrarily small energy. 

In \eqref{Pi_phi}, if the energy $p_0=\omega$ is smaller than $2 M_\chi + M_\eta$, 
then the decay term  $\mathcal{D}$ doesn't contribute as the decay process is not allowed by the energy conservation. 
Therefore let us focus on Landau damping terms  \eqref{S1} and \eqref{S2}. The integral over $\Omega_\k$ does not  vanish, only if there exists an overlap between the integral region of $\Omega_\k$ and the region allowing $a_j^2<1$ (with $j=1, 2$) due to the step function $\theta(1- a_j^2)$ from the angle integral in \eqref{angle-zeromode} (now with $a=a_j$).  

Let us find the range of $\Omega_\k$ that allows $a_j^2<1$. Using the definition of $a_j$ given in \eqref{abc2} (with $\p=0$) and introducing dimensionless variables as
\begin{equation}
t= \frac{\Omega_\k}{M_\chi}\geq1,\,\, s= \frac{\Omega_\l}{M_\chi}\geq1,\,\, \tilde{\omega}= \frac{\omega}{M_\chi},\,\, u=\frac{M_\eta^2}{2 M_\chi^2},
\end{equation}
the range of $t$ for $a_j^2<1$ reads  
\begin{equation}
t_{j-} < t<t_{j+}   
\end{equation}  
with $t_{j\pm}$ being solutions of $a_j^2=1$, given (up to the first order in infinitesimal $\tilde{\omega}$) by
\begin{eqnarray}
t_{1\pm}&=& (u-1) s \pm \sqrt{u(u-2) (s^2-1)} \nonumber \\
 && \,\,  +\,\tilde{\omega} \left(1-u +(2u-1)s^2  \pm s \sqrt{u(u-2)(s^2-1)}\left(\frac{2u-3}{u-2} \right) \right) + \mathcal{O}(\tilde{\omega}^2), \\
t_{2\pm}&=& - (u-1) s \pm \sqrt{u(u-2) (s^2-1)} \nonumber \\
 && \,\,  +\,\tilde{\omega} \left(-1+u - (2u-1)s^2  \pm s \sqrt{u(u-2)(s^2-1)}\left(\frac{2u-3}{u-2} \right) \right) + \mathcal{O}(\tilde{\omega}^2).
 \nonumber \\
\end{eqnarray}

First of all, in order to have real $t_{j\pm}$, it should be that $u\geq2$ (i.e. $M_\eta\geq2 M_\chi$). If $u < 2$ (i.e. $M_\eta < 2 M_\chi$), then  it follows that $a_j^2>1$  and  integrals vanish. When $u\geq2$, one can see that 
$t_{2\pm}<0$ for arbitrarily small $\tilde{\omega}$ since  $(u-1) s>\sqrt{u(u-2) (s^2-1)}$, so there is no overlap between the integral range of $\Omega_k$ and the range allowing $a_2^2<1$. $\mathcal{S}_2$ thus vanishes due to the angle integral\footnote{This can be physically understood in the following way: $\mathcal{S}_2$ terms correspond to Landau damping $\eta \chi \leftrightarrow \eta \chi$, which effectively becomes decay $\chi \leftrightarrow \eta \chi$ for $\omega \to 0$ and this process is kinematically forbidden for on-shell particles.}. 

For $\mathcal{S}_1$ corresponding to Landau damping $\eta \eta \leftrightarrow \chi \chi$, which is the only potential non-vanishing contribution, one can show that $M_\chi t_{1\pm}>\Omega_1^+$ for arbitrarily small $\tilde{\omega}$, so that the integral over $\Omega_\k$ becomes
\begin{eqnarray}
 && \int_{\Omega_1^+}^\infty d\Omega_{\k}\, F_{s1}(\Omega_{\k},\Omega_{\l},A_1)\,\theta(1-a_1^2) = 
 \int_{M_\chi t_{1-}}^{M_\chi t_{1+}} d\Omega_{\k}\, F_{s1}(\Omega_{\k},\Omega_{\l},A_1)    \\
&& \,\, = T \Big[f_B(\Omega_\l) - f_B(\Omega_\l-\omega) \Big]  
  \log\left[\frac{f_B(-M_\chi t_{1+}) f_B(-M_\chi t_{1-}-\Omega_\l+\omega)}{f_B(-M_\chi t_{1-}) f_B(-M_\chi t_{1+}-\Omega_\l+\omega)} \right] \\
&& \,\, \stackrel{\omega \to 0}{=} \omega\, T\, \frac{d\, f_B(\Omega_\l)}{d\, \Omega_\l} 
\log\left[\frac{f_B(-\Omega_{1+}) f_B(-\Omega_{1-}-\Omega_\l)}{f_B(-\Omega_{1-}) f_B(-\Omega_{1+}-\Omega_\l)} \right]. \label{int-k}
\end{eqnarray}
In the second and third equality we have analytically performed the integral and taken limit $\omega\to0$ to obtain a result up to the first order in $\omega$. Here
\begin{eqnarray}
\Omega_{1\pm} = M_\chi t_{1\pm} (\omega=0)= M_\chi \Big[(u-1) s \pm \sqrt{u(u-2) (s^2-1)}\Big],
\end{eqnarray} 
which depend on $\Omega_\l$ and $M_\eta$ through $s$ and $u$. 

Let us  now consider the $\Omega_\l$-integration of \eqref{int-k}, i.e.
\begin{equation}
\omega\, T\,\int_{M_\chi}^\infty \,d\Omega_{\l}\, \frac{d\, f_B(\Omega_\l)}{d\, \Omega_\l} 
\log\left[\frac{f_B(-\Omega_{1+}) f_B(-\Omega_{1-}-\Omega_\l)}{f_B(-\Omega_{1-}) f_B(-\Omega_{1+}-\Omega_\l)} \right].  \label{S1-integral-l}
\end{equation}
This integral does not seem to allow an exact analytical result, so we  try to get an approximate estimate of this integral.   

\begin{figure}
	\center
	\includegraphics[scale=0.7]{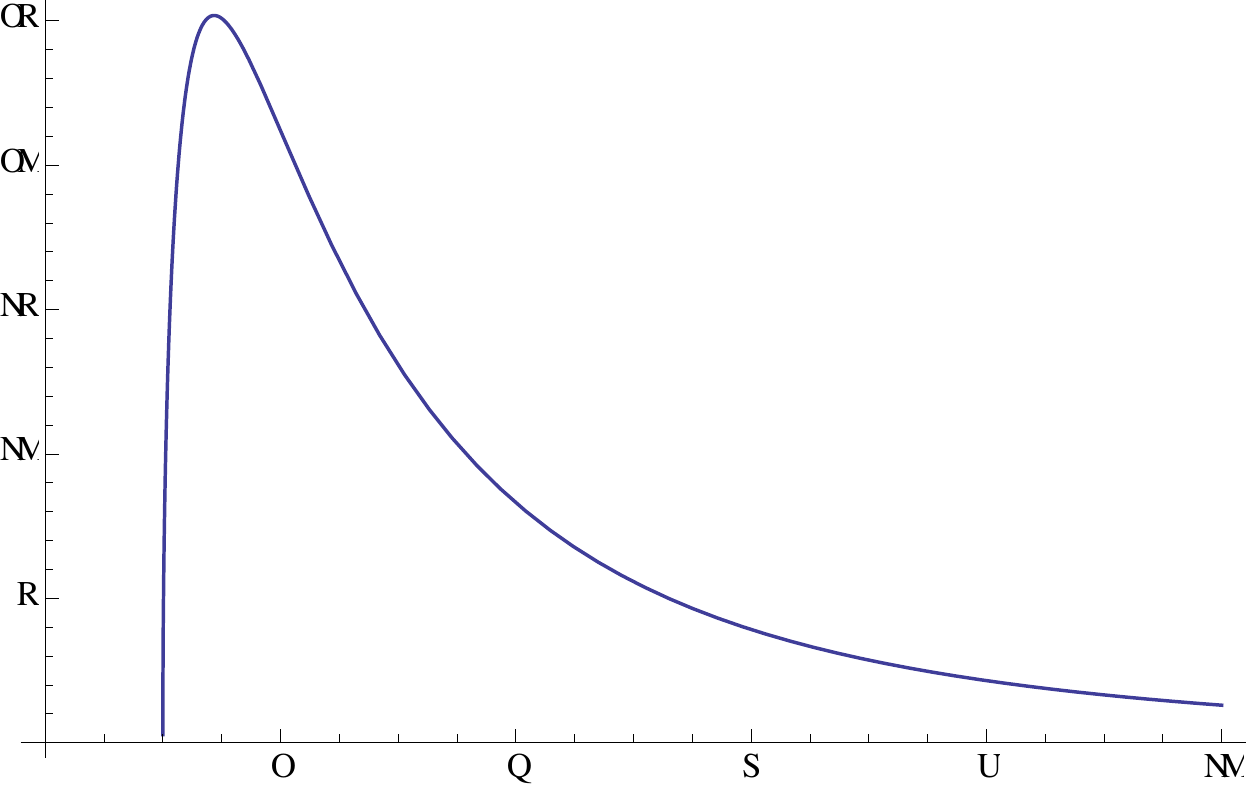}
	\caption{Integrand in \eqref{S1-integral-l} (multiplied by $T$) as a function of $s$ for $u=4$ (i.e. $M_\eta \simeq 3 M_\chi $) and $T=10 M_\chi$. Note that the function plotted here (the vertical axis) is dimensionless.
}
	\label{peak}
\end{figure}

First, when $u$ is close to 2 we use the fact that the integrand has a peak\footnote{To see that the integrand has a peak, note that the logarithm is zero at $s=1$, at which $\Omega_{1+}=\Omega_{1-}$, while $\left|\frac{d\, f_B(\Omega_\l)}{d\, \Omega_\l}\right|$ falls with $s\geq 1$. Furthermore, we numerically verified that the position of peak $1<s\lesssim2$ little changes with temperature $T$.}
around $s=2$ 
(i.e. $\Omega_\l = 2 M_\chi$) as shown in figure \ref{peak}
and fix the logarithm at the peak to pull it out of the integral. Then we can perform the integral (with lower limit $M_0$ instead of $M_\chi$) to get  
\begin{eqnarray}
- 2 \omega\, T  \left. \log\left[\frac{f_B(-\Omega_{1+}) f_B(-\Omega_{1-}-\Omega_\l)}{f_B(-\Omega_{1-}) f_B(-\Omega_{1+}-\Omega_\l)} \right]
 \right|_{\Omega_\l = M_0} f_B(M_0),
\end{eqnarray} 
where we have put an overall factor 2 to compensate the change of the lower limit of the integral.   
Indeed numerical analysis shows that  the above expression with $M_0=2 M_\chi$ is a good approximation of the integral for $2\leq M_\eta / M_\chi  \leq 3$.

\begin{figure}
	\center
	\includegraphics[scale=0.5]{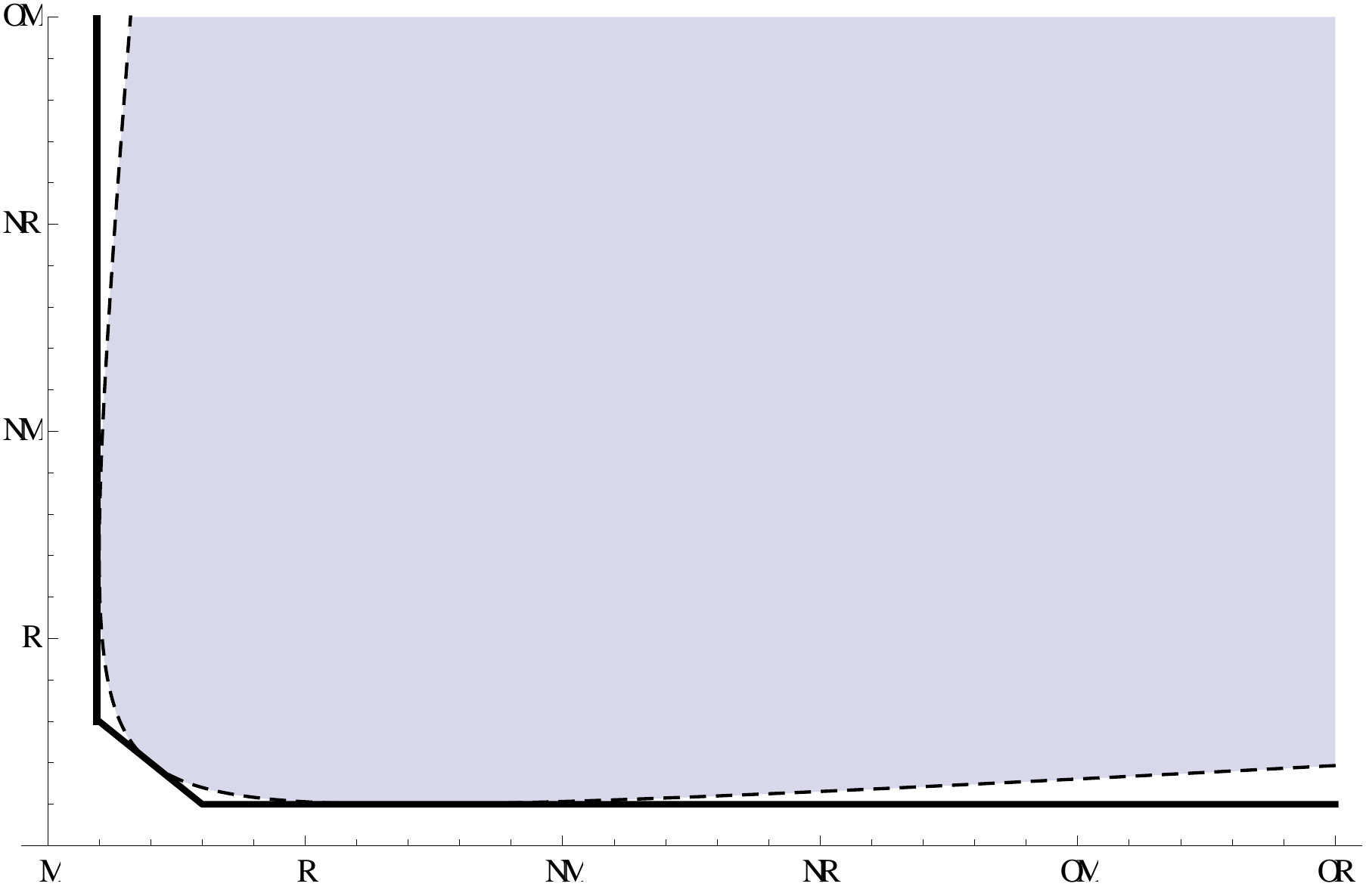}
	\caption{Integration region (shaded) in $s$-$t$ plane restricted by the angle integral (i.e. $a_1^2<1$) for $u=4$ (i.e. $M_\eta \simeq 3 M_\chi $). 
	The dashed curve is the solution of  $a_1^2=1$ consisting of $t_{1\pm}$, 
	while the solid lines correspond to the original boundary of the integral region when not taking into account the angle integral (see figure \ref{int_region} with $\Omega_\p = M_\eta$) .}
	\label{int-phi2sigma2}
\end{figure}

On the other hand, if $u$ becomes larger, then $\Omega_{1+}$ and $\Omega_{1-}$ approach the original integration boundary (see figure \ref{int-phi2sigma2}). Therefore for sufficiently big $u$, one can take $\Omega_{1+} \to \infty$ and $\Omega_{1-}\to\Omega_1^+= {\rm max}\big[M_\eta -\Omega_{\l},M_\chi\big]$  and then it is possible to analytically perform the integral giving
\begin{eqnarray}
&& \omega\, T \left[\log\left(1- e^{-\frac{M_\eta}{T}} \right) - 2 f_B(M_\eta) \left( \frac{M_\eta-2 M_\chi}{2 T} +\log\frac{f(M_\eta-M_\chi)}{f(M_\chi)}\right) \right. \nonumber \\  
&& \hspace{6cm}  \left. + 2 f_B(M_\chi)\, \log\left( \frac{1+ f_B(M_\eta-M_\chi)}{1+ f_B(M_\eta)}  \right) \right]. 
\end{eqnarray} 
We numerically checked that the above result agrees with the exact integral to a factor of $\mathcal{O}(1) $ for $M_\eta / M_\chi > 3$. 

Altogether we obtain\footnote{So far we have considered only $\mathcal{S}_1(\omega)$ in \eqref{Pi_phi} and have to include $\mathcal{S}_1(-\omega)$  to get $\Pi^-$. Inclusion of this term gives rise to an overall factor 2 since $\mathcal{S}_1(\omega)=-\mathcal{S}_1(-\omega)$}
an approximation of $\Pi_0^-(\omega)$ for infinitesimally small $\omega$ (to a factor of $\mathcal{O}(1)$)
\begin{eqnarray}
\lim_{\omega \to 0} \frac{i\, \Pi_0^-(\omega)}{\omega} \approx \frac{h^2\, T}{8 (2\pi)^3}
\begin{cases}
\left. 2 \log\left[\frac{f_B(-\Omega_{1+}) f_B(-\Omega_{1-}-\Omega_\l)}{f_B(-\Omega_{1-}) f_B(-\Omega_{1+}-\Omega_\l)} \right]
 \right|_{\Omega_\l \simeq 2 M_\chi} f_B(2 M_\chi)  & \text{ if }  2 M_\chi \lesssim M_\eta \lesssim 3 M_\chi,    \vspace{0.5cm} \\
  \log\left(1- e^{-\frac{M_\eta}{T}} \right) - 2 f_B(M_\eta) \left( \frac{M_\eta-2 M_\chi}{2 T} +\log\frac{f(M_\eta-M_\chi)}{f(M_\chi)}\right)  & \\
+ 2 f_B(M_\chi)\, \log\left( \frac{1+ f_B(M_\eta-M_\chi)}{1+ f_B(M_\eta)}  \right) 
  & \text{ if }  3 M_\chi \lesssim  M_\eta      \\
\end{cases}
\end{eqnarray}
and zero otherwise (i.e.  if $M_\eta \le 2 M_\chi$).  When $M_\chi \ll M_\eta \ll T$, the above expression (i.e. the second one) can be approximated, to leading order, as
\begin{equation} \label{Pi_A}
\lim_{\omega \to 0} \frac{i\, \Pi_0^-(\omega)}{\omega} \approx \frac{h^2\, T}{8 (2\pi)^3} \left(\frac{2 T}{M_\eta} \log\frac{M_\eta}{M_\chi}\right) \mathcal{O} (1).
\end{equation}

\section{The self-energy $\Pi_\chi^-$ from setting-sun diagram} \label{On-shell}

In this appendix we calculate the spectral self-energy $\Pi^-$ for the self-interacting scalar particle $\chi$ from the setting-sun diagram shown in figure \ref{fig:diagsig4}. In particular we find approximate estimates of $\Pi^-$ for the on-shell case (i.e. when the external momentum is on-shell) in appendix \ref{chi4}. This is needed to determine $\Gamma_\chi$ in \eqref{tildePieq} in section \ref{tildePi}. For $\p=0$,  $\Pi^-$ has already been calculated in \cite{Drewes:2013iaa} and here we generalize the results to arbitrary $\p\neq0$.

\subsection{General expression}
\begin{figure}
	\center
	\includegraphics[scale=0.2]{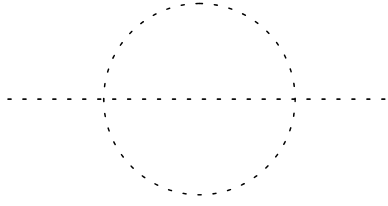}
	\caption{Setting sun diagram for $\lambda_\chi\chi^4/4!$ interaction. Dashed line is $\chi$}
	\label{fig:diagsig4}
\end{figure}

The spectral self-energy $\Pi_\chi^-$ from the setting-sun diagram in figure \ref{fig:diagsig4} has a general expression  
\begin{eqnarray}
\Pi^-_{\chi \p}(p_0)&=& -\frac{i \lambda_\chi^2}{6} \int \frac{d^4q}{(2 \pi)^4} \frac{d^4k}{(2 \pi)^4} \frac{d^4 l}{(2 \pi)^4} (2 \pi)^4 \delta^{(4)}(p-q-k-l) 
\rho_{\chi}(q) \rho_{\chi}(k) \rho_{\chi}(l) \nonumber \\
&& \times \Big[\left(1+f_B(q_0)\right) \left(1+f_B(k_0)\right) \left(1+f_B(l_0)\right)
-f_B(q_0)f_B(k_0)f_B(l_0)\Big],\label{SettingSunB}  
\end{eqnarray}
which can be derived in the same way as \eqref{SettingSunA}.
After integrations over $q$ and some angles of momenta $\k$ and $\l$, we get
\begin{eqnarray}
\Pi^-_{\chi \p}(p_0)=- i \big(\mathcal{D}^{[\chi]}_{\p}(p_0)+\mathcal{S}^{[\chi]}_\p(p_0)-\mathcal{D}^{[\chi]}_\p(-p_0)-\mathcal{S}^{[\chi]}_{\p}(-p_0)\big) \label{SettingSunB1}
\end{eqnarray}
with
\begin{eqnarray}
\mathcal{D}^{[\chi]}_{\p}(p_0)&=&\pi\frac{\lambda_\chi^2}{24(2\pi)^5}\theta(p_0)
\int_{M_\chi}^{p_0-2M_\chi} d\Omega_{\l} \int_{M_\chi}^{\Omega^-} d\Omega_{\k} \, F_d(\Omega_{\k},\Omega_{\l},-A)\,\mathcal{I}(a,b,c) \label{D} \\
\mathcal{S}^{[\chi]}_{\p}(p_0)&=&3\pi\frac{\lambda_\chi^2}{24(2\pi)^5}\theta(p_0)
\int_{M_\chi}^\infty d\Omega_{\l} \int_{\Omega^+}^\infty d\Omega_{\k}\, F_s(\Omega_{\k},\Omega_{\l},A)\,\mathcal{I}(a,b,c). \label{S}
\end{eqnarray}
Here
\begin{eqnarray}
F_d(\Omega_{\k},\Omega_{\l},A)&=&
\big(1+f_B(\Omega_{\k})\big) \big(1+f_B(\Omega_{\l})\big)
\big(1+f_B(A)\big)-
f_B(\Omega_{\k}) f_B(\Omega_{\l})  f_B(A), \\
F_s(\Omega_{\k},\Omega_{\l},A)&=&
\big(1+f_B(\Omega_{\k})\big) \big(1+f_B(\Omega_{\l})\big)
f_B(A)-
f_B(\Omega_{\k}) f_B(\Omega_{\l})  \big(1+f_B(A)\big), \label{F_s} \\
\Omega^\pm&=&{\rm max}\big[p_0-\Omega_{\l}\pm M_\chi,M_\chi\big], \\
A&=&\Omega_{\k}+\Omega_{\l}-p_0, \label{A} \\
\mathcal{I}(a,b,c)&=&\int_{-1}^{1}dx \int_{-1}^1 dy\, \frac{\theta\left(I(x,y,z)\right)}{\sqrt{I(x,y,z)}}, \label{I_B}\\
z&=& a + b\, y + c\, x, \\
I(x,y,z)&=&(1-x^2)(1-y^2)-(z-xy)^2, \\
a&=& \frac{ A^2 - \p^2-\k^2-\l^2-M_\chi^2}{2 |\k| |\l|}, \,\, b= \frac{|\p|}{|\k|}, \,\, c= \frac{|\p|}{|\l|}. 
\end{eqnarray}
The notations here are the same as in appendix \ref{appendixA1}.
The integration limits imply that $\mathcal{D}^{[\chi]}_{\p}(p_0)$ vanishes for $p_0<3M_\chi$, due to the energy-momentum conservation in the decay and inverse decay processes $\chi\leftrightarrow \chi\chi\chi$, which are not allowed for on-shell particles.

\subsection{Approximate estimates for on-shell quasiparticles} \label{chi4}
\begin{figure}
	\center
	\includegraphics[scale=0.5]{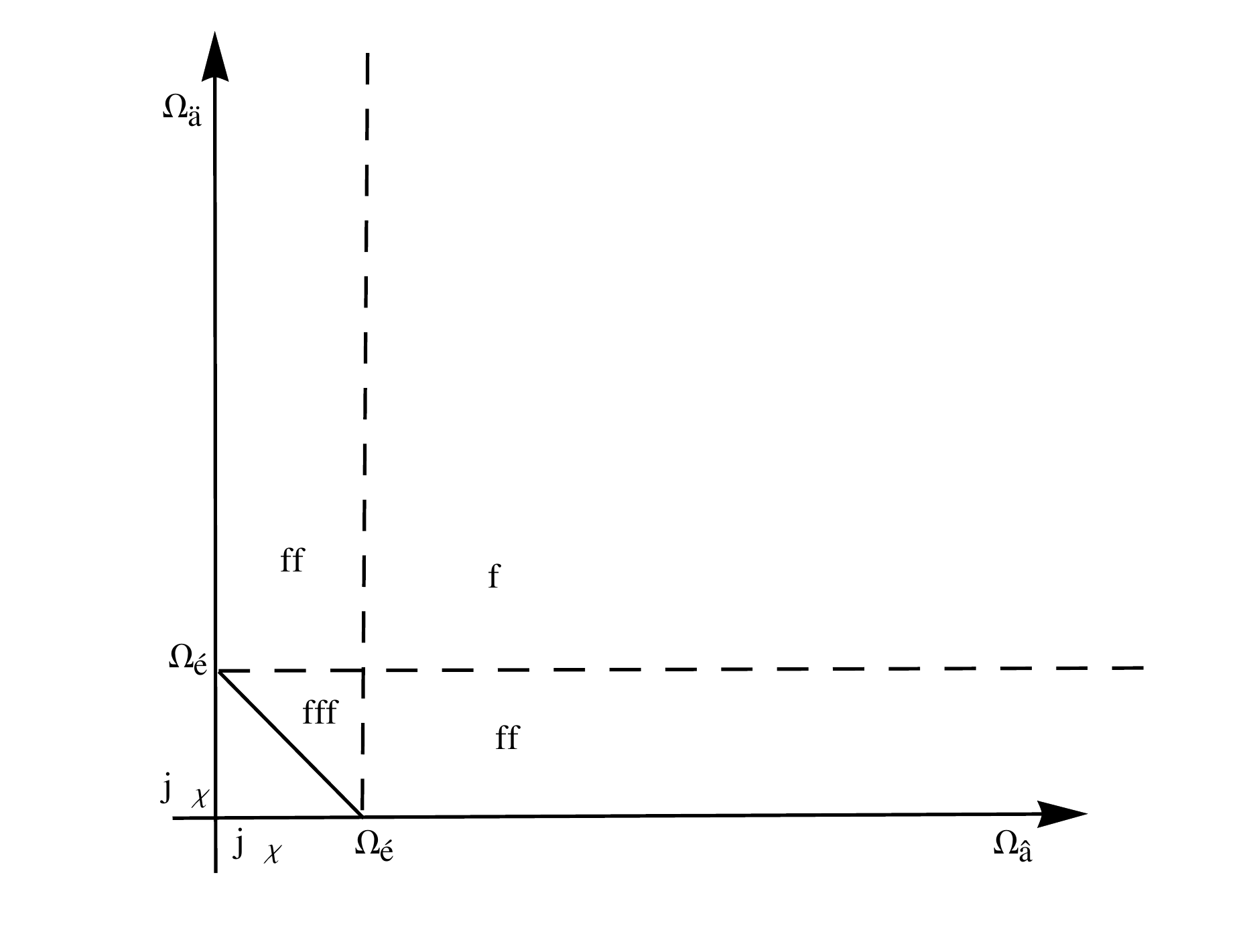}
	\caption{Division of the integral region in \eqref{Fs}. }
	\label{int_region}
\end{figure}

Here we derive approximate estimates of the self-energy \eqref{SettingSunB1} for the on-shell particle with arbitrary spatial momentum.
In the on-shell case (i.e. $p_0=\Omega_\p$), we have\footnote{Note that $\mathcal{D}$ terms and $\mathcal{S}^{[\chi]}_{\p}(-p_0)$ vanish for on-shell.}
\begin{eqnarray} \label{SE-onshell}
\Pi^-_{\chi \p}(\Omega_\p)= - 3\pi\frac{i\, \lambda_\chi^2}{24(2\pi)^5}
\int_{M_\chi}^\infty d\Omega_{\l} \int_{\Omega^+}^\infty d\Omega_{\k} \, F_s(\Omega_{\k},\Omega_{\l},A)\,\mathcal{I}(a,b,c).
\end{eqnarray}
Using the result on the angle integral given in  \eqref{I-sigma} in appendix \ref{onshell}, the above integral becomes
\begin{eqnarray}
\mathcal{F}_s&\equiv&\int_{M_\chi}^\infty d\Omega_{\l} \int_{\Omega^+}^\infty d\Omega_{\k} \, F_s(\Omega_{\k},\Omega_{\l},A)\,\mathcal{I}(a,b,c) \\
&=&2\pi\int_{\Omega_\p}^\infty d\Omega_{\l} \int_{\Omega_\p}^\infty d\Omega_{\k}\, F_s(\Omega_{\k},\Omega_{\l},A) \hspace{4.8cm} {\rm I}  \nonumber \\
&&+2 \left(\frac{2\pi}{|\p|}\right) \int_{M_\chi}^{\Omega_\p} d\Omega_{\l}\,|\l| \int_{\Omega_\p}^\infty d\Omega_{\k}\, F_s(\Omega_{\k},\Omega_{\l},A) \hspace{3cm} {\rm II}  \nonumber \\
&&+ \frac{2\pi}{|\p|} \int_{M_\chi}^{\Omega_\p} d\Omega_{\l} \int_{M_\chi+\Omega_\p-\Omega_\l}^{\Omega_\p} d\Omega_{\k}\, 
\sqrt{A^2-M_\chi^2}\,F_s(\Omega_{\k},\Omega_{\l},A). \hspace{1cm} {\rm III} \label{Fs}
\end{eqnarray}
In the second line, symmetry with respect to $\Omega_\k \leftrightarrow \Omega_\l$ has been taken into account, resulting in overall factor 2.  Here I, II and III refer to the corresponding integral regions in $\Omega_\k$-$\Omega_\l$ plane as shown in figure \ref{int_region}. 

The first line (integral over region I) can be analytically performed, giving
\begin{eqnarray} \label{I}
\mathcal{F}_s^{[{\rm I}]}&=&2\pi\int_{\Omega_\p}^\infty d\Omega_{\l} \int_{\Omega_\p}^\infty d\Omega_{\k}\, F_s(\Omega_{\k},\Omega_{\l},A)  \\
&=&2\pi\, T^2 \left[-{\rm Li}_2\left(\frac{1}{1-e^{\frac{\Omega_\p}{T}}}\right) - \frac{\left(\log\left(1-e^{-\frac{\Omega_\p}{T}}\right) \right)^2}{2} \right]  \label{F1}\\
&=& 
\begin{cases} 
2\pi\, T^2 \left[\pi^2/6\right] & \text{ if } \frac{\Omega_\p}{T} \to 0,  \\
0  & \text{ if } \frac{\Omega_\p}{T} \to \infty.
\end{cases} \label{I0}
\end{eqnarray}

When $\Omega_\p=M_\chi$, $\mathcal{F}_s^{[{\rm I}]} =\mathcal{F}_s$ and \eqref{F1} corresponds to the result of the entire $\mathcal{F}_s$ for zero mode, so that 
\begin{equation}
\Pi^-_{\chi 0}(M_\chi)= - 3\pi\frac{i\, \lambda_\chi^2\,(2\pi\, T^2)}{24(2\pi)^5} 
 \left[-{\rm Li}_2\left(\frac{1}{1-e^{\frac{M_\chi}{T}}}\right) - \frac{\left(\log\left(1-e^{-\frac{M_\chi}{T}}\right) \right)^2}{2} \right].  \label{Pi0}
\end{equation}

Now let's turn to integrals over region II and III. Concerning II, at most one integral can be performed, giving
\begin{eqnarray}
\mathcal{F}_s^{[{\rm II}]}&=&2 \left(\frac{2\pi}{|\p|}\right) \int_{M_\chi}^{\Omega_\p} d\Omega_{\l}\,|\l| \int_{\Omega_\p}^\infty d\Omega_{\k}\, F_s(\Omega_{\k},\Omega_{\l},A) \label{II0}\\
&=&\frac{4\pi\,T}{|\p|} \int_{M_\chi}^{\Omega_\p} d\Omega_{\l}\,|\l| \Big(f_B(\Omega_\l)-f_B(\Omega_\l-\Omega_\p)\Big) 
\log\left[\frac{f_B(-\Omega_\l)}{f_B(-\Omega_\p)}\right]. \label{II}
\end{eqnarray} 
The above integral and the one over region III,
\begin{eqnarray}
\mathcal{F}_s^{[{\rm III}]}=\frac{2\pi}{|\p|} \int_{M_\chi}^{\Omega_\p} d\Omega_{\l} \int_{M_\chi+\Omega_\p-\Omega_\l}^{\Omega_\p} d\Omega_{\k}\, 
\sqrt{A^2-M_\chi^2}\,F_s(\Omega_{\k},\Omega_{\l},A),  \label{III}
\end{eqnarray} 
can not be done analytically. Therefore we try to obtain approximations of the integrals over II and III by considering cases of $\Omega_\p<T$ and $\Omega_\p>T$ separately, under the assumption $M_\chi<T$.   

\subsubsection{$\Omega_\p<T$}

\paragraph{- By expansion of exponential in distribution functions:}

For $\Omega_\p<T$ one can expand the exponentials inside $f_B$ in \eqref{II}-\eqref{III}, i.e. $f_B(\Omega_\l) \approx T/\Omega_\l$ , and after some algebra we get 
\begin{eqnarray}
\mathcal{F}_s^{[{\rm II}]}&\approx&4\pi\,T^2\,\frac{\Omega_\p}{|\p|} \, G\left(\frac{M_\chi}{\Omega_\p} \right),  \label{FII}  \\
\mathcal{F}_s^{[{\rm III}]}&\approx& 2\pi\,T^2\, \frac{\Omega_\p}{|\p|} \, \left[G_1\left(\frac{M_\chi}{\Omega_\p} \right)
+ \frac{\Omega_\p}{T}\, G_2\left(\frac{M_\chi}{\Omega_\p} \right)\right] \label{FIII}
\end{eqnarray}  
with
\begin{eqnarray}
G(x) &=& - \int_x^1 dt \, \frac{\log t}{1-t} \, \sqrt{1-\frac{x^2}{t^2}}, \label{G0} \\ 
G_1(x) &=& \int_x^1 dt \int_{x+1-t}^1 ds \, \frac{\sqrt{(t+s-1)^2-x^2}}{t\,s\,(t+s-1)}, \label{G1}\\
G_2 (x) &=&  \int_x^1 dt \int_{x+1-t}^1 ds \, \frac{\sqrt{(t+s-1)^2-x^2}}{t+s-1}. \label{G2}
\end{eqnarray}
Here we have introduced variables $t=\Omega_\l/\Omega_\p$, $s=\Omega_\k/\Omega_\p$ and $x=M_\chi/\Omega_\p$. From above expressions for the $G$ functions it follows  that 
\begin{eqnarray}
& G(0)=\frac{\pi^2}{6}, & G(1)=0,  \\
& G_1(0)=\frac{\pi^2}{6}, & G_1(1)=0,  \\
& G_2(0)=\frac{1}{2}, & G_2(1)=0.  
\end{eqnarray}
We have numerically checked that for $x<1/2$ (i.e. $\Omega_\p>2 M_\chi$)  the $G$ functions in \eqref{G0}-\eqref{G2} change rather slow with $x$ and that for  $x<1/3$ (i.e. $\Omega_\p>3 M_\chi$) they are little different from the values at $x=0$. 
Thus for $3 M_\chi\lesssim \Omega_\p < T$ we can replace the $G$ functions in \eqref{FII}-\eqref{FIII} by their values at zero, giving
\begin{eqnarray}
\mathcal{F}_s^{[{\rm II}]}&\approx&4\pi\,T^2\,\frac{\Omega_\p}{|\p|} \, \frac{\pi^2}{6}\,\mathcal{O}(1),   \label{FsII} \\
\mathcal{F}_s^{[{\rm III}]}&\approx& 2\pi\,T^2\, \frac{\Omega_\p}{|\p|} \, \left[\frac{\pi^2}{6}\,\mathcal{O}(1)
+ \frac{\Omega_\p}{2T}\, \mathcal{O}(1)\right].  \label{FsIII}
\end{eqnarray}  
Together with \eqref{F1}, the entire result for the self-energy \eqref{SE-onshell} reads 
\begin{eqnarray}
\Pi^-_{\chi \p}(\Omega_\p) \approx  3\pi\frac{i\, \lambda_\chi^2\,(2\pi\,T^2)}{24(2\pi)^5} \, 
\left[ {\rm Li}_2\left(\frac{1}{1-e^{\frac{\Omega_\p}{T}}}\right) + \frac{\left(\log\left(1-e^{-\frac{\Omega_\p}{T}}\right) \right)^2}{2} 
 - \frac{\Omega_\p}{|\p|} \left(\frac{\pi^2}{2}
+ \frac{\Omega_\p}{2 T}\right)  \right]. \nonumber \\
\end{eqnarray}

\paragraph{- Another way:}

The function $F_s(\Omega_{\k},\Omega_{\l},A)$, given in \eqref{F_s}, has a maximum at $\Omega_\l=M_\chi$ and $\Omega_\k=\Omega_\p$ (also at $\Omega_\k=M_\chi$ and $\Omega_\l=\Omega_\p$, by  symmetry $\Omega_\l\leftrightarrow\Omega_\p$). 
Around this point, $F_s$ is dominated by the term $f_B(\Omega_\l)^2$ for $\Omega_\l<T$. 
Then one can use  $f_B(\Omega_\l) \approx T/\Omega_\l $ to see that $|\l|\, f_B(\Omega_\l)^2 $ has a peak at $|\l|\sim M_\chi$. 
Therefore the integrand in \eqref{II0} has a maximum at  $|\l|\sim M_\chi$ and in a similar way one can show that the integrand in  \eqref{III} has a peak when $\sqrt{A^2-M_\chi^2} \sim M_\chi$.
Using these facts, we can obtain approximations to \eqref{II0} and \eqref{III} by replacement $|\l| \rightarrow M_\chi$ in \eqref{II0} and $\sqrt{A^2-M_\chi^2}  \rightarrow M_\chi$ in \eqref{III}. Then the integrals can be analytically performed, giving
\begin{eqnarray}
\mathcal{F}_s^{[{\rm II}]}&=&2 \left(\frac{2\pi}{|\p|}\right) \int_{M_\chi}^{\Omega_\p} d\Omega_{\l}\,|\l| 
\int_{\Omega_\p}^\infty d\Omega_{\k}\, F_s(\Omega_{\k},\Omega_{\l},A)\nonumber  \\
&\approx& 2 \left(\frac{2\pi M_\chi}{|\p|}\right) \int_{M_\chi}^{\Omega_\p} d\Omega_{\l}\,\int_{\Omega_\p}^\infty d\Omega_{\k}\, F_s(\Omega_{\k},\Omega_{\l},A) \nonumber \\
&=& \left(\frac{4 \pi M_\chi T^2}{|\p|}\right) \left[ \frac{\pi^2}{6} + \frac{\left(\log\left(1 - e^{-M_\chi/T}\right)\right)^2}{2} 
- \frac{\left(\log\left(1 - e^{-\Omega_\p/T}\right)\right)^2}{2}  \right. \nonumber \\
&& \left.+ \log\left(\frac{1 - e^{-\Omega_\p/T}}{1 - e^{-M_\chi/T}}\right) \log\left(e^{-M_\chi/T} - e^{-\Omega_\p/T}\right) - 
 {\rm  Li}_2\left(\frac{1 - e^{-M_\chi/T}}{1 - e^{-\Omega_\p/T}}\right)
\right]
\end{eqnarray} 
and 
\begin{eqnarray}
\mathcal{F}_s^{[{\rm III}]}&=&\frac{2\pi}{|\p|} \int_{M_\chi}^{\Omega_\p} d\Omega_{\l} \int_{M_\chi+\Omega_\p-\Omega_\l}^{\Omega_\p} d\Omega_{\k}\, 
\sqrt{A^2-M_\chi^2}\,F_s(\Omega_{\k},\Omega_{\l},A), \\
&\approx&\frac{2\pi M_\chi}{|\p|} \int_{M_\chi}^{\Omega_\p} d\Omega_{\l} \int_{M_\chi+\Omega_\p-\Omega_\l}^{\Omega_\p} d\Omega_{\k}\, 
F_s(\Omega_{\k},\Omega_{\l},A), \nonumber \\
&=&\frac{2\pi M_\chi T^2}{|\p|} \left[
-\frac{\pi^2}{6} + \frac{\left(\log\frac{1 - e^{-M_\chi/T}}{1 - e^{-\Omega_\p/T}}\right)^2}{2} 
- (\Omega_\p + M_\chi) \log\left(\frac{1 - e^{-M_\chi/T}}{1 - e^{-\Omega_\p/T}}\right)  \right.  \nonumber \\
&& \left.+ 
 \log\left(\frac{1 - e^{-M_\chi/T}}{1 - e^{-\Omega_\p/T}}\right) \log\left(\frac{1 - e^{(M_\chi - \Omega_\p)/T}}{1 - 
      e^{(-M_\chi - \Omega_\p)/T}}\right) - {\rm Li}_2 \left(\frac{e^{-M_\chi/T} - e^{-\Omega_\p/T}}{e^{-M_\chi/T} - 1}\right)\right. \nonumber \\
 && \left. + 
 {\rm Li}_2\left(\frac{1 - e^{-M_\chi/T}}{1 - e^{(-M_\chi - \Omega_\p)/T}}\right) - {\rm Li}_2\left(\frac{1 - e^{-\Omega_\p/T}}{1 - e^{(-M_\chi - \Omega_\p)/T}}\right) + 
 {\rm Li}_2 \left(\frac{1 - e^{-M_\chi/T}}{1 - e^{-\Omega_\p/T}}\right)
\right]. \nonumber \\
\end{eqnarray} 
These expressions have more involved forms than \eqref{FsII}-\eqref{FsIII}, which can be seen as approximations to the above expressions. 

\subsubsection{$\Omega_\p\gtrsim T $}

In this case the integrands in \eqref{II0} and \eqref{III} are maximal at $|\l| \sim T$ and $\sqrt{A^2-M_\chi^2}\sim T$, respectively.
Therefore we can obtain approximations by replacing $|\l|\rightarrow T$ in \eqref{II0} and $\sqrt{A^2-M_\chi^2}\rightarrow T$ in \eqref{III}.  
Then we can analytically perform the integrals to obtain
\begin{eqnarray}
\mathcal{F}_s^{[{\rm II}]}&=&2 \left(\frac{2\pi}{|\p|}\right) \int_{M_\chi}^{\Omega_\p} d\Omega_{\l}\,|\l| 
\int_{\Omega_\p}^\infty d\Omega_{\k}\, F_s(\Omega_{\k},\Omega_{\l},A)\nonumber  \\
&\approx& 2 \left(\frac{2\pi T}{|\p|}\right)  \int_{M_0}^{\Omega_\p} d\Omega_{\l}\,\int_{\Omega_\p}^\infty d\Omega_{\k}\, F_s(\Omega_{\k},\Omega_{\l},A) \nonumber \\
&=& \left(\frac{4 \pi T^3}{|\p|}\right) \left[ \frac{\pi^2}{6} + \frac{\left(\log\left(1 - e^{-M_0/T}\right)\right)^2}{2} 
- \frac{\left(\log\left(1 - e^{-\Omega_\p/T}\right)\right)^2}{2}  \right. \nonumber \\
&& \left.+ \log\left(\frac{1 - e^{-\Omega_\p/T}}{1 - e^{-M_0/T}}\right) \log\left(e^{-M_0/T} - e^{-\Omega_\p/T}\right) - 
 {\rm  Li}_2\left(\frac{1 - e^{-M_0/T}}{1 - e^{-\Omega_\p/T}}\right)
\right]
\end{eqnarray} 
and 
\begin{eqnarray}
\mathcal{F}_s^{[{\rm III}]}&=&\frac{2\pi}{|\p|} \int_{M_\chi}^{\Omega_\p} d\Omega_{\l} \int_{M_\chi+\Omega_\p-\Omega_\l}^{\Omega_\p} d\Omega_{\k}\, 
\sqrt{A^2-M_\chi^2}\,F_s(\Omega_{\k},\Omega_{\l},A) \\
&\approx&\frac{2\pi T}{|\p|} \int_{M_0}^{\Omega_\p} d\Omega_{\l} \int_{M_0+\Omega_\p-\Omega_\l}^{\Omega_\p} d\Omega_{\k}\, 
F_s(\Omega_{\k},\Omega_{\l},A) \nonumber \\
&=&\frac{2\pi T^3}{|\p|} \left[
-\frac{\pi^2}{6} + \frac{\left(\log\frac{1 - e^{-M_0/T}}{1 - e^{-\Omega_\p/T}}\right)^2}{2} 
- (\Omega_\p + M_0) \log\left(\frac{1 - e^{-M_0/T}}{1 - e^{-\Omega_\p/T}}\right)  \right.  \nonumber \\
&& \left.+ 
 \log\left(\frac{1 - e^{-M_0/T}}{1 - e^{-\Omega_\p/T}}\right) \log\left(\frac{1 - e^{(M_0 - \Omega_\p)/T}}{1 - 
      e^{(-M_0 - \Omega_\p)/T}}\right) - {\rm Li}_2 \left(\frac{e^{-M_0/T} - e^{-\Omega_\p/T}}{e^{-M_0/T} - 1}\right)\right. \nonumber \\
 && \left. + 
 {\rm Li}_2\left(\frac{1 - e^{-M_0/T}}{1 - e^{(-M_0 - \Omega_\p)/T}}\right) - {\rm Li}_2\left(\frac{1 - e^{-\Omega_\p/T}}{1 - e^{(-M_0 - \Omega_\p)/T}}\right) + 
 {\rm Li}_2 \left(\frac{1 - e^{-M_0/T}}{1 - e^{-\Omega_\p/T}}\right)
\right].  \nonumber \\
\end{eqnarray} 
Here we have replaced $M_\chi$ in the low limits of the integral by some constant $M_0$, which we choose $M_0\sim T/2$. For  $\Omega_\p \gg T$, from above we get
\begin{eqnarray}
\mathcal{F}_s^{[{\rm II}]}+ \mathcal{F}_s^{[{\rm III}]} &\approx& 2 \pi T^2 \left(\frac{M_0}{T}-\log(e^{M_0/T}-1)\right)  \\
&\simeq& 2 \pi T^2\, \mathcal{O}(1), 
\end{eqnarray} 
and this is the result of the whole $\mathcal{F}_s$ since $\mathcal{F}_s^{[{\rm I}]}\simeq 0$ for $\Omega_\p \gg T$ (see \eqref{I0}).

To sum up, when $M_\chi \ll T$ the spectral self-energy for on-shell $\chi$-particle with any momentum $\p$ can be approximated as 
\begin{equation} \label{On-shell-Pichi}
\Pi^-_{\chi \p}(\Omega_\p) \approx - 3\pi\frac{i\, \lambda_\chi^2\,(2\pi\,T^2)}{24(2\pi)^5}\, \mathcal{O}(1).
\end{equation}

\section{The angle integral in setting-sun diagrams} \label{Angle}
In this appendix we estimate the angle integrals for setting-sun diagrams $\mathcal{I}(a,b,c)$, see \eqref{I_A} and \eqref{I_B}. 
The angle integral has a general form
\begin{eqnarray}
\mathcal{I}(a,b,c)&=&\int_{-1}^{1}dx \int_{-1}^1 dy\, \frac{\theta\left(I(x,y,z)\right)}{\sqrt{I(x,y,z)}}, \\
I(x,y,z)&=&(1-x^2)(1-y^2)-(z-xy)^2, \\
z&=& a + b\, y + c\, x, \\
a&=& \frac{ A^2 - \p^2-\k^2-\l^2-M^2}{2 |\k| |\l|}, \,\, b= \frac{|\p|}{|\k|}, \,\, c= \frac{|\p|}{|\l|}. \label{abc1}
\end{eqnarray}
Here $A$ is a function of $p_0$, $\Omega_\k$ and $\Omega_\l$, see e.g. \eqref{A}. First we consider the cases with zero external momentum $\p=0$ (on-shell or off-shell) and then turn to the on-shell cases with non-zero momentum $\p\neq0$. 

\subsection{Zero mode ($\p=0$)}

\begin{figure}
	\center
	\includegraphics[scale=.5]{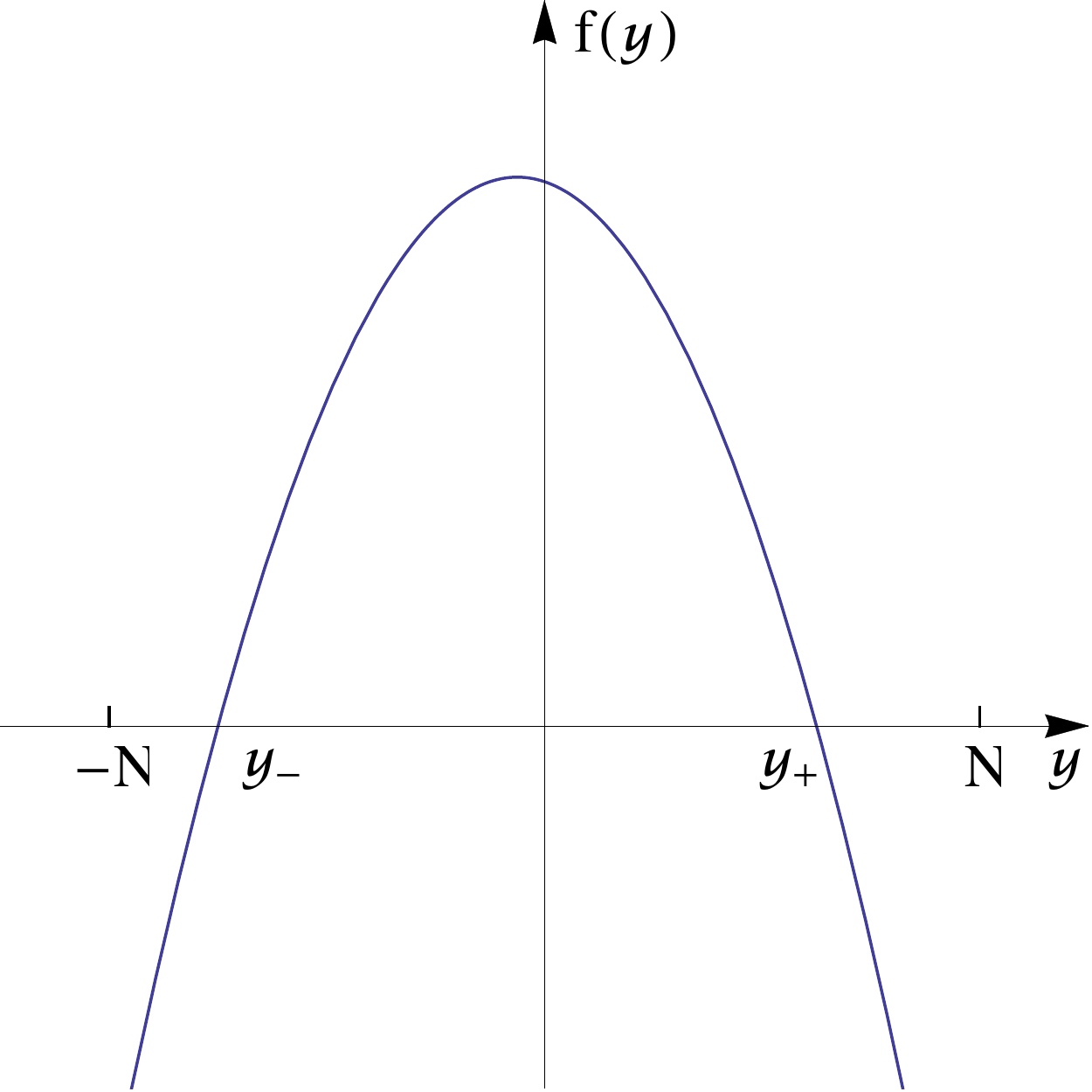}
	\caption{$I$ as a function of $y$ when $-1<y_-, y_+<1$.}
	\label{Iintegral}
\end{figure}

In this case we have $b=0=c$ and $z=a$, so that
\begin{eqnarray}
I(x,y,z)= (1-x^2)(1-y^2)-(a-x\,y)^2.
\end{eqnarray}
If $a^2<1$ and $x^2<1$, there exists $-1<y_{\pm}<1$ that solve $I(x,y)=0$, so that one can write (see figure \ref{Iintegral})
\begin{eqnarray}
I(x,y,z)= -(y-y_-)(y-y_+)
\end{eqnarray}
with
\begin{eqnarray}
y_{\pm}=  a\,x \pm\sqrt{(1-x^2)(1-a^2)}
\end{eqnarray}
One can easily see that $|y_{\pm}|<1$ for $a^2<1$ and $x^2<1$.  
Then assuming $a^2<1$ the angle integral can be performed, giving
\begin{eqnarray}
\mathcal{I}&=&\int_{-1}^{1}dx \int_{-1}^1 dy\, \frac{\theta\left(I(x,y)\right)}{\sqrt{I(x,y)}} \\
&=&\int_{-1}^{1}dx \int_{y_-}^{y_+} dy\, \frac{1}{\sqrt{-(y-y_-)(y-y_+)}} \\
&=&\int_{-1}^{1}dx\, \pi \\
&=& 2 \pi.
\end{eqnarray} 
Note that in the second line the $y$-integral gives $\pi$ irrespectivly of $x$, though $y_\pm$ depends on $x$.
If $a^2>1$, there exists no real $y_\pm$ for $x^2<1$. This implies $I(x,y)<0$ for $x^2<1$ and $y^2<1$, giving vanishing result for the integral due to $\theta(I(x,y))$ in the integrand.
 
The final result for zero mode is thus 
\begin{eqnarray} \label{angle-zeromode}
\mathcal{I}= 2 \pi\, \theta(1-a^2).
\end{eqnarray}
Note that $a$ depends on $p_0$, $|\k|$, $|\l|$ and $M_\chi$ by \eqref{abc1}, so that the step-function $\theta(1-a^2)$ in \eqref{angle-zeromode} restricts the integral region on $\Omega_{\k}$-$\Omega_{\l}$ plane to the one allowing $a^2<1$. 

\subsection{Non-zero mode ($\p\neq0$) }

Here we generalize the zero-mode result obtained above to the cases of non-zero mode. First we consider generic cases, which include off-shell as well as on-shell, and then focus on the latter, which is needed to evaluate $\Gamma_\chi$ in \eqref{tildePieq} in  section \eqref{tildePi}. 

\subsubsection{Generic cases}
For $|\p|\neq 0$,  $z$ is given by $z=a+b\, y+c\,x$ with $b\neq0\neq c$ and $I(x,y,z)$ can be written as 
\begin{eqnarray}
I(x,y,z) = - (1+b^2-2 b\,x)(y-y_+)(y-y_-),
\end{eqnarray}
where
\begin{eqnarray}
y_\pm= \frac{(x-b) (a+c\,x)\pm\sqrt{(1-x^2)\big(1+b^2-2 b\,x-(a+c\,x)^2\big)}}{1+b^2-2 b\,x}.
\end{eqnarray}
Let us find the conditions, under which there exists $-1<y_\pm<1$ assuming $|x|<1$.
The conditions that $y_\pm$ be real and that it satisfy $-1<y_\pm<1$ require 
\begin{eqnarray}
(a+c\,x)^2 &<& (1+b^2-2b\,x), \\
(x-b)^2 (a+c\,x)^2&<&(1+b^2-2b\,x)^2,
\end{eqnarray}
respectively. The second condition above is obtained\footnote{Since $I(y=\pm1)\leq0$, $y_-$ and $y_+$ must lie in $\{y_-, y_+<-1\}$ or $\{-1<y_-, y_+<1\}$ or $\{1<y_-, y_+\}$,  see figure \ref{Iintegral}. Among these possibilities,  only when $-1<y_-, y_+<1$, the integral does not vanish and only this case fulfills $-1<(y_-+y_+)/2<1$.} by requiring $-1<\frac{y_-+y_+}{2}<1$. 
Since $0<(x-b)^2<b^2-2b\,x+1$ for $|x|<1$, the second condition also follows from the first one with $|x|<1$. 
Thus the condition for $-1<y_\pm<1$ is 
\begin{equation}
(a+c\,x)^2 < (1+b^2-2b\,x) \quad{\rm and}\quad |x|<1.
\end{equation}  
The range of $x$ satisfying this condition is
\begin{eqnarray}
X\equiv (-1,1) \cap (x^-,x^+) 
\end{eqnarray}
where $x^\pm$ are solutions of $(a+c\,x)^2 =(1+b^2-2b\,x)$,
\begin{equation}
x^\pm= \frac{-b-ac \pm \sqrt{b^2+2 a b c + c^2 +b^2 c^2}}{c^2}, \label{x^pm}
\end{equation}
 and they are real when 
\begin{equation} \label{Rex}
b^2+2 a b c + c^2 +b^2 c^2>0.
\end{equation} 
Furthermore it follows from \eqref{abc1} that
\begin{eqnarray} \label{A-M}
b^2+2abc+c^2+b^2c^2=\frac{(A^2-M^2)b^2c^2}{|\p|^2}.
\end{eqnarray}
The right hand side is positive over the integral region in \eqref{D} and \eqref{S} since it has been chosen in such a way that $A^2>M_\chi^2$ holds, by using $\Omega^{\pm}$ for the integral limits. Thus $x^\pm$ are always real on the integral region. 
 

If $X=\emptyset$, then the angle integral vanishes. If $X\neq \emptyset$,  we define $x_-$ and $x_+$ as a lower and upper bound of $X$, respectively and then the angle integral can be performed as
\begin{eqnarray}
\mathcal{I}(a,b,c)&=&\int_{-1}^{1}dx \int_{-1}^1 dy\, \frac{\theta\left(I(x,y)\right)}{\sqrt{I(x,y)}} \\
&=& \int_{X}\frac{dx}{\sqrt{1+b^2-2b\,x}} \int_{y_-}^{y_+} \frac{dy}{\sqrt{-(y-y_+)(y-y_-)}} \\
&=& \int_{x_-}^{x_+}\frac{dx}{\sqrt{1+b^2-2b\,x}} \,\pi \\
&=&  \,\frac{\pi\left(\sqrt{1+b^2-2b\,x_-}-\sqrt{1+b^2-2b\,x_+}\right)}{b}. \label{int_b}
\end{eqnarray}
Here $x_-={\rm max}[-1, x^-]$ and $x_+={\rm min}[1, x^+]$ provided that $X\neq \emptyset$. 
Since $x^\pm$ are functions of $a$, $b$ and $c$, 
the condition $X\neq\emptyset$ can be reduced 
in terms of these parameters, which are functions of energy, momenta and masses of particles through \eqref{abc1}. In the following we consider on-shell cases and obtain analytical results for the angle integral.  

\subsubsection{On-shell cases} \label{onshell}
As a specific example, we apply the results obtained in the previous subsection to the diagram given in figure \ref{fig:diagsig4}, when the external four-momentum is on-shell (i.e. $p_0= \Omega_{\p}= \sqrt{\p^2+M_\chi^2}$).
By introducing $\gamma\equiv \frac{M_\chi}{|\p|}$ the parameter  $a$ can be expressed as (using \eqref{abc1} and \eqref{A})
\begin{equation}
a= \gamma^2 bc+\sqrt{(1+\gamma^2b^2)(1+\gamma^2c^2)}- \sqrt{1+\gamma^2}\left(c\sqrt{1+\gamma^2 b^2}+b\sqrt{1+\gamma^2c^2}\right),
\end{equation}
which has properties 
\begin{equation}
a(b=1)=-c,\quad a(c=1)=-b.
\end{equation}
Using the second property above in \eqref{x^pm}, we get
\begin{equation}
x^\pm(c=1)=\pm1
\end{equation} 
and one can show that 
\begin{eqnarray}
&& |x^\pm| >1  \quad {\rm if }\quad c<1,\\
&& |x^\pm| < 1 \quad {\rm if} \quad c >1. 
\end{eqnarray}
Therefore
\begin{eqnarray}
X=(-1,1)\cap(x^-,x^+) =(x_-,x_+)=
\begin{cases} (-1,1) & \text{if } c \leq 1,\\ (x^-,x^+) & \text{if }  c > 1.
\end{cases}
\end{eqnarray}
Then in \eqref{int_b} we have 
\begin{equation}
\sqrt{1+b^2-2b\, x_\pm}=
\begin{cases}  |b\mp1| & \text{if } c \leq 1,\\ |a+c x^\pm| & \text{if }  c > 1.
\end{cases}
\end{equation}
In the second line on the right hand side we have used the fact that $x^\pm$ solve $1+b^2-2b\, x= (a+c\,x)^2$.
Noting that  $a+c x^- <0$ and using properties for  $c>1$,
\begin{eqnarray}
a+ c x^+ \begin{cases}  < 0 & \text{if }  b > 1  \\ = 0 & \text{if }  b = 1  \\ > 0 &\text{if }  b < 1,
\end{cases}
\end{eqnarray}
we obtain the final expression for the angle integral \eqref{int_b}, 
\begin{eqnarray}
\mathcal{I}
&=&  \,\frac{\pi\left(\sqrt{1+b^2-2b\,x_-}-\sqrt{1+b^2-2b\,x_+}\right)}{b}  \nonumber \\
&=& 
\begin{cases}  2\pi & \text{if } b\leq 1   \text{ and } c\leq1  \\ 
 2\pi\, b^{-1} & \text{if } b > 1   \text{ and } c\leq1  \\ 
2\pi\, c^{-1} & \text{if } b\leq 1   \text{ and } c >1 \\ 
   2\pi \frac{\sqrt{A^2-M_\chi^2}}{|\p|} & \text{if } b> 1   \text{ and } c> 1.  \\ 
\end{cases}
\end{eqnarray} 
In the last line we have used \eqref{A-M} and assumed\footnote{As mentioned before, $A\geq M_\chi$ holds over the integral region on $\Omega_{\k}$-$\Omega_{\l}$ plane in \eqref{S} by imposing the integral limit $\Omega^{+}$ } $A\geq M_\chi$ (i.e. $\Omega_{\k}+\Omega_{\l}>\Omega_{\p}+M_\chi$ in  \eqref{S}). 
In terms of energies and momenta, the result above reads
\begin{eqnarray} \label{I-sigma}
\mathcal{I}
&=& 
\begin{cases}  2\pi & \text{if } \Omega_{\p}\leq \Omega_{\k}, \Omega_{ \l }  \\ 
 2\pi \frac{ |\k|}{|\p|} & \text{if }  \Omega_{\k}  < \Omega_{\p}\leq \Omega_{\l}  \\ 
2\pi \frac{ |\l|}{|\p|} & \text{if }  \Omega_{\l}< \Omega_{\p}\leq \Omega_{\k}    \\ 
   2\pi \frac{\sqrt{\left(\Omega_{\k}+ \Omega_{\l} -\Omega_{\p}\right)^2-M_\chi^2}}{|\p|} & \text{if }  \Omega_{\k}, \Omega_{\l}< \Omega_{\p} \text{ and }   
  M_\chi \leq \Omega_{\k}+ \Omega_{\l} -\Omega_{\p}\\ 
   0 & \text{ otherwise}. 
\end{cases}
\end{eqnarray}

\end{appendix}

\bibliographystyle{JHEP}

\bibliography{all}

\end{document}